\documentclass[twoside,twocolumn,9pt]{article}

\usepackage[T1]{fontenc}
\usepackage[english]{babel}

\usepackage{extsizes}
\usepackage[super,sort&compress,comma]{natbib} 
\usepackage[version=3]{mhchem}
\usepackage[left=1.5cm, right=1.5cm, top=1.785cm, bottom=2.0cm]{geometry}
\usepackage{balance}
\usepackage{mathptmx}
\usepackage{sectsty}
\usepackage{graphicx} 
\usepackage{lastpage}
\usepackage[format=plain,justification=justified,singlelinecheck=false,font={stretch=1.125,small,sf},labelfont=bf,labelsep=space]{caption}
\usepackage{float}
\usepackage{fancyhdr}
\usepackage{fnpos}

\addto{\captionsenglish}{%
  
}
\usepackage{array}
\usepackage{droidsans}
\usepackage{charter}
\usepackage{multirow}

\usepackage[usenames,dvipsnames]{xcolor}
\usepackage{setspace}
\usepackage[compact]{titlesec}
\usepackage{hyperref}
\usepackage{comment}
\usepackage{bm}
\usepackage{adjustbox}

\definecolor{cream}{RGB}{222,217,201}

\fancypagestyle{plain}{

}

\makeFNbottom
\makeatletter
\renewcommand\LARGE{\@setfontsize\LARGE{15pt}{17}}
\renewcommand\Large{\@setfontsize\Large{12pt}{14}}
\renewcommand\large{\@setfontsize\large{10pt}{12}}
\renewcommand\footnotesize{\@setfontsize\footnotesize{7pt}{10}}
\makeatother

\makeatletter 
\renewcommand\@biblabel[1]{#1}            
\renewcommand\@makefntext[1]%
{\noindent\makebox[0pt][r]{\@thefnmark\,}#1}
\makeatother 
\renewcommand{\figurename}{\small{Fig.}~}
\sectionfont{\sffamily\Large}
\subsectionfont{\normalsize}
\subsubsectionfont{\bf}
\titlespacing*{\section}{0pt}{4pt}{4pt}
\titlespacing*{\subsection}{0pt}{15pt}{1pt}

\makeatletter 
\newlength{\figrulesep} 
\newcommand{\topfigrule}{\vspace*{-1pt}%
\noindent{\color{cream}\rule[-\figrulesep]{\columnwidth}{1.5pt}} }

\newcommand{\botfigrule}{\vspace*{-2pt}%
\noindent{\color{cream}\rule[\figrulesep]{\columnwidth}{1.5pt}} }

\newcommand{\dblfigrule}{\vspace*{-1pt}%
\noindent{\color{cream}\rule[-\figrulesep]{\textwidth}{1.5pt}} }

\makeatother

\begin{document}

\twocolumn[
\begin{@twocolumnfalse}
{\includegraphics[height=30pt]{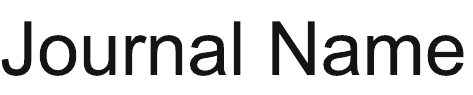}\hfill\raisebox{0pt}[0pt][0pt]{\includegraphics[height=55pt]{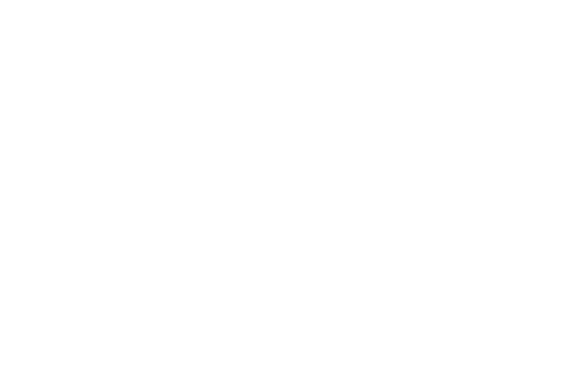}}\\[1ex]
\includegraphics[width=18.5cm]{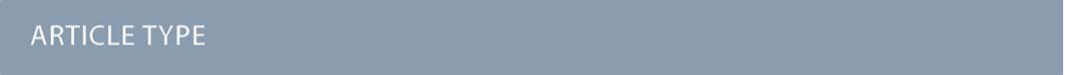}}\par
\vspace{1em}
\sffamily
\begin{tabular}{m{4.5cm} p{13.5cm} }
\includegraphics{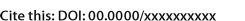} & \noindent\LARGE{\textbf{Excited-state nonadiabatic dynamics in explicit solvent using machine learned interatomic potentials$^\dag$}} \\
\vspace{0.3cm} & \vspace{0.3cm} \\
 & \noindent\large{Maximilian X. Tiefenbacher,\textit{$^{a,b}$} Brigitta Bachmair,\textit{$^{a,b,c}$} Cheng Giuseppe Chen,\textit{$^{c,d}$}  Julia Westermayr,\textit{$^{e,f}$} Philipp Marquetand,\textit{$^{a,c}$} Johannes C. B. Dietschreit,\textit{$^{\ddagger c}$} Leticia González\textit{$^{\ast a,c}$}} \\
\includegraphics{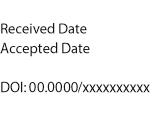} & \noindent\normalsize{%
Excited-state nonadiabatic simulations with quantum mechanics/molecular mechanics (QM/MM) are essential to understand photoinduced processes in  explicit environments. 
However, the high computational cost of the underlying quantum chemical calculations limits its application in combination with trajectory surface hopping methods.
Here, we use FieldSchNet, 
a machine-learned interatomic potential capable of incorporating electric field effects into the electronic states, to replace traditional QM/MM electrostatic embedding  with its ML/MM counterpart for nonadiabatic excited state trajectories.
The developed method is applied to furan in water, including five coupled singlet states.
Our results demonstrate that with sufficiently curated training data, the ML/MM model reproduces the electronic kinetics and structural rearrangements of QM/MM surface hopping reference simulations. 
Furthermore, we identify performance metrics that provide robust and interpretable validation of model accuracy. 
}
\end{tabular}
\end{@twocolumnfalse} 
\vspace{0.6cm}
]

\renewcommand*\rmdefault{bch}\normalfont\upshape
\rmfamily
\section*{}
\vspace{-1cm}


\footnotetext{\textit{$^{a}$~Research Platform on Accelerating Photoreaction Discovery (ViRAPID), University of Vienna, Währinger Straße 17, 1090 Vienna, Austria.}}
\footnotetext{\textit{$^{b}$~Vienna Doctoral School in Chemistry, University of Vienna, Währinger Straße 42, 1090 Vienna, Austria.}}
\footnotetext{\textit{$^{c}$~Institute of Theoretical Chemistry, Faculty of Chemistry, University of Vienna, Währinger Straße 17, 1090 Vienna, Austria.}}
\footnotetext{\textit{$^{d}$~Department of Chemistry, Sapienza University of Rome, Piazzale Aldo Moro, 5, Rome, 00185, Italy.}}
\footnotetext{\textit{$^{e}$~Wilhelm-Ostwald Institute, University of Leipzig, Linnéstraße 2, 04103 Leipzig.}}
\footnotetext{\textit{$^{f}$~Center for Scalable Data Analytics and Artificial Intelligence (ScaDS.AI), Dresden/Leipzig, Humboldtstraße 25, 04105 Leipzig, Germany.}
}
\footnotetext{\textit{$^{\ddagger}$~\url{johannes.dietschreit@univie.ac.at}}}
\footnotetext{\textit{$^{\ast}$~\url{leticia.gonzalez@univie.ac.at}}}

\footnotetext{\dag~Electronic Supplementary Information (ESI) available: [details of any supplementary information available should be included here]. See DOI: 00.0000/00000000.}



\section{Introduction}

Photochemistry lies at the heart of essential natural processes. One example is photosynthesis, where the absorption of light by chlorophyll triggers a cascade of electron transfers, ultimately capturing solar energy and storing it in chemical bonds.\cite{Ort1996, Nugent1996, Stirbet2019}
Beyond sustaining life, light-driven reactions hold promise for advancements in energy conversion,\cite{Devens2009} molecular electronics\cite{Huang2020}, and the development of photonic materials.\cite{Butt2021} 
Understanding photochemical mechanisms is thus important to inspire new technologies that harness the power of light. 
To this end, computational approaches play a key role, offering time-resolved insights that complement and enhance modern ultrafast experimental techniques.\cite{Mai2020,Lindh2020} 

A popular method to carry out excited-state dynamics simulations of molecular systems is trajectory surface hopping (TSH).\cite{Tully1990}
In this so-called mixed quantum-classical approach, electrons --responsible for electronic transitions and excited-state properties-- are treated quantum mechanically, while the heavier nuclei are described using classical mechanics.
At the expense of neglecting nuclear quantum effects, TSH effectively captures the quantum nature of electrons, even if it requires propagating numerous independent classical nuclear trajectories to accurately simulate the behavior of a nuclear wave packet as it splits during non-adiabatic events.\cite{Barbatti2011,Mai2018,Mai2020}  
Despite being attractive, the need for many trajectories makes TSH simulations computationally expensive, especially when the underlying on-the-fly calculations of the coupled potential energy surfaces (PESs) are performed using accurate quantum mechanical methods.

The situation becomes more challenging when simulating photochemical processes in the condensed phase.\cite{Nogueira2018}
For example, chromophores in nature are rarely isolated; rather, they are typically embedded within complex, heterogeneous environments that include a variety of molecular interactions, including solvent effects, hydrogen bonding, and intricate structural features. 
These environmental factors can significantly influence the photochemical behavior of the system, altering deactivation pathways. 
Implicit solvation models,\cite{Decherchi2015} which treat the solvent as a continuous medium, are often insufficient to capture detailed interactions.  
A more accurate treatment is achieved by  modeling the environment explicitly. 
One efficient way to do that is by using hybrid quantum mechanics/molecular mechanics (QM/MM) methods, where the region of interest, such as the excited chromophore, is treated with QM, while the surrounding environment, including solvent or complex heterogeneous scaffold, is modeled with MM.\cite{QMMMreview, Groenhof2012,Hai2024} 
This approach offers a balance between accuracy and computational cost by restricting expensive QM calculations to the critical region of the system that actually requires a quantum mechanical treatment. 
The advantages of QM/MM strategies are indisputable, and in combination with TSH, have enabled impressive simulations of time-resolved photochemical processes in large complex systems.\cite{Brunk2015, Schnedermann2018, Yang2022, Bondanza2022, Toldo2023, Coffman2023}
However, because the computational expense of a QM/MM calculation is largely determined by the level of theory employed in the QM region, QM/MM dynamical simulations suffer from comparable (or greater) costs than simulations of the isolated chromophore. 
The situation becomes more prohibitive the more trajectories are needed to achieve statistically meaningful results.\cite{Brunk2015} 

Over the years, several strategies have been developed to reduce the cost of TSH simulations.
One notable example is the use of vibronic coupling models,\cite{Marian2012} which replace the expensive on-the-fly calculations of the PESs by pre-parameterized potentials that are approximated by scaled harmonic oscillators in the simplest case.\cite{Plasser2019,Zobel2021b} 
Recently, TSH simulations using linear vibronic coupling PESs have been extended to include a classical MM environment,\cite{Polonius2024} enabling efficient time-resolved analysis of three-dimensional solvent-solute interactions.\cite{Polonius2024a}
While this approach reduces the computational cost significantly, it is only applicable to rather rigid molecules, where anharmonic effects, such as large amplitude motions or bond rearrangements, do not play a role in the relaxation dynamics.

An alternative and highly flexible approach to reduce the cost of the underlying electronic structure problem is the use of machine learned (ML) potentials.\cite{Dral2020,Noé2020,Keith2021,Unke2021}
Trained on high-quality quantum mechanical data, ML potentials have demonstrated their ability to replicate the accuracy of \textit{ab initio} calculations at a fraction of the computational cost.\cite{Yinuo2024}
ML potentials have already shown considerable success in modeling dynamics in the electronic ground-state.\cite{Behler2017, Yang2021, Guan2021,Dong2024,batatia2024foundationmodelatomisticmaterials,Unke2024}
In contrast, their application to excited-state dynamics, which is significantly more expensive than ground state simulations, is currently only feasible for small molecular systems like organic chromophores and is limited by the availability of accurate reference data.\cite{Dral2018, Westermayr2019, Westermayr2020a, Westermayr2020c, Dral2021,Westermayer2022, Axelrod2022}  
This is in contrast to ground-state ML potentials, of which many are transferable between molecular systems and can thus simulate large biomolecules or materials using training data of smaller building blocks. 
As a consequence, the simulation of large systems in their excited state also requires methods like mixed ML/MM (machine learning/molecular mechanics).\cite{Westermayr2020rev}
However, the integration of ML potentials with MM for both ground- and excited-state dynamics remains in its infancy.

One of the key challenges in developing an ML/MM approach is to accurately describe the interaction between the ML potential and the surrounding MM environment. 
In QM/MM simulations, the interactions between the quantum region and the classical region are clearly defined by the Hamiltonian, which ensures that the treatment of the two regions is physically correct.\cite{QMMMreview} 
For ML/MM, the interaction needs to be carefully modeled to ensure that the combined system behaves correctly; however, there appears to be little consensus on the best way to do so.\cite{Yao2018,Smith2019,Ko2021,BURNN,Gastegger2021,Hofstetter2022,Kirill2023,Thrlemann2023,Kirill2024,Semelak2024,Grassano2024,Lei2024}

\begin{figure}[tb]
\centering
\includegraphics[width=\linewidth]{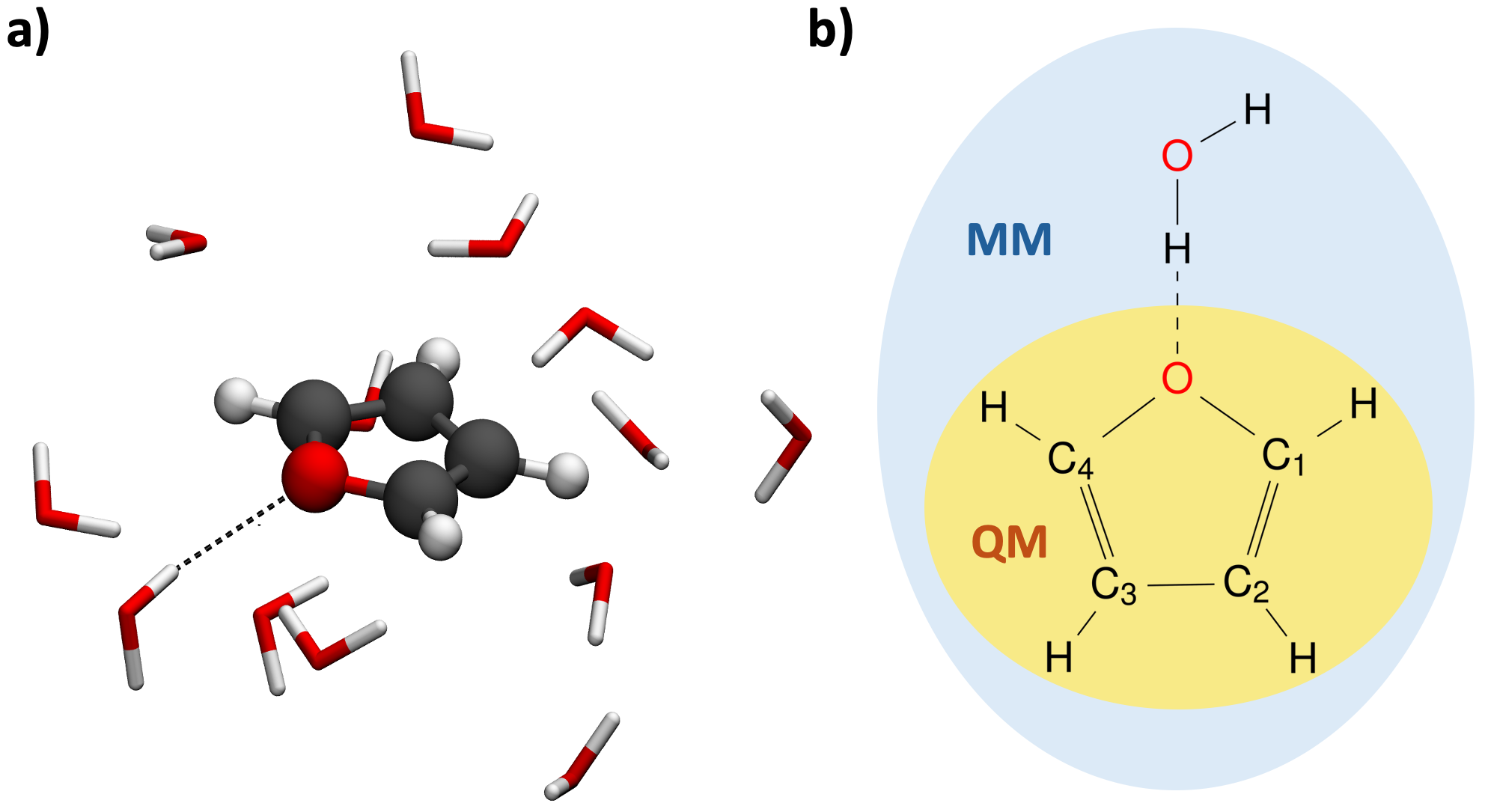}
    \caption{
    Depiction of the investigated system, furan, treated quantum mechanically (QM), solvated in water, which is described with molecular mechanics (MM). 
    a) 3D rendering of furan, depicted as balls and sticks, surrounded by the water molecules of the first solvation shell, shown as sticks. 
    The hydrogen bond present in this configuration is highlighted by a black dashed line.
    Hydrogen, carbon, and oxygen atoms are colored in white, black, and red, respectively.
    b) Lewis structure of furan with numbered carbon atoms. 
    }  
    \label{fig:system}
\end{figure}

A recent study by Mazzeo et al.\cite{Mazzeo2024} demonstrates the use of Gaussian process regression to describe the excited-state dynamics of a solvated molecule by learning ML/MM energies and forces with kernel models in a two-step process. 
First, they fit the vacuum PESs and then subsume the differences between pure QM and QM/MM under polarization, which is described by a second model.
Their approach is restricted to purely adiabatic dynamics in the excited state, neglecting any coupling between states.
In contrast, to the best of our knowledge, ML/MM implementations for nonadiabatic excited-state dynamics using an electrostatic embedding do not exist. 
In this work, we propose the first ML/MM nonadiabatic excited-state dynamics using electrostatic embedding and a general number of electronic states. 
We use the FieldSchNet architecture of Gastegger et al.,\cite{Gastegger2021} which allows the inclusion of the electric field via an additional ML input.
The electric field is generated by point charges of the MM environment and alters the different excited states.
As an application, we investigate the excited-state dynamics of furan in water (Figure~\ref{fig:system}a), including three coupled electronic singlet states.
Furan is a small heterocyclic organic molecule that serves as a building block in biologically relevant systems, such as DNA and proteins, and has long been the focus of theoretical and experimental studies.\cite{Roebber1980, Pastore2006, Gavrilov2008, Fuji2010, Stenrup2011, Gromov2011, Gromov2013, Spesyvtsev2015, Oesterling2017} 
Few also considered the explicit interaction of furan with water forming hydrogen bonds,\cite{Brupbacher1998, Lockwood2018} highlighting the need of QM/MM studies able to capture these interactions and their impact on the excited state relaxation dynamics explicitly.

The remainder of the paper is organized as follows.
In Section~2, we present the theory behind a QM/MM setup with electrostatic embedding and describe all the necessary terms when using an ML model.
Next, in Section~3 we outline the data collection method, architecture, and training process of the ML interatomic potentials and summarize the numerical experiments performed.
The results of various training settings, the quality of the obtained PESs, and the ML-driven dynamics are compared to on-the-fly TSH using the quantum chemical reference method used to train our ML potentials in Section~4.

\section{Theory}

\subsection{Electrostatic Embedding}

In QM/MM simulations, the most relevant region of the chemical system is described with quantum mechanics (QM region), while the surroundings are modeled with computationally more efficient classical force fields (MM region).
The total Hamiltonian $H_\text{tot}$ and total energy $E_\text{tot}$ of the system within an electrostatic embedding framework can be expressed as,\cite{QMMMreview}
\begin{align}
    H_\text{tot} & = H_\text{MM} + H_\text{QM} + H_\text{QM-MM}, \label{eq:QMMM_general} \\
    E_\text{tot} & = 
    E_{MM} + E^\text{vdW}_\text{QM-MM}
    + \underbrace{E_\text{QM} + E_{\text{QM-MM}}^{\text{Coulomb}}}_{E_\text{QM}^\text{el.-embed.}}, \label{eq:QMMM_elecembed} 
\end{align}
where the indices MM and QM represent the contributions of each region individually, and QM-MM indicates terms that depend on both regions.
Assuming that the border between the QM and MM regions does not cut through any chemical bonds, the only interactions between the QM and MM regions are the van der Waals (vdW) and Coulomb potentials (eq.~(\ref{eq:QMMM_elecembed})).
In the context of electrostatic embedding, the vdW interactions are computed at the level of the MM region, while the Coulomb interactions are represented by an additional term in the Hamiltonian of the QM region.\cite{Bakowies1996}
The term $E_{\text{QM-MM}}^{\text{Coulomb}}$ includes the influence of the surrounding classical region through electrostatic interactions in the description of the QM region.
The partial atomic charges of the classical environment $\{q^\text{MM}\}$ located at $\{\mathbf{R}^\text{MM}\}$ generate an electrostatic potential at position $\mathbf{r}$:
\begin{equation}
    V_{\text{MM}}(\mathbf{r}) = \sum_j \frac{q_j^\text{MM}}{|\mathbf{r} - \mathbf{R}^\text{MM}_j|} 
    \label{eq:electrostatic_pot}
\end{equation}
Equation~\ref{eq:electrostatic_pot} is part of $H_\text{QM-MM}$ and is included in the calculation of the quantum subsystem, thus coupling the QM and MM regions.\cite{QMMMreview,Kirill2023}

\subsection{Adapted Machine-Learned Gradients for Electrostatic Embedding}

The QM calculation is set to find the wave function, \textit{i.e.}, the electron density, corresponding to the given Hamiltonian, which in the case of electrostatic embedding includes the electrostatic potential defined in eq.~(\ref{eq:electrostatic_pot}).\cite{QMMMreview}
Hence, as implied by eq.~(\ref{eq:QMMM_elecembed}), the energy and subsequently also the forces computed for the electrostatic embedding Hamiltonian cannot be separated into the vacuum and the MM polarization contributions without additional QM calculations of the system in vacuum.
This is important, because it means that the energy (and forces) of the same arrangement of atoms in the QM region will vary depending on the configuration of the MM region.
Therefore, to perform electrostatic embedding simulations using an ML interatomic potential, it is essential to model the QM region and its interaction with the surrounding MM region simultaneously. 
This can be accomplished by either passing the positions and charges of the MM atoms to the ML architecture, $\hat{E}_\text{QM}^\text{el.-embed.} = E_\text{ML}(\{\mathbf{R}^\text{QM}\}, \{\mathbf{R}^\text{MM}\}, \{q^\text{MM}\})$, or by considering the electric field $\bm{\epsilon}$ created by the surrounding MM atoms at the position of each QM atom, $\hat{E}_\text{QM}^\text{el.-embed.} = E_\text{ML}(\{\mathbf{R}^\text{QM}\}, \{\bm{\epsilon}\})$. 
The latter approach offers the benefit of obviating the need to provide the positions of all MM atoms to the model. 
Instead, one can conveniently use precomputed electric field values, substantially minimizing the data set file sizes.
In this work, we use FieldSchNet,\cite{Gastegger2021} one of the first ML interatomic potentials capable of reproducing QM/MM calculations with electrostatic embedding by using the electric field as an additional input besides the atomic positions. 
FieldSchNet is based on the SchNet continuous-filter convolutional neural\cite{Schutt2018} network architecture that takes the electric field as an additional input to learn a representation of the system.

The electric field $\bm{\epsilon}_{i}$ at position $\mathbf{R}^\text{QM}_i$ of QM atom $i$ is the sum over all atoms in the MM region ($N_\text{MM}$) and is defined as 
\begin{equation}
     \boldsymbol{\epsilon}_{i} = \sum_{j}^{N_\text{MM}} \boldsymbol{\epsilon_{ij}} = \sum_{j}^{N_\text{MM}} q_j^\text{MM} \frac{ \mathbf{R}^\text{QM}_i-\mathbf{R}^\text{MM}_j}{|\mathbf{R}^\text{QM}_i-\mathbf{R}^\text{MM}_j|^3} \ ,
     \label{equ:field}
\end{equation}
where $q_j^\text{MM}$ is the partial charge of MM atom $j$ with coordinates $\mathbf{R}^\text{MM}_j$. 
Equation~(\ref{equ:field}) highlights that the electric field depends on both the QM and MM atoms.\cite{Gastegger2021}
Hence, energies that depend on this field induce forces that act on both the QM and MM atoms. 
However, if only the electric field value is passed to the ML model and not the positions and charges of every MM atom, the gradient of the ML energy with respect to the position of ML (or QM) atoms does not include the field-dependent terms. 
To circumvent this problem, we add the respective derivative to those forces that are computed via automatic differentiation. 
The forces applied to the ML region are then given by
\begin{equation}
    -\frac{\mathrm{d} E_\text{ML}}{\mathrm{d}\mathbf{R}^\text{ML}} = - \left(\frac{\partial E_\text{ML}}{\partial\mathbf{R}^\text{ML}} + \frac{\partial E_\text{ML}}{\partial\bm{\epsilon}} \frac{\partial \bm{\epsilon}}{\partial\mathbf{R}^\text{ML}} \right) \ ,
    \label{eq:QMderiv_EML}
\end{equation}
whereas the contribution of the ML region to the forces acting on the MM atoms is
\begin{equation}
    - \frac{\mathrm{d} E_\text{ML}}{\mathrm{d}\mathbf{R}^\text{MM}} = - \frac{\partial E_\text{ML}}{\partial\bm{\epsilon}} \frac{\partial \bm{\epsilon}}{\partial\mathbf{R}^\text{MM}}.
    \label{eq:MMderiv_EML}
\end{equation}
The partial derivatives in eqs.~(\ref{eq:QMderiv_EML}) and~(\ref{eq:MMderiv_EML}) of the ML energy with respect to the ML atoms and the field are computed via back-propagation. 
The derivatives of the field with respect to the nuclear positions (MM or ML) are obtained analytically, based on eq.~(\ref{equ:field}).

%

\subsection{Augmented Loss for Training ML Interatomic Potentials}
When training ML interatomic potentials on QM data, the loss function is usually a linear combination of the mean squared error for energies and forces.
Accordingly, the loss function for a single configuration and electronic state can be given by
\begin{align}
    \mathcal{L} & = w_\text{E} \left(E - \hat{E}\right)^2 + \frac{w_\text{F}}{3N_\text{ML}} \sum_{i=1}^{N_\text{ML}}\sum_{j \in xyz}\left(F_{i,j} - \hat{F}_{i,j}\right)^2 
    \nonumber \\
    & = w_\text{E} \left(E - \hat{E}\right)^2 + \frac{w_\text{F}}{3N_\text{ML}} \sum_{i=1}^{N_\text{ML}}\sum_{j \in xyz}\left(F_{i,j} +  \left[\frac{\partial \hat{E}}{\partial\mathbf{R}}\right]_{i,j}\right)^2 
    \ ,
    \label{eq:old_loss}
\end{align}
where $N_\text{ML}$ is the number of ML atoms in a configuration, $w_\alpha$ is the weight for the loss of property $\alpha$, and $\hat{\ }$ indicates the prediction of the ML model.
However, since the ML prediction does not include the field-dependent gradients, we amend the loss to be
\begin{align}
    \mathcal{L}
    & = w_\text{E} \left(E_\text{QM}^\text{el.-embed.} - \hat{E}_\text{ML}\right)^2 \nonumber \\
    & \quad + \frac{w_\text{F}}{3N_\text{ML}} \sum_{i=1}^{N_\text{ML}}\sum_{j \in xyz}\left(F_{i,j}  + \left[\frac{\partial \hat{E}_\text{ML}}{\partial\mathbf{R}^\text{ML}} + \frac{\partial \hat{E}_\text{ML}}{\partial\bm{\epsilon}} \frac{\partial \bm{\epsilon}}{\partial\mathbf{R}^\text{ML}}\right]_{i,j}\right)^2 
    \ .
    \label{eq:new_loss}
\end{align}
Since the contribution of the field to the nuclear gradient is expected to be small, this term was neglected in the original FieldSchNet paper.\cite{Gastegger2021}
We call eq.~(\ref{eq:new_loss}) the "augmented loss" and use it for all ML trainings in this work, to be consistent with the forces used for dynamics simulations, as we found the field-dependent term to be crucial for stable excited-state molecular dynamics simulations.

\subsection{Excited-State Dynamics Simulations using Trajectory Surface Hopping} 

The excited-state dynamics are carried out in the context of TSH,~\cite{Tully1971} 
where the nuclei are propagated classically on a single PES at each time step, solving Newton's equations of motion.
Still, TSH couples several PESs by allowing stochastic transitions (or "hops") between electronic states, which are described by the time-dependent Schrödinger equation,
\begin{equation}
\frac{d}{dt} c_k(t) = - \sum_\ell \left[ \frac{i}{\hbar} H_{k\ell} +   \mathbf{d}_{k\ell} \cdot \mathbf{v} \right]  c_\ell(t).
\end{equation}
Here, $c_k(t)$ is the time-dependent coefficient of the electronic wavefunction for state $k$ and $H_{k\ell}$ is an element of the electronic Hamiltonian matrix in the molecular Coulomb Hamiltonian basis, where the electronic states are ordered by energy and can change their character.\cite{Richter2011,Mai2015}  
In this representation, $\mathbf{d}_{k\ell} = \langle \phi_k | \nabla_{\mathbf{R}} | \phi_\ell \rangle$ is the nonadiabatic coupling vector between the states $k$ and $\ell$, and $\mathbf{v}$ is the velocity vector of the nuclei.
The explicit calculation of the nonadiabatic couplings is often circumvented by using the local diabatization scheme,~\cite{Granucci2001, Plasser2012} which relies on calculating the overlap matrix between two sequential wavefunctions.\cite{Plasser2012,Plasser2016} 
Alternatively, when neither explicit nonadiabatic coupling vectors nor wavefunctions are available ---as is the case when using an adiabatic machine learning model--- it is possible to approximate the coupling vectors using the time derivative of the nuclear forces, as in the so-called curvature-driven Tully surface hopping.\cite{Shu2022}

The probability of hopping between the electronic states is determined by the fewest switches algorithm,\cite{Tully1990} which minimizes the number of state transitions. 
The transition probability between two states $n$ and $m$ is given by
\begin{equation}
P_{n \to m} = \max \left(0, \frac{2 \, \text{Re}\left\{ c_n^* c_m \left[ \frac{i}{\hbar} H_{nm} +  \mathbf{d}_{nm} \cdot \mathbf{v} \right] \right\}}{|c_n|^2} \Delta t \right),
\end{equation}
where $\Delta t$ is the nuclear time step.
After each time step, a uniform random number $\xi \in [0,1]$ is drawn. 
If $\xi < P_{n \to m}$, the system attempts to switch from state $n$ to state $m$. 
If a hop occurs, the nuclear kinetic energy is adjusted to conserve total energy.\cite{Jasper2003}
However, if the kinetic energy after a hop is insufficient to sustain the motion, the hop is rejected, and the system remains on the original PES.\cite{Jasper2001}

\section{Computational Details}

\subsection{Molecular Dynamics Simulations}

Initial molecular dynamics (MD) simulations of furan in water were carried out with AMBER2022.\cite{Case2023}
Partial charges for furan were calculated with antechamber using bond charge corrected charges derived from the semi-empirical Austin Model 1 (AM1-BCC charges),\cite{Dewar1985, Jakalian2002} which were combined with GAFF2 parameters using parmchk.
The furan molecule was solvated in a cubic box of TIP3P water molecules with a side length of 15~\AA~containing 1365 water molecules. 

\begin{figure}
    \centering
    \includegraphics[width=\linewidth]{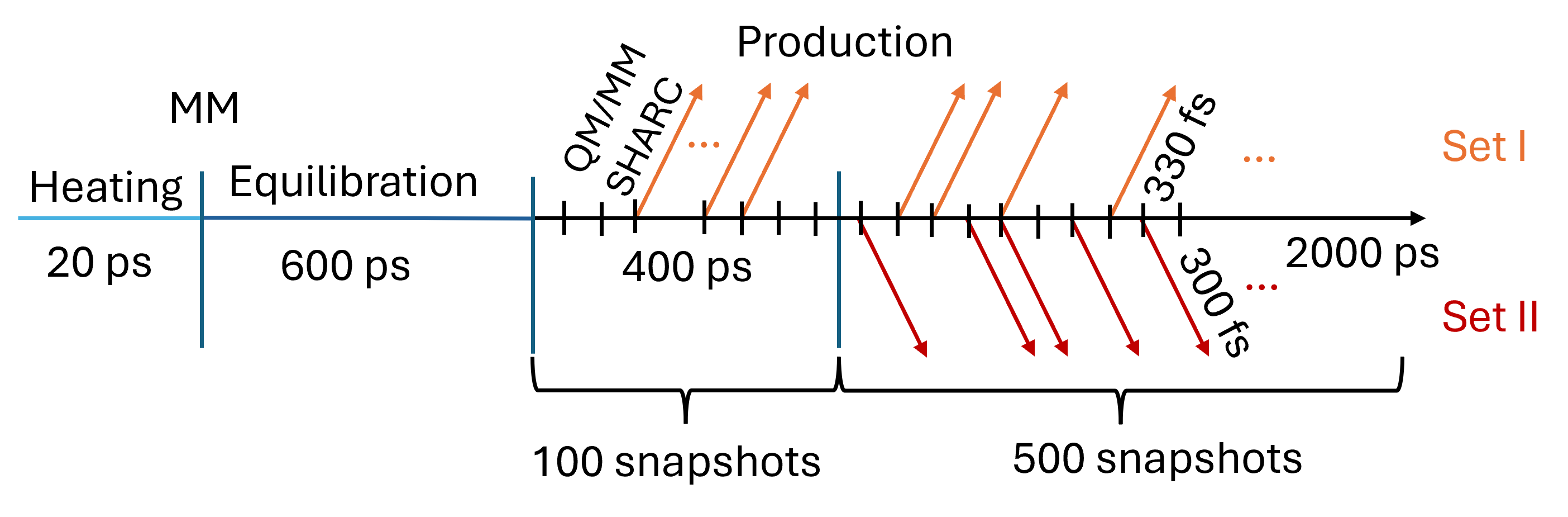}
    \caption{Timeline of the different molecular dynamics simulations performed for furan in water. 
    Molecular mechanics (MM) is first used for heating (20 ps) and equilibration (600 ps) steps. 
    The production run is divided into two parts, 400~ps and 2000~ps long.
    Frames are saved every 4~ps, resulting in 100 and 500 snapshots, respectively, which serve as potential initial conditions for two sets of QM/MM-SHARC trajectories. 
    The number of selected trajectories (colored arrows) depends on whether an excitation can occur within a specific energy window (7.3 to 7.5~eV for set I and 6 to 7~eV for set II.).
    For furan in water, this results in 46 (set I, out of 600 initial conditions) and 66 (set II, out of 500 initial conditions) trajectories.
    }
    \label{fig:simulation_timeline}
\end{figure}

For energy minimization, heating, equilibration, and production runs, we used the MD engine sander.\cite{Case2023} 
Figure~\ref{fig:simulation_timeline} shows a schematic timeline for the MM-MD simulations and how snapshots for the subsequent excited-state QM/MM simulations were collected.

The MM-MD simulations employ a time step of 2~fs, a Langevin thermostat with a friction constant of 2~ps$^{-1}$, and constrained hydrogen-heavy atom bond distances by means of the SHAKE\cite{Ryckaert1977} algorithm.
After the initial setup, the energy of the system was minimized for 2000 steps using steepest descent.
After minimization, the system was heated for 20~ps to 300~K through continuous heat transfer from the thermostat (the bath temperature is always at 300~K). 
Subsequently, the system was equilibrated for 600~ps in the isobaric-isothermal ensemble with the Berendsen barostat.\cite{Berendsen1984}
Using the same settings as for the equilibration, we performed a production run with a total of 2.4~ns.
Snapshots were recorded every 4~ps, resulting in a total of 600 equidistant frames, which were used to build two sets of initial conditions (sets~I and~II in Figure \ref{fig:simulation_timeline}) for subsequent excited-state QM/MM simulations -- explained next. 

\subsection{QM/MM Trajectory Surface Hopping Simulations} \label{sec:qm/mmsurfacehopping}

Two sets of nonadiabatic QM/MM-TSH simulations were performed using the QM/MM interface of the SHARC 3.0\cite{Richter2011,Mai2015,Mai2018} package to the TINKER MM engine.\cite{Rackers2018} 
Set~I was used for model training, validation, and testing. 
Set~II was used to provide reference QM/MM-TSH simulations to compare with ML/MM-TSH dynamics carried out under the exact same conditions (see details below). 

The computational details specified here are common to both sets, unless stated otherwise. 
The interactions of QM and MM regions were described via electrostatic embedding.
RATTLE\cite{Andersen1983} was used to constrain the bond vibrations of the water molecules of the MM region, whereas the hydrogen atoms of furan were left unconstrained. 
Furan was described using time-dependent density functional theory (TD-DFT) at the BP86/def2-SVP\cite{Becke1988,Perdew1986,Weigend2005,Weigend2006} level of theory, as implemented in Orca~5.0.\cite{Nesse2020}
This level of theory was chosen based on a benchmark performed by Hieringer et al.,~\cite{Hieringer2002} in which a total of 12 different methods were compared, including other DFT functionals, and wave function methods such as CASPT2 or CC3.
BP86 showed the best agreement with the experimental excitation energies from Kamada et al.\cite{Kamada1998}

\begin{figure}[ht]
    \centering
    \includegraphics[width=0.8\linewidth]{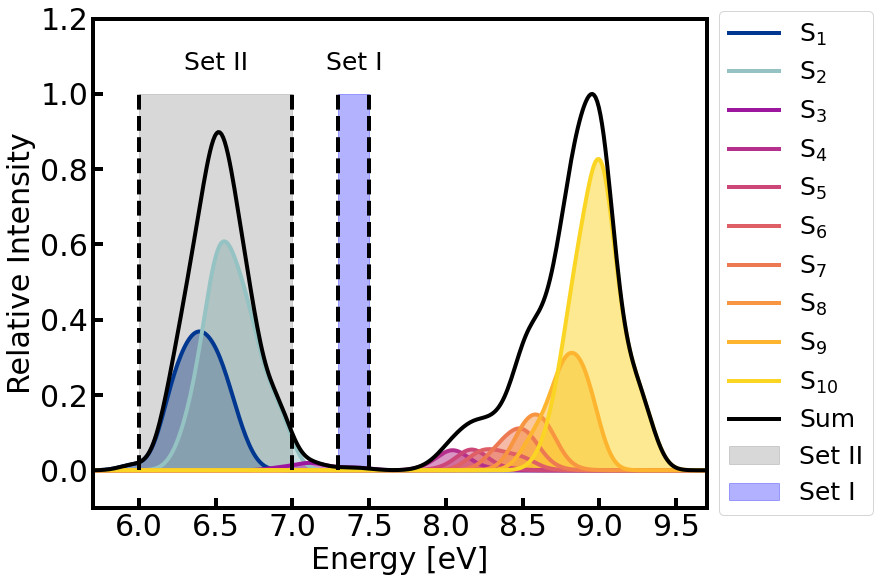}
    \caption{Absorption spectrum of furan in water (solid black line) calculated at BP86/def2-SVP level of theory from the first 100 snapshots of the MM-MD production run. 
    Contributions from different states are indicated by colors as indicated. 
    Energies and oscillator strengths were convoluted  with a Gaussian with a full width at half-maximum of 0.2~eV. 
    Shaded vertical areas denote the two energy windows chosen to initiate the QM/MM-TSH dynamics, using the snapshots of set~I and~II, respectively.}
    \label{fig:spectrum}
\end{figure}

The absorption spectrum of furan in water was calculated using the first 100 frames from the production run of the MM-MD simulations (recall Figure~\ref{fig:simulation_timeline}).
The spectrum was convoluted from the lowest ten singlet excited states, covering an energy range up to 9.5~eV, see Figure~\ref{fig:spectrum} and Table~S1 of the Supporting Information\dag. 

In the simulations set~I, we use all 600 snapshots generated from the MM simulation as possible initial conditions (position and velocities) for the subsequent QM/MM-TSH excited-state dynamics, which included the lowest eleven singlet states. 
Furan is then excited within the energy range of 7.3--7.5~eV, primary targeting the S$_3$ state.
Based on the associated excitation energies and oscillator strengths of the sampled 600 geometries, the initially excited electronic states were selected stochastically,\cite{Barbatti2007} resulting in 46 initial conditions. 
Out of those, 39 started in S$_3$ (the state with a large dipole moment, see Table S1) and 7 started in S$_4$.

Following previous studies~\cite{Fuji2010,Oesterling2017}, which indicate that the relaxation of furan to the ground state occurs within approximately 200~fs, the trajectories of set~I were propagated for 330~fs.
A nuclear time step of 0.5~fs and an electronic time step of 0.025~fs were employed.  
Nonadiabatic couplings between singlet states were obtained from the local diabatization algorithm by Granucci \textit{et al.},\cite{Granucci2001} computing the overlap of the wave function between two consecutive time steps.\cite{Plasser2012,Plasser2016}
To conserve energy, the velocities of the QM particles were uniformly rescaled after each hop to account for the potential energy difference between the new and old states (no explicit nonadiabatic coupling vectors were used for a projection of the velocity vectors). 
Since TD-DFT cannot describe conical intersections between S$_1$ and S$_0$ due to the degeneracy of the reference state (ground state) and the first excited state,\cite{Plasser2014} trajectories were forced to hop to the ground state whenever the energy difference between the states S$_1$ and S$_0$ states was smaller than 0.1~eV. 
Once in the ground state, no back transitions were allowed until the end of the propagation time.
The data (set~I) were then used for model training, validation, and testing.

Since the local diabatization scheme cannot be used with ML/MM simulations because of the unavailability of the wavefunction, a second set of QM/MM-TSH simulations was performed.  
In this, so-called, set~II, the nonadiabatic couplings between the singlet states were determined using the curvature-driven TSH scheme recently developed by Zhao et al.\cite{Zhao2023}, which relies on the second time derivative of the energies only and thus is accessible in conjunction with ML potentials. 
In this way, the resulting ML/MM-TSH dynamics are directly comparable with the reference QM/MM curvature-driven TSH ones.
The initial conditions for set~II of QM/MM simulations were taken from the last 2~ns of the MM trajectory, corresponding to 500 possible initial conditions (see Figure \ref{fig:simulation_timeline}).  
In order to investigate the transferability of the ML potential, we excite between 6 and 7~eV in order to have a different excitation window and thus obtain a different set of trajectories with initial conditions and energies different from those used in the training set (set~I). 
The chosen window resulted in 66 excitations to the bright S$_2$ state.
The 26 excitations to the S$_1$ were not propagated, in order to have all trajectories starting from the same state, making the dynamics and kinetic fits easier to interpret. 
All other settings were identical to those employed in set~I.

\subsection{Machine-Learning Setup}

\begin{figure}[tb]
\centering
\includegraphics[width=\linewidth]{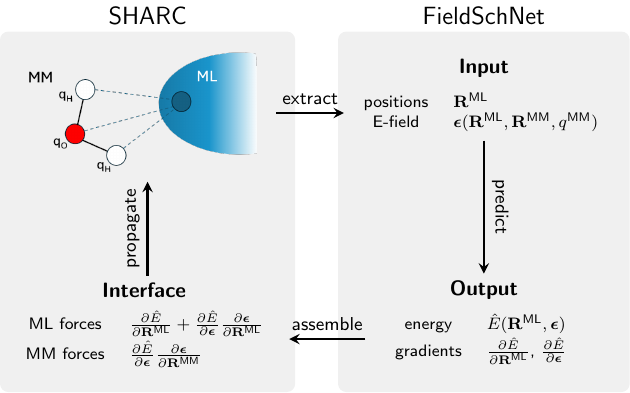}
    \caption{Schematic representation of the communication between the driver of the nonadiabatic dynamics, the SHARC engine,\cite{SHARC3} and the ML model, FieldSchNet, needed to perform excited state ML/MM dynamics. 
    The SHARC interface extracts the value of the electric field at the position of every ML atom and passes both the field value and the atom positions to the ML model. 
    The FieldSchNet model, predicts the field-dependent energies and gradients and passes them back to the SHARC interface. 
    The system is then propagated in time by the SHARC driver. 
    Here, none of the steps are based on file I/O further accelerating the dynamics.}  
    \label{fig:interface}
\end{figure}    

To perform ML/MM simulations, we interfaced the graph convolutional neural network FieldSchNet\cite{Gastegger2021} with the SHARC engine\cite{SHARC3}, as schematically depicted in Fig.~\ref{fig:interface}. 
This neural network models atoms in its chemical and structural environment within a cutoff region.
The radial cutoff for the construction of the graph around the central atom was set to 10~\AA\, and we used 50 equidistantly spaced Gaussians in the filter-generating network,\cite{Schutt2017a} which is used in SchNet to featurize the interatomic distances. 
The atomic features were updated in six message-passing layers, and the length of the atomic feature vectors was 256.
Since we did not observe any hops to higher-lying states than S$_4$ in the QM/MM TSH simulations of set~I, only the five lowest-lying singlet states were considered for learning. 
For the prediction of the lowest five adiabatic energies, we used one dense read-out block with four hidden layers (256-128-64-32-16-5), each halving the length of the previous layer up to a final representation of size 16, followed by a linear output layer that gives the five energies.
We used the shifted softpuls function\cite{Schutt2017} for all nonlinearities, as done in Gastegger et al.\cite{Gastegger2021}
The forces are computed as the gradient of the energy with respect to the input coordinates (eqs.(\ref{eq:QMderiv_EML}) and~(\ref{eq:MMderiv_EML})).

\begin{figure*}[ht]
\centering
\includegraphics[width=\linewidth]{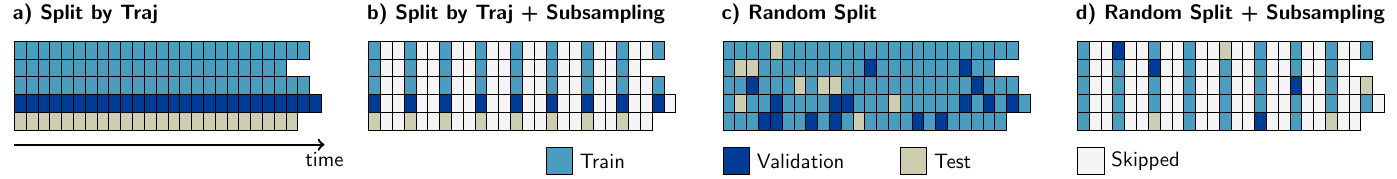}
    \caption{Strategies to partition data from QM/MM-TSH dynamics, shown for 5 example trajectories of varying lengths.
    a) Split by trajectory means that all frames belonging to one trajectory are allocated to the same subset (train, validation, or test), reducing overlap particularly between the train and test sets.
    b) To avoid temporal proximity and high correlation of frames, time-based subsampling prior to splitting into train, test, and validation sets is done  (exemplary shown for retaining every third frame, i.e. 33~\% data usage). 
    c) Random split means that the trajectory frames are pooled and randomly allocated to any of the three, train, validation, or test subtests.
    d) Random split combined with time-based subsampling, here also using every third frame.}
    \label{fig:SamplingStrategies}
\end{figure*}

Data set~I comprises all points collected from the 46 QM/MM-TSH trajectories, each 330~fs long with a 0.5~fs time step (i.e. $46\times 330 \times 2$ points). 
However, because a few trajectories ended prematurely, the total amounts to 28,935 data points. 
This training set was then divided into training, validation and testing using 37, 5 and 4 trajectories, respectively. 
The validation and test sets were used solely for analyzing the training error on energies and forces, but not for comparing dynamics results, as this cannot be done under identical conditions. 
We call this approach "split by trajectory", see Fig.~\ref{fig:SamplingStrategies}a.
Furthermore, we performed an alternative set of ML trainings, in which we randomly assigned frames from the 46 trajectories to the three different tasks (train, validation, and test) in a ratio of 80:10:10.
We call this way of mixing the data  "random split", see Fig.~\ref{fig:SamplingStrategies}c. 
The advantage of "split by trajectory" is that the test error is more likely to reflect the true performance, as it ensures that the model has not seen any frames of the trajectories belonging to the test set.
The disadvantage is that the number of trajectories from which the model can learn is smaller than in the "random split" scheme. 
To further investigate the impact of data partitioning in the dynamics, we also examined the effect of subsampling in time for each partitioning scheme, as consecutive frames that are 0.5~fs apart are likely to be very similar and thus add little new information to the data set. 
We therefore tested the effect of using only every second (50~\% of data usage), third (33~\%), etc, data point. 
This is schematically indicated in panels b and d of Fig.~\ref{fig:SamplingStrategies} for the example of 33~\% of data usage.

The ML trainings were performed using the Adam optimizer.\cite{Diederik2017}
We set the starting learning rate to a value of 10$^{-5}$, and the parameters for the exponential averages of past gradients (momentum term) and squared gradients (raw second moment) to 0.9 and 0.999, respectively. 
Higher initial learning rates were consistently found to cause numerical instabilities during training.
A learning rate scheduler was used to adjust the learning rate. 
If the model's performance on the validation set did not improve for 20 consecutive epochs, the learning rate was decreased by 20~\% to avoid overstepping.
The maximum number of epochs was set to 5000; alternatively, training was stopped early if the learning rate fell below 10$^{-6}$.
The batch size was 10.
We used the augmented mean squared error from eq.~(\ref{eq:new_loss}) as the loss function with $w_E = 1$ and $w_F = 10$.

\section{Results and Discussion}\label{sec:results}

\subsection{Learning Curves and Test Performance}

\begin{figure}
    \centering
    \includegraphics[width=\linewidth]{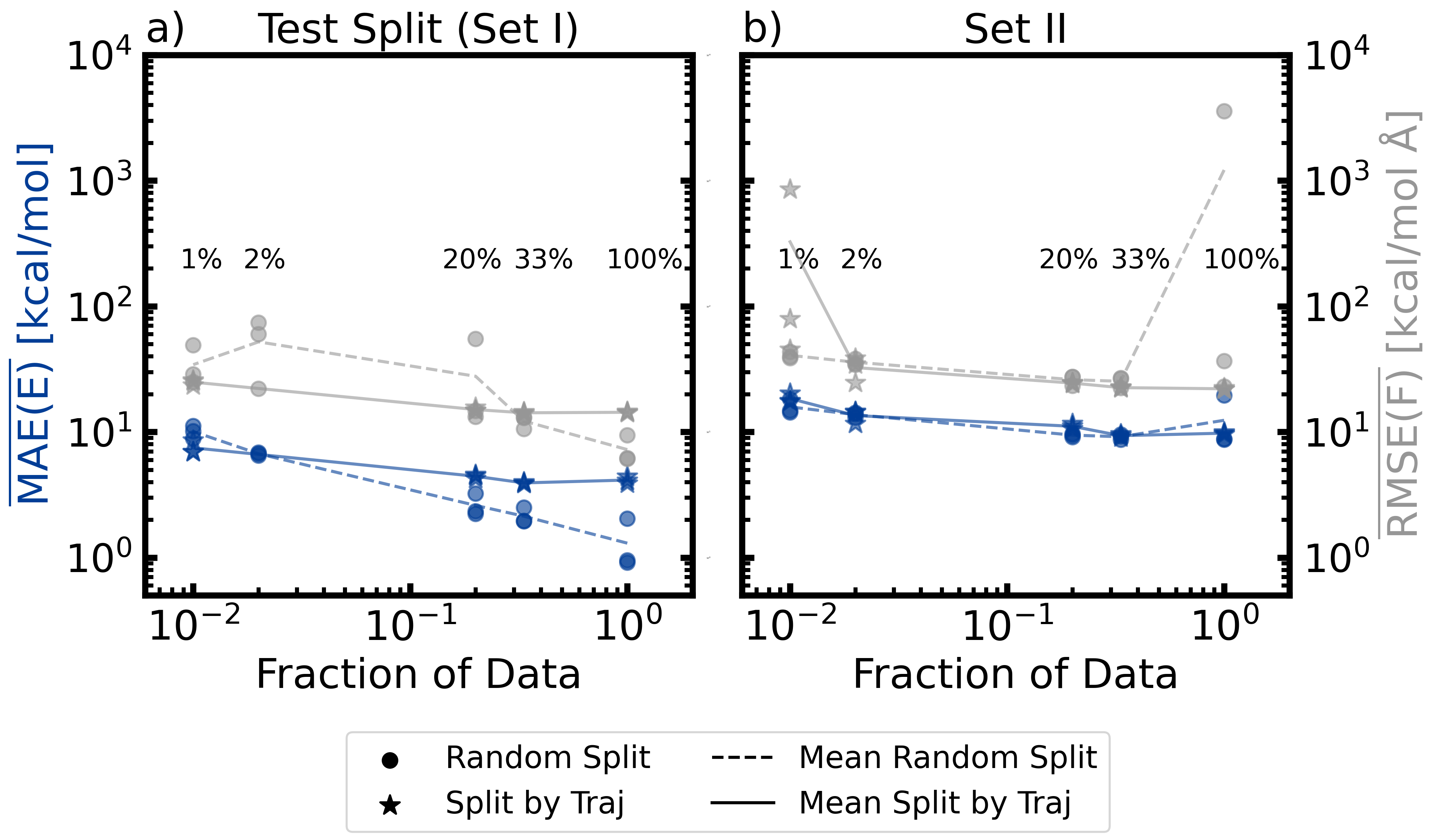}
    \caption{
    Change in model test performance on a) the test set taken from set~I and b) trajectories from set~II, as a function of the fraction of  data used during model training.
    Results of "Random Split" models are shown as circles, "Split by Trajectory" as stars. 
    Mean absolute error (MAE) of the energy and root mean square error (RMSE) in the forces are shown in blue and grey, respectively.
    All values correspond to averages over all five electronic states, indicated by the bars over MAE and RMSE.
    The three symbols for each combination of split type and fraction of data correspond to the three models trained with different random initializations
    To guide the eye, lines connect the averages of the models with identical hyperparameters.
       }
    \label{fig:lcurve}
\end{figure}

The learning curves (change in test performance) of the two partitioning schemes, "random split" and "split by trajectory", are shown in Fig.~\ref{fig:lcurve}a, as a function of the equidistant subsampling in time. 
To that end, the errors of energies and forces are plotted against the amount of data included in the training, averaged over all five electronic states. 
For every specific combination of split and fraction of data that was retained after subsampling in time, we trained three independent models, which differ with respect to the randomly initialized parameter values (a list of the random seeds together with the data sets can be found in the Zenodo archive associated with this manuscript).  
As a result, no training with the same hyperparameters produces the exact same result, but is generally assumed to converge to a similarly deep minimum (see Section~S2).
In general, the performance of the model seems to improve as the training data becomes denser, \textit{i.e.} with more frequent temporal sampling.

The learning curve (Fig.~\ref{fig:lcurve}a) from the "random split" (circles) forms a nearly perfect line on a log-log graph, suggesting a power-law relationship between the sample count and the error size. 
This partitioning scheme often places training and testing frames in close temporal proximity. 
When using the entire data set, the training set probably includes the surrounding time frames of the test frames. 
This introduces a strong dependency between training and testing errors, raising doubts about whether a small testing error truly reflects the model's ability to generalize (i.e. the power to correctly extrapolate).

In contrast, the "split by trajectory" training approach shows a different pattern. 
The errors initially decrease before stabilizing at around 33~\% data usage, which corresponds to using every third time step. 
This behavior arises because the training and test sets are less correlated in this scheme. 
The "split by trajectory" method ensures that test errors directly evaluate the model's generalization capabilities. 
Beyond a certain point, adding more closely spaced data does not further improve the test performance. 
This is due to minimal configuration changes within a time step of 0.5~fs, offering little additional information for extrapolation.

The discrepancy between the "split by trajectory" and random split test errors when using closely spaced trajectory frames is noteworthy. 
The scatter (parity) plots shown in Figs.~S2 and~S4 for the models using 100~\% of the training data do not exhibit any trends that would point to a general difference between those models, except that the "split by trajectory" models perform worse, which one would expect based on their relative average test performance from Fig.~\ref{fig:lcurve}a.
However, one should note that the models trained using the "split by trajectory" exhibit a much more homogeneous test performance in comparison to the "random split" models,  where some appear to be much better or worse than the other two.
If one were to choose settings based on Fig.~\ref{fig:lcurve}a, then "ramdom split" with 100~\% and "split by trajectory" with 33~\% of the data are expected to perform best. 

\subsection{ML/MM-TSH Nonadiabatic  Dynamics}

\begin{figure}[!h]
    \centering
    \includegraphics[width=1\linewidth]{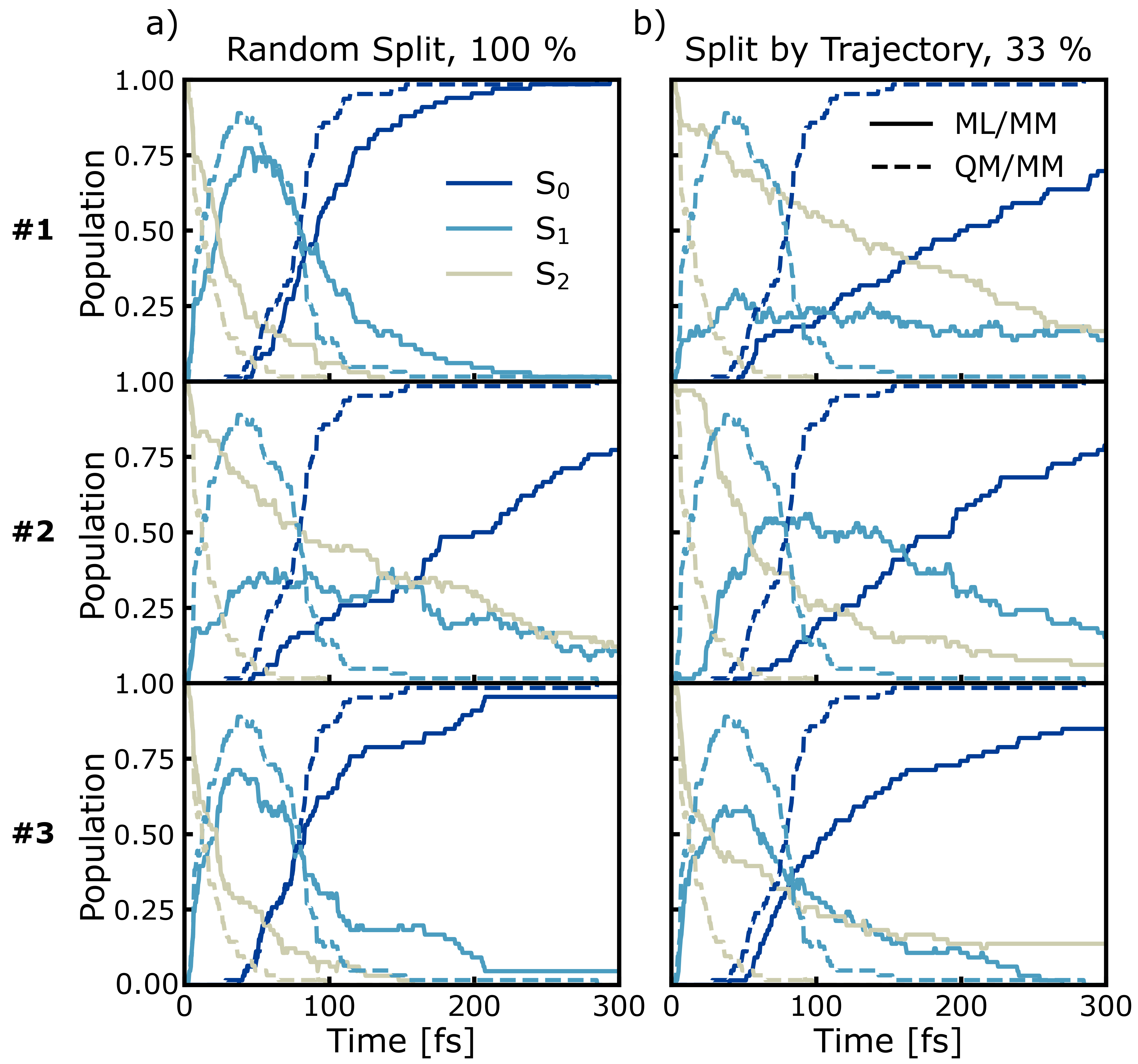}
    \caption{Excited state occupation population dynamics of furan in water using ML/MM (solid lines) trajectories, trained with random split 100~\% (a) or split by trajectory 33~\% (b) partition sets, compared to reference QM/MM (dashed lines) trajectories.
    The three plots in each partition scheme correspond to three different FieldSchNet models trained with different random seeds.}
    \label{fig:occs}
\end{figure}

To further assess the ML models, we performed a second set of nonadiabatic QM/MM dynamics using curvature-driven TSH (set~II), which allows us to directly compare both the reference QM/MM and ML/MM dynamics. 
The initial conditions for the ML/MM trajectories, including geometries, velocity vectors, and random number seeds are the same as those employed for the QM/MM trajectories of set~II. 
Figure~\ref{fig:occs} shows the time-resolved electronic populations based on the active state of the reference trajectories, the QM/MM dynamics (dashed lines in each panel).
The furan relaxes rapidly from the initially populated S$_2$ state to the S$_1$ state within the first 100~fs, followed by a decay to the electronic ground state S$_0$. 
Within 300~fs, nearly all 66 trajectories have reached S$_0$.
These populations are compared to different ML/MM TSH simulations (solid lines), conducted using the three models trained with the same splitting and subsampling settings.
Specifically, we carried out simulations for 100~\%, 33~\% and 1~\% for both random and trajectory split (from which Figure~\ref{fig:occs} shows only random split with 100~\% and  split by trajectory 33~\%, as these were the best hyperparameters deduced from Figure\ref{fig:lcurve}a). 
The electronic populations for all ML/MM dynamics can be found in Figures~S16 and S17.

As it can be seen, the electronic populations derived from ML/MM show significant differences depending on which model was used to generate the random split 100~\% and split by trajectory 33~\% trajectories, even if those models differ only by their random initialization. 
The random split 100~\% models \#1 and \#3 produce dynamics with very similar population curves as the reference QM/MM dynamics, whereas the dynamics of model \#2 show much slower internal conversions. 
For the split by trajectory 33~\% models, the visual agreement of the predicted populations increases from model \#1 to \#3.

The agreement of the ML/MM electronic populations with the corresponding QM/MM reference cannot easily be explained with the test statistics presented in the previous section. 
Inspection of the parity plots for the energy gap between neighboring PESs (Figs.~S2-S4 of the ESI\dag) reveals that the split by trajectory models appear to be worse at predicting this energy difference.
This is problematic, as the energy gap is used to force hops into the ground state from S$_1$.   
To probe the model performance in regions with small energy gaps further we employ an energy-gap-weighted error measure, 
\begin{equation}
\begin{split}
    \text{wRMSE}(X^j) = \sqrt{\frac{\sum_i^{N_\text{frames}} \left( X_i^j - \hat{X}_i^j \right)^2 w_i^j}{\sum_i^{N_\text{frames}} w_i^j}} \quad \\
    \text{with} \quad w_i^j = \frac{1}{\min(|E_i^j - E_i^{j-1}|, |E_i^j - E_i^{j+1}|)} \ ,
\end{split}
    \label{eq:weighted_error}
\end{equation}
where $E_i^j$ represents the ground truth energy for configuration $i$ in state $j$ and $X^j_i$ denotes either energy or forces for this configuration and state.

To facilitate a more effective comparison across the various models, in addition to visually evaluating the population curves, we fitted a kinetic model to the different steps of the relaxation process. 
This model includes only two time constants: $\tau_{2\rightarrow 1}$ for the internal conversion from S$_2$ to S$_1$ and $\tau_{1\rightarrow 0}$ for the transition from S$_1$ to S$_0$. 
The QM/MM reference simulations predict the first internal conversion to be about five times faster ($\tau_{2\rightarrow 1}^\text{QM/MM} = 16.8$~fs) than the second one ($\tau_{1\rightarrow 0}^\text{QM/MM} = 64.9$~fs). 
These time constants are comparable to those predicted in the gas phase (9.2~fs and 60~fs, for the lifetimes of the  $S_2$  and  $S_1$ states, respectively), by Fuji et al.~\cite{Fuji2010} using similar TSH simulations and TD-DFT, which in turn are also consistent with time-resolved photoelectron spectra, recorded by the same authors. 
The similarity of the time constants indicates that the effect of the solvent is not very pronounced, slightly decelerating the decay of the bright S$_2$ state.

The different numerical values of the kinetic fits and (weighted) errors, obtained by the ML/MM simulations are collected in Table~\ref{tab:split}.
We observe that none of the ML/MM models produces dynamics that relax as fast as observed in the QM/MM simulations.
Furthermore, the relative error for $\tau_{2\rightarrow 1}$ is almost always larger than for $\tau_{1\rightarrow 0}$, because of their difference in magnitude.
Intriguingly, the kinetics of the ML/MM simulations can differ greatly even for those models that have the same training settings and only differ by weight initialization.
In general, models trained using the random split appear to perform better, especially when the training frames are taken at smaller intervals. 
The reason for this is likely the small set of trajectories available for training, such that splitting by trajectory imposed a stronger limit on the phase space available for training than the random splitting.
More trajectories in the training set are expected to remove this trend.
Indeed, random split with 100~\% of the training data delivers the best results.

While it is generally a problem to have a confidence measure for ML-based MD simulations where the behavior of the system is not known, it is gratifying to see that the weighted error, as specified in eq.~(\ref{eq:weighted_error}), serves as a metric to predict which of two models  with identical hyperparameters will outperform.
Comparing the normal and weighted RMSE, it can be seen that the energy errors change by a maximum of 1~kcal/mol, except for the models trained with only 1~\% of the data.
We can therefore assume that energy values close to the intersection seams are learned with a similar accuracy as points further away from these important regions.
However, the same does not appear to be true for the gradients of the PESs.
The increase from non-weighted to energy-gap-weighted RMSE varies, but is approximately a factor of 5 to 10.
Weighted energy and force \mbox{RMSEs} taken together suggest that the distance between the different electronic PESs is roughly correct; however, the topology close to the avoided crossings is not.
We therefore conclude that the general feature of the avoided crossing is represented correctly, but that the individual ML-PESs are significantly less smooth in these regions, leading to a strong increase in the force errors, albeit not in the energies.
Surprisingly, a model with a weighted force RMSE of 50~kcal/(mol~\AA) predicts the relaxation dynamics qualitatively correctly (random split, 100~\%, models no.~1 and~3).
Since the weighted error decreases with larger training set sizes, increasing the training set size even further should lead to smaller errors.
Focusing on the correct reconstruction of the slope near the intersection seams seems to be especially important.
This interpretation is further supported by test calculations where we used the time-derivative of the gradients instead of the energies, which lead to significantly decreased agreement between the ground truth and the ML models.
Hence, the computed ML transition rates have the correct order of magnitude, because the couplings that were used in our simulations only depend on the energies and not on the gradients.

\begin{table*}[!h]
    \centering
    \caption{Time constants for the sequential S$_2\rightarrow$S$_1$ and S$_1\rightarrow$S$_0$ internal conversions of furan in water, as obtained from ML/MM models trained with 100~\%, 33~\%, and 1~\% of the data from set~I (subsampled in time) within the random split and trajectory split schemes.
    The simulations were started from the initial conditions of set~II and are compared with the QM/MM trajectories of this set.
    The test errors in energies (E) and forces (F) are averaged over the five electronic states (S$_0$ - S$_4$).  
    The weighted RMSEs were computed according to eq.~(\ref{eq:weighted_error}).
    All the trainings were performed with the augmented loss. 
    The time constants obtained from the reference QM/MM curvature-driven TSH simulations are $\tau_{2\rightarrow 1}^\text{QM/MM} = 17 \pm 2$~fs and $\tau_{1\rightarrow 0}^\text{QM/MM} = 65 \pm 3$~fs.
    }
\begin{tabular}{c|c|cr@{ $\pm$ }rr@{ $\pm$ }rcccc}
         \multirow{2}{*}{split type} & \multirow{2}{*}{~\% of data} & \multirow{2}{*}{model \#} & \multicolumn{2}{c}{$\tau_{2\rightarrow 1}$} & \multicolumn{2}{c}{$\tau_{1\rightarrow 0}$} & $\overline{\text{RMSE(E)}}$ & $\overline{\text{wRMSE(E)}}$ & $\overline{\text{RMSE(\textbf{F})}}$ & $\overline{\text{wRMSE(\textbf{F})}}$ \\
         & & & \multicolumn{2}{c}{[fs]} & \multicolumn{2}{c}{[fs]} & [kcal/mol] & [kcal/mol] & [kcal/(mol \AA )] & [kcal/(mol \AA )] \\
         \hline
         \multirow{9}{*}{random} & \multirow{3}{*}{100} & 1 & 29 & 4 & 73 &  4 & 1.4 & 1.4 & 5.9 & 56.3\\
         && 2 & 138 & 28 & 92 & 12 &  2.7 & 3.2 & 9.1 & 76.2\\
         && 3 & 28  & 5 & 73 & 6 & 1.3 & 1.3 & 6.0 & 54.9 \\
         \cline{2-11}
         & \multirow{3}{*}{33} & 1 & 63 & 16 & 90 &	8 & 3.3 & 3.8 & 10.8 & 95.3\\
         && 2& 36 & 8 & 97 & 8 & 2.6 & 2.7 & 9.8 & 69.5\\
         && 3& 66& 11 & 84 & 6 & 5.0 & 5.8 & 19.0 & 126.0 \\\cline{2-11}
        & \multirow{3}{*}{1} & 1 & 240 & 41 & 164 & 29 & 28.5 & 22.9 & 26.0 & 168.0\\
         && 2 &134&	18 & 130 & 16 & 9.5 & 10.0 & 23.6 & 126.7\\
         && 3 & 97 & 	10 & 116	& 14 & 11.1 & 11.3 & 23.2 & 124.9 \\
        \hline \hline
         \multirow{9}{*}{by~traj} & \multirow{3}{*}{100} & 1 &153	& 26 & 95	& 10 & 6.3 & 6.3 & 13.7  & 81.7\\
         && 2 & 37	& 6 & 251	& 33 & 6.0 & 6.0 & 14.1 & 83.9  \\
         && 3 & 292 & 61 & 109 & 17 & 6.0 & 5.9 & 13.9 & 82.6
\\\cline{2-11}
         & \multirow{3}{*}{33} & 1 & 67	& 17 & 67	& 6  & 5.6 & 5.6 & 13.6 & 80.4 \\
         && 2 & 86	& 11 & 134	& 18 & 5.9 & 6.0 & 14.1 & 82.7 \\
         && 3 & 173	& 32 & 76	& 9 & 5.7 & 5.8 & 14.0 & 82.0 
\\\cline{2-11}
        & \multirow{3}{*}{1} & 1 & 231	& 23 & 70	& 11 & 10.0 & 10.7 & 23.2 & 122.8 \\
         && 2 & 393	& 70 & 100	& 22 & 10.0 & 11.2 & 24.8 & 132.2\\
         && 3 &	292	& 51 & 109	& 18 & 12.1 & 13.4 & 25.4 & 132.6\\
        \hline
    \end{tabular}

    \label{tab:split}
\end{table*}

To further understand the performance of the different models, we also performed an error analysis regarding energies, gaps, and forces for the data in set~II.
Parity plots for set~II are shown in Section~S2.2 of the ESI\dag and the average errors in Fig.~\ref{fig:lcurve}b.
One can see that on this data --unseen by all models-- "random split" and "split by trajectory" models perform similarly. 
This underlines that test statistics based on the random splitting (see Fig.~\ref{fig:lcurve}a) provide a strong underestimation of the true error. 
Furthermore, the parity plots of Section~S2.2 in the ESI\dag show that all models (random and trajectory split) appear to strongly overestimate $E_{S_0}$ for configurations with a ground state energy above 75~kcal/mol (likely geometries with strongly elongated or broken bonds), which leads to a significant underestimation of the energy gap to the first excited state in these regions. 
It may appear confusing that the models tend to underestimate this gap when their dynamics is always slower than the QM/MM reference. 
However, this only happens for configurations with a high ground-state energy, which are those occurring near the end of the 300~fs trajectories, where molecules that have already relaxed to the ground state are prevented from hopping back up. 

In general, the errors of the models based on the set~II are so large that it is surprising that they can produce reasonable population decays.
These errors are based on geometries from the entire 300~fs, where furan displayed ring opening and subsequent bond rearrangements (discussed in detail in Section~S4 of the ESI\dag), which might be difficult to describe with (TD-)DFT. 
The initial part of the dynamics is much better reproduced by the ML/MM than the latter parts, since it corresponds to regions in configuration space that are reasonably described by (TD-)DFT.
To test this hypothesis, we performed a second round of error analysis on set~II, where we only included frames from the first 75~fs. 
The much better performance of the models on these configurations becomes apparent when looking at the parity plots in Section~S2.3 of the ESI\dag.
The errors (MAE and RMSE) of forces, energies, and energy gaps are significantly lower than those on the entire set~II.
For split by trajectory they are basically identical to the test statistics obtained from set~I.
This means that the models fit the PESs well in the part of the configuration space that is explored right after irradiation, which is the part needed for the relaxation back to the electronic ground state. 
The subsequent distortions due to the excess energy of 6--7~eV, are not well described, but these deficiencies are less relevant to the change in electronic populations. 
This is why most models were able to reproduce the relaxation dynamics of set~II even though they cannot extrapolate correctly to the configurations visited in later stages of the QM/MM simulations.

\subsection{Structural Analysis of Trajectories}

\begin{figure}
    \centering
    \includegraphics[width=\linewidth]{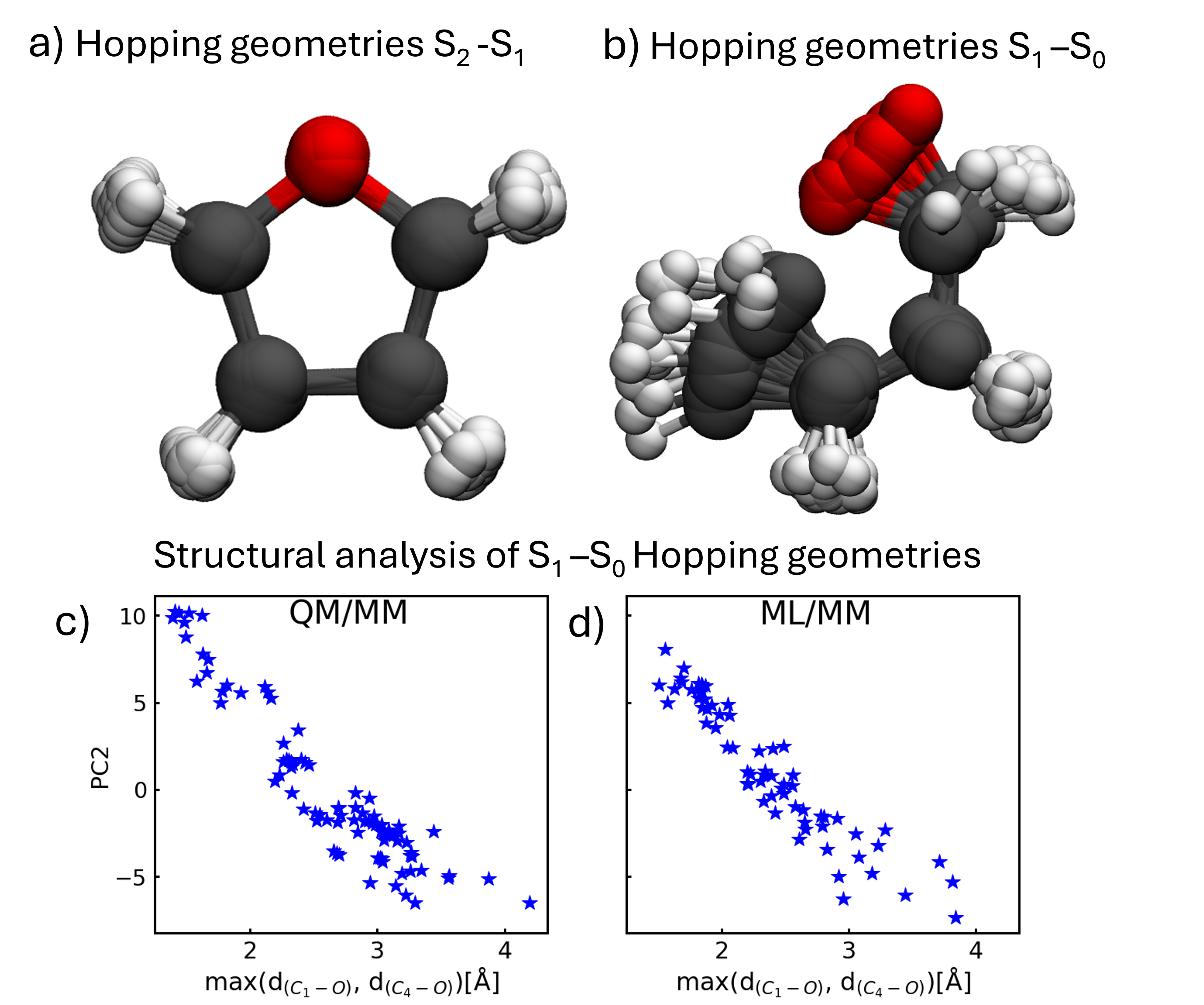}
    \caption{Aligned QM/MM S$_2$--S$_1$ (a) and S$_1$--S$_0$ (b) hopping geometries obtained from trajectories in set~II.
    Comparison of S$_1$--S$_0$ hopping geometries obtained from QM/MM (c) and ML/MM (d) (random split, 100~\% data, model \#1)  by projecting them onto the maximum of either the C$_1$-O and C$_4$-O bond distance (see Fig.~\ref{fig:system}b) as well as the PC2 from a PCA performed solely on the QM/MM frames (Section~S5).
    }
    \label{fig:hoppings}
\end{figure}

In order to analyze whether the ML/MM simulations show the same structural changes during the dynamics as the reference QM/MM simulations, we compare the hopping geometries from the QM/MM simulations with those encountered in the ML/MM simulations, here done exemplary for model \#1 with 100~\% of the data and the random split procedure (for a comparison of all 18 ML/MM models, see Section~S4.2 of the ESI\dag).
The hopping geometries from the QM/MM simulations (set~II) for the S$_2$--S$_1$ and S$_1$--S$_0$ transitions are shown in Figure~\ref{fig:hoppings}a and~b, respectively.
The geometries responsible of the first internal conversion (Figure~\ref{fig:hoppings}a) are very similar because the hops occur shortly after excitation, so they closely resemble the ground-state MM geometries of the Franck-Condon ensemble. 
The geometries corresponding to the  S$_1$--S$_0$ deactivation (Figure~\ref{fig:hoppings}b) are more diverse.
They encompass a smooth interpolation between configurations with a closed and an open ring. 
For easier visualization, the geometries were aligned so that only one bond appears to open, however, both bonds C$_1$-O and C$_4$-O (following the naming convention of Figure~\ref{fig:system}b) are equally likely to break.

To analyze the similarities of the hopping structures found in the QM/MM and ML/MM simulations, we performed a principal component analysis (PCA) on the set of QM/MM geometries for the transitions between S$_1$ and S$_0$ using a Coulomb matrix representation\cite{Rupp2012} of furan (Section~S4 of the ESI\dag). 
We found that the first principal component (PC1), which recovers about 89~\% of the overall variance, focuses on the distance between the oxygen atom and carbon atoms~1 and~4 , whereas PC2, which reflects 9~\% of the variance, is a linear combination of several interatomic distances.
In Figure~\ref{fig:hoppings}c we replaced PC1 with the maximum length of the two bonds C$_1$-O and C$_4$-O for easier visualization, as PC1 creates an inverted V due to the symmetry of the breaking bonds (see Figure~S19 in Section~S4 of the ESI\dag).
One can see that the geometries form one continuous hopping seam.

When applying the same linear transformation to the hopping geometries of the ML/MM simulation (here shown for random split, 100~\%, model \#1) and projecting them onto PC2 and the maximum of the C-O bond distances, they form the same diagonal line, indicating that hops occur in the same region of configuration space. 
This is not only true for the single model analyzed here, but for most models (Fig.~S20 in Section~S4.2.1 of the ESI\dag).
Deviations from the rather even distribution seen in Figure~\ref{fig:hoppings}c correlate with significantly different relaxation dynamics, e.g. some models hop only at certain points along the seam creating visual clusters in the 2D-projection.
As the S$_2$--S$_1$ hopping geometries are all very similar, the projection onto the first two principal components forms a single cluster (Fig.~S21 in Section~S4.2.2 of the ESI\dag).
Trajectories with slow internal conversion from S$_2$ to S$_1$ show outliers in the 2D-projection (Section S4.2.2 and Fig.~S22 of the ESI\dag).

\section{Conclusions}

We have integrated the FieldSchNet machine learning interatomic potential into the SHARC software package to enable nonadiabatic  ML/MM dynamics simulations in an electrostatic embedding framework, in analogy to the traditional QM/MM counterpart. 
By developing a training loss function that incorporates the consistent gradients required during simulation, we ensured the inclusion of electric field-dependent components in the nuclear forces, enhancing the accuracy of our approach.
Our method was applied to furan in water, trained with almost 30,000 data points obtained from QM/MM nonadiabatic trajectories at the BP86/def2-SVP level of theory.
We compared the training errors derived from using two distinct data splitting strategies -- "random sampling" and "split by trajectory", depending on whether the data points were sampled randomly from any trajectory, or entire trajectories are used for training, validation and testing. 
Although "split by trajectory" offers cleaner separation and more realistic error statistics, its performance is limited by the smaller number of available trajectories. 
In general, we observe a wide range in the performance of the ML/MM models when trying to reproduce the nonadiabatic QM/MM dynamics of a set of held out trajectories, even for ML models with the same hyper-parameters.
Strong sensitivities of excited state kinetics generated by ML models were already noted in earlier studies.\cite{Westermayr2020a}
However, well-chosen test statistics served as reliable indicators of model performance.  
Energy gap-weighted errors help to highlight discrepancies near the intersection seams.
Projecting hopping geometries onto two dimensions provides valuable insights to understand the difference between ground truth dynamics and those derived from the ML model.

Based on our findings, we recommend against using costly nonadiabatic QM/MM dynamics simulations to generate training data. 
Most of the computational effort is spent on geometries that are far away from critical regions of the PESs (the intersection seams). 
Furthermore, the produced geometries are highly correlated as they are closely spaced in time.
Future applications should therefore aim to collect training data through active learning schemes,\cite{Tan2025} which has the added benefit that such data are uncorrelated, which means that no computational effort is wasted on obtaining frames that might later be discarded by subsampling in time.
We expect models trained on such data to generate better forces near the intersection seams and to reproduce the dynamics of the QM method more reliably.



\section{Data and Code Availability}
QM/MM data generated for furan in water is deposited in a Zenodo archive (\url{https://doi.org/10.5281/zenodo.14536036}) together with all trained ML models.
Additionally, the archive also contains code to interface FieldSchNet with an upcoming release of SHARC, which is published under the GPL license at \url{https://github.com/sharc-md/}. 

\section*{Author Contributions}
MXT (Data curation, Formal analysis, Investigation, Methodology, Software, Visualization, Validation, Writing - original draft). 
BB (Software and Writing - review). 
CGC (Investigation).
JW (Methodology, Writing - review).
PM (Conceptualization, Methodology, Supervision, Funding acquisition) 
JCBD (Methodology, Validation, Supervision, Visualization, Writing - original draft).
LG (Conceptualization, Supervision, Funding acquisition, Project administration, Writing - review).

\section*{Conflicts of interest}
There are no conflicts to declare.

\section*{Acknowledgements}
This work is funded from the University of Vienna in the framework of the research platform ViRAPID. 
M.X.T. and L.G. appreciate additional support provided by the Austrian Science Fund, W 1232 (MolTag). 
The Vienna Scientific Cluster is thanked for generous allocation of computer resources. 
The authors thank the SHARC development team, the ViRAPID members, and Michael Gastegger for fruitful discussions. 



\balance


\bibliography{main} 
\bibliographystyle{rsc} 

\clearpage
\onecolumn

\setcounter{section}{0}
\setcounter{figure}{0}
\renewcommand{\thesection}{S\arabic{section}}
\renewcommand{\theequation}{S\arabic{equation}}
\renewcommand\figurename{Figure S\hspace*{-4px}}
\renewcommand\tablename{Table S\hspace*{-4px}}

\setlength{\figrulesep}{0.5\textfloatsep} 
\renewcommand{\topfigrule}{\vspace*{-1pt} }
\renewcommand{\botfigrule}{\vspace*{-2pt}}
\renewcommand{\dblfigrule}{\vspace*{-1pt}}

\begin{center}
    \LARGE
    Supporting Material for:\\ 
    Excited-state nonadiabatic dynamics in explicit solvent using machine learned interatomic potentials
\end{center}

\section{Excitation analysis}

\begin{figure}[h!]
    \centering
    \includegraphics[width=\linewidth]{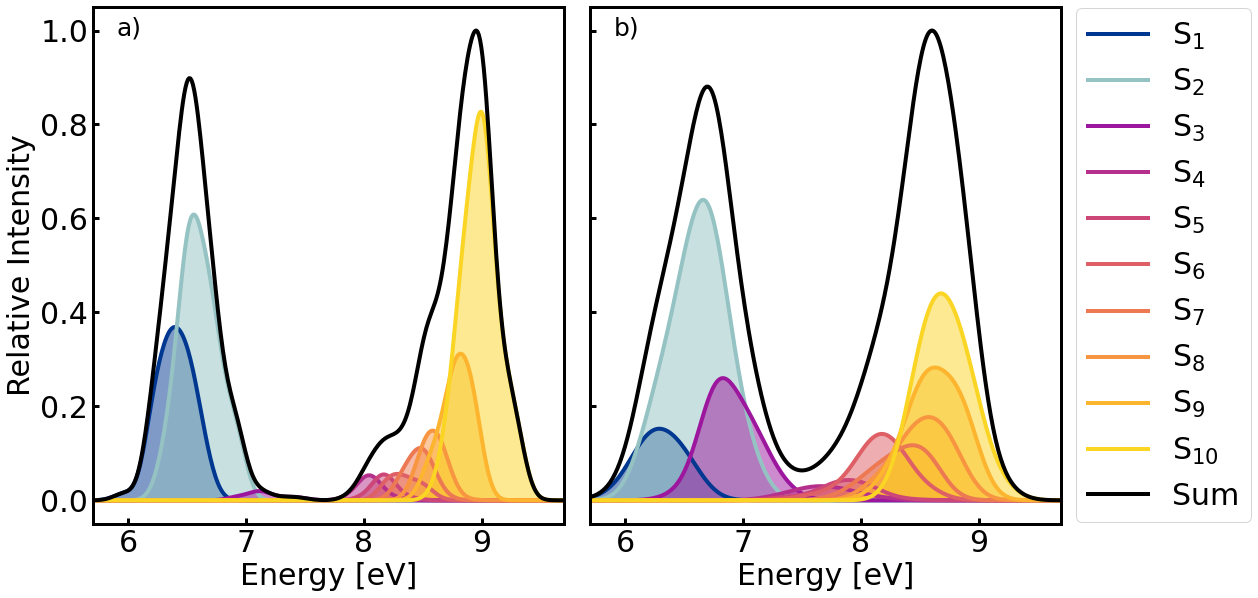}
    \caption{Absorption spectrum of furan based on the lowest 10 excited singlet states for a) solvated in explicit water obtained as average over 100 MM-MD snapshots and b) in the gas phase obtained from 100 Wigner samples.}
    \label{fig:both_spectra}
\end{figure}

Figure~\ref{fig:both_spectra} shows the simulated spectra of furan solvated in water (a) and gas phase (b).
While the two peaks of the total intensity are quite similar between the two spectra, the solvated furan shows slimmer peaks.
Hence, in the case of the solvated furan the area between 7 and 8~eV is almost completely dark.
This is mainly caused by the S$_3$ being a dark state in solution, while it contributes significantly to the signal in the gas phase.
The S$_1$ shows the opposite behavior, having less intensity in the gas phase than in solution. 
However, the reduction is not as large.
Furthermore, S$_6$ is also brighter in the gas phase, while S$_{10}$ has a lower peak.

The peak between 6 and 7~eV is created by transition to S$_1$, S$_2$, and S$_3$, with S$_2$ having almost triple the height of the S$_1$ and S$_3$ in vacuum.
In solution, the relative intensity of the transition to S$_2$ remains largely unchanged, however, S$_1$ increases in intensity and S$_3$ is an almost dark state.

\begin{table}[h!]
    \centering
    \caption{TDA TD-DFT excitation energies, oscillator strengths, character of the transitions, and the dipole moment of the first 10 excited states are shown. 
    Obtained for furan optimized in the gas phase using the BP86/def2-SVP.
    }
    \label{tab:excitations}
    \begin{tabular}{l|rrcr}
State	&	Excitation energy [eV]	&	Oscillator strength [10$^{-3}$]	&	Character	&	Dipole moment [Debye]	\\\hline
S$_0$	&	-	&	-	&	-	&	0.19	\\
S$_1$	&	6.49	&	0.0	&	$\pi \rightarrow \pi^*$	&	0.27	\\
S$_2$	&	6.78	&	236.5	&	$\pi \rightarrow \pi^*$	&	0.19	\\
S$_3$	&	7.24	&	0.0	&	$\pi \rightarrow$ ryd	&	0.93	\\
S$_4$	&	8.13	&	0.0	&	$\pi \rightarrow$ ryd	&	0.42	\\
S$_5$	&	8.16	&	4.2	&	$\pi \rightarrow$ ryd	&	1.46	\\
S$_6$	&	8.37	&	1.0	&	n$\rightarrow\pi^*$	&	0.17\\
S$_7$	&	8.45	&	0.2	&	$\pi \rightarrow$ ryd	&	0.46	\\
S$_8$	&	8.56	&	0.3	&	$\pi \rightarrow$ ryd	&	2.03	\\
S$_9$	&	9.05	&	1.1	&	$\sigma \rightarrow \pi^*$	&	0.11	\\
S$_{10}$	&	9.22	&	0.0	&	$\sigma \rightarrow \pi^*$,$\pi \rightarrow$ ryd	&	0.75	\\
    \end{tabular}
\end{table}

Table~\ref{tab:excitations} shows an analysis of the first ten excitations of furan in the gas phase.
The first two excitations are $\pi \rightarrow \pi^*$ transitions.
While the first excited state is dark for the equilibrium geometry, the second one is has the strongest oscillator strength.
All other states with an oscillator strength of zero in this table are 0 because of selection rules, which are do not apply to distored geometries obtained in the Wigner sampling (Fig.~\ref{fig:both_spectra}b).


\section{Parity Plots}

In this section, we have collected the parity plots for all trained models. 
The plots show the scatter of predicted vs. ground-truth label for energies and forces of the five electronic states predicted by all models as well as the indirectly predicted energy gap between neighboring levels (four differences).
The gap is not a direct output of the models, but rather the difference of two adjacent energy levels.

The first subsection shows the model performance on the original test set, i.e. frames taken from Set~I (training and validation data is always taken from this set). 
We only show parity plots for models trained on 100~\% and 33~\% of the data, as when retaining only 1~\% of the data, the parity plots are almost empty. 

In contrast, the tests performed on Set~II always use exactly the same frames. 
Therefore, the number of points does not change with the type of split or the amount of subsampling used during the training. 

\clearpage
\subsection{Performance on Set~I}

\subsubsection{Split by Trajectory; 100~\% of Available Data}

\begin{figure}[h!]
    \centering
    \begin{tabular}{ll}
         Model 1 &
         \adjustbox{valign=t}{\includegraphics[width=0.64\linewidth]{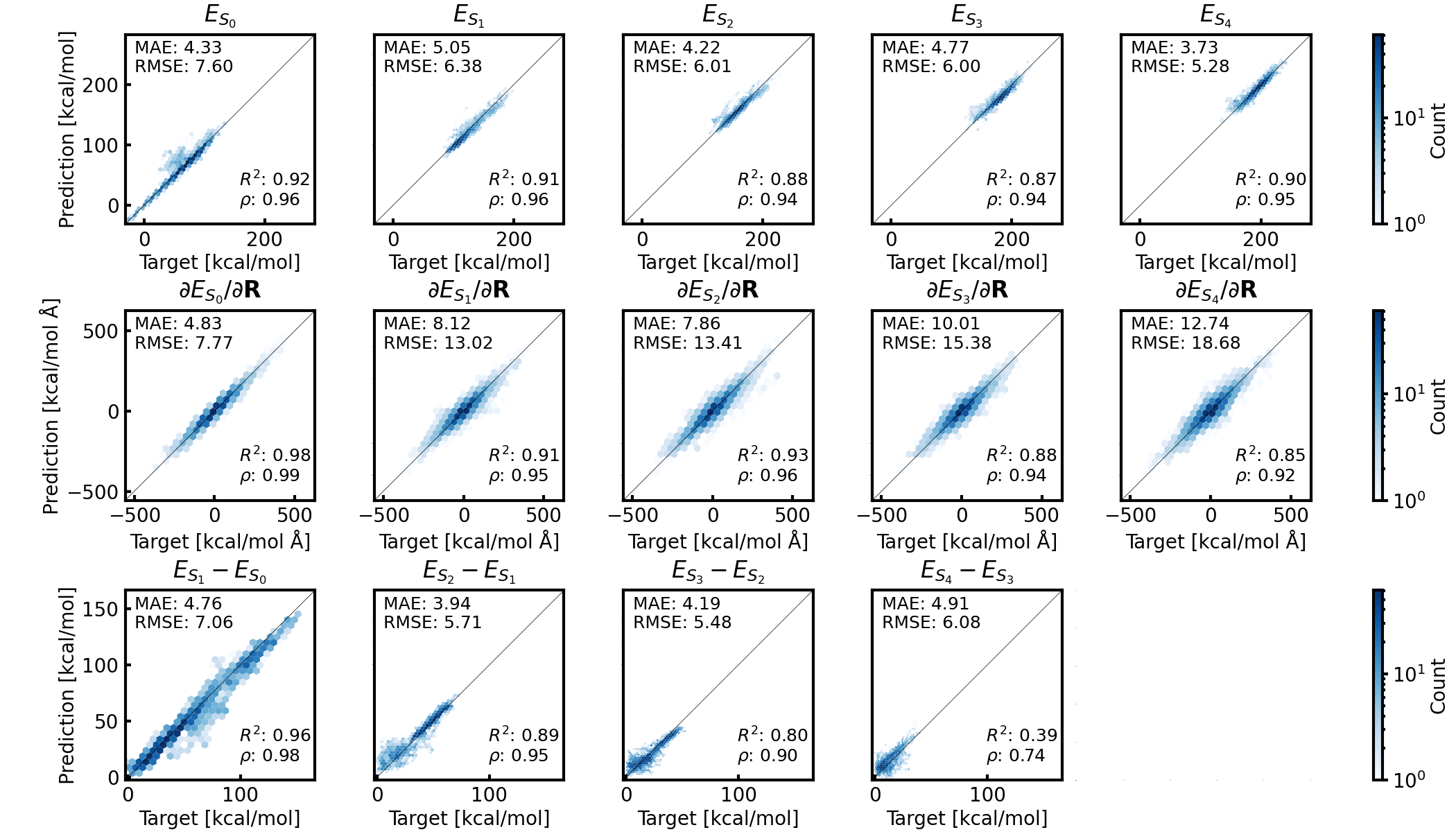} }\\
         \\
         Model 2 &
         \adjustbox{valign=t}{\includegraphics[width=0.64\linewidth]{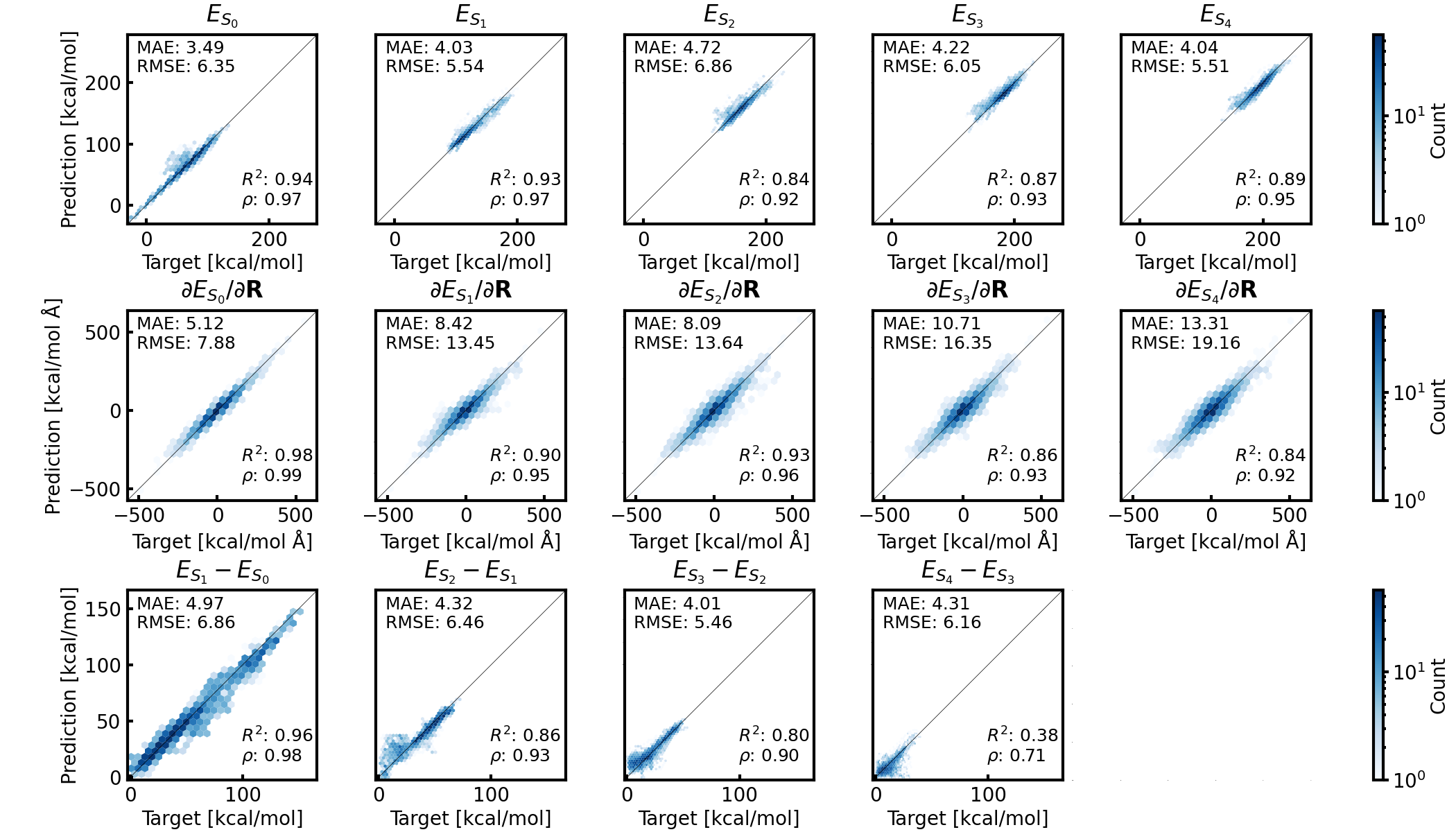} }\\
         \\
         Model 3 &
         \adjustbox{valign=t}{\includegraphics[width=0.64\linewidth]{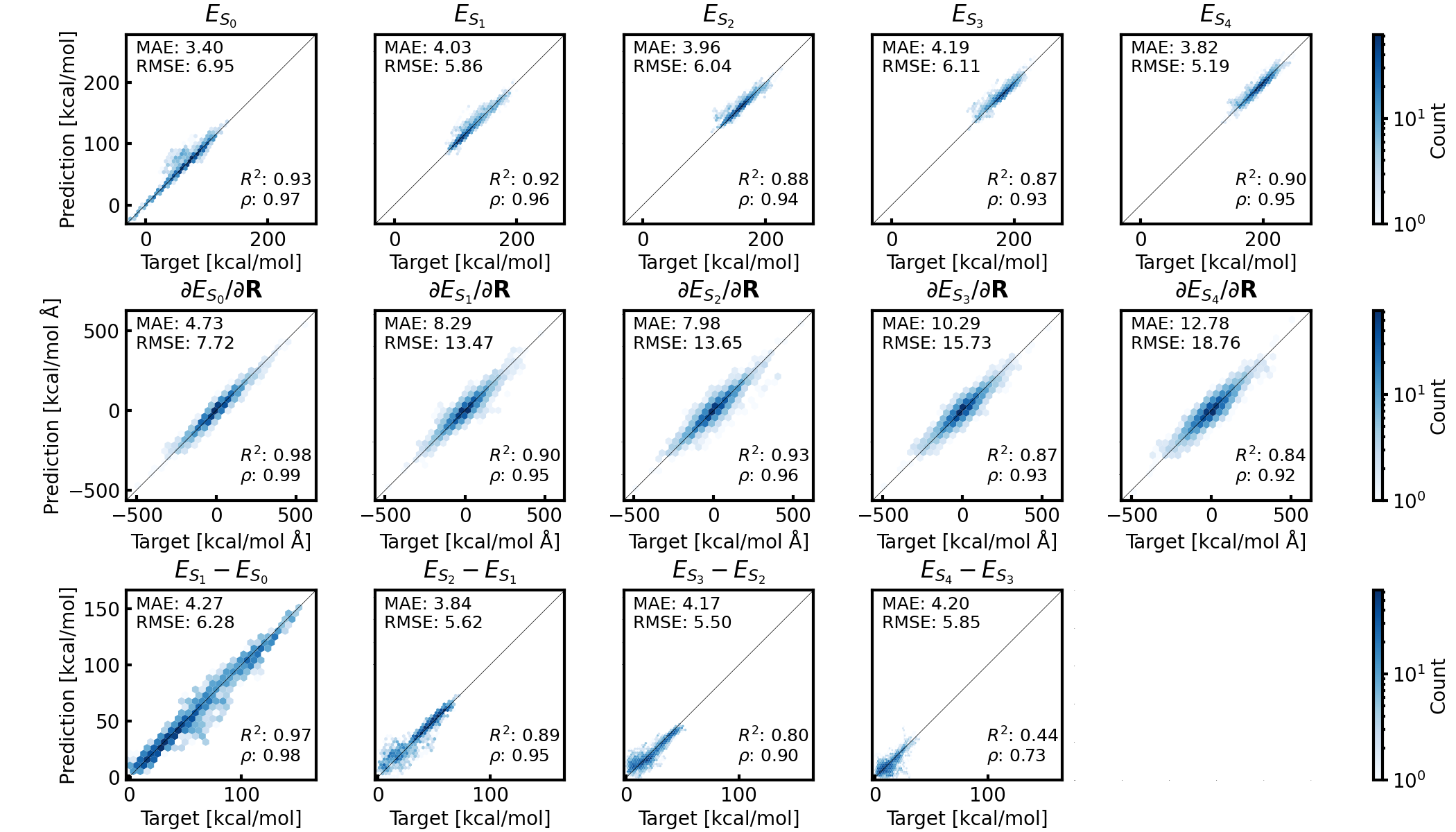} }
    \end{tabular}
    \caption{Parity plots for the set~I test set for the three "Split by Trajectory" models trained on 100~\% of the available frames (every 0.5~fs) from the 36 training trajectories.}
    \label{fig:SplitbyTraj_SetI_skip1}
\end{figure}

\clearpage
\subsubsection{Split by Trajectory; 33~\% of Available Data}

\begin{figure}[h!]
    \centering
    \begin{tabular}{ll}
         Model 1 &
         \adjustbox{valign=t}{\includegraphics[width=0.66\linewidth]{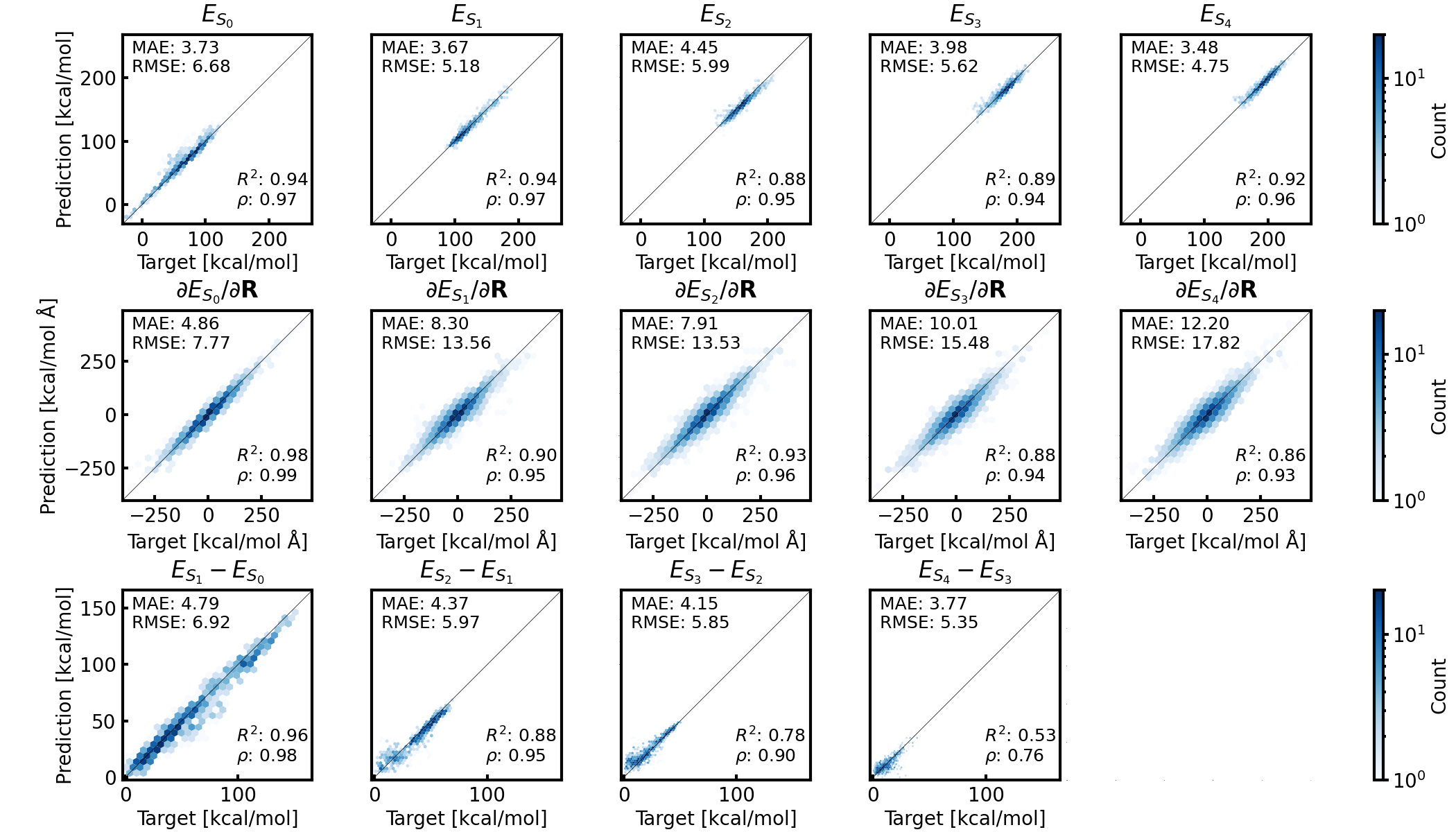} }\\
         \\
         Model 2 &
         \adjustbox{valign=t}{\includegraphics[width=0.66\linewidth]{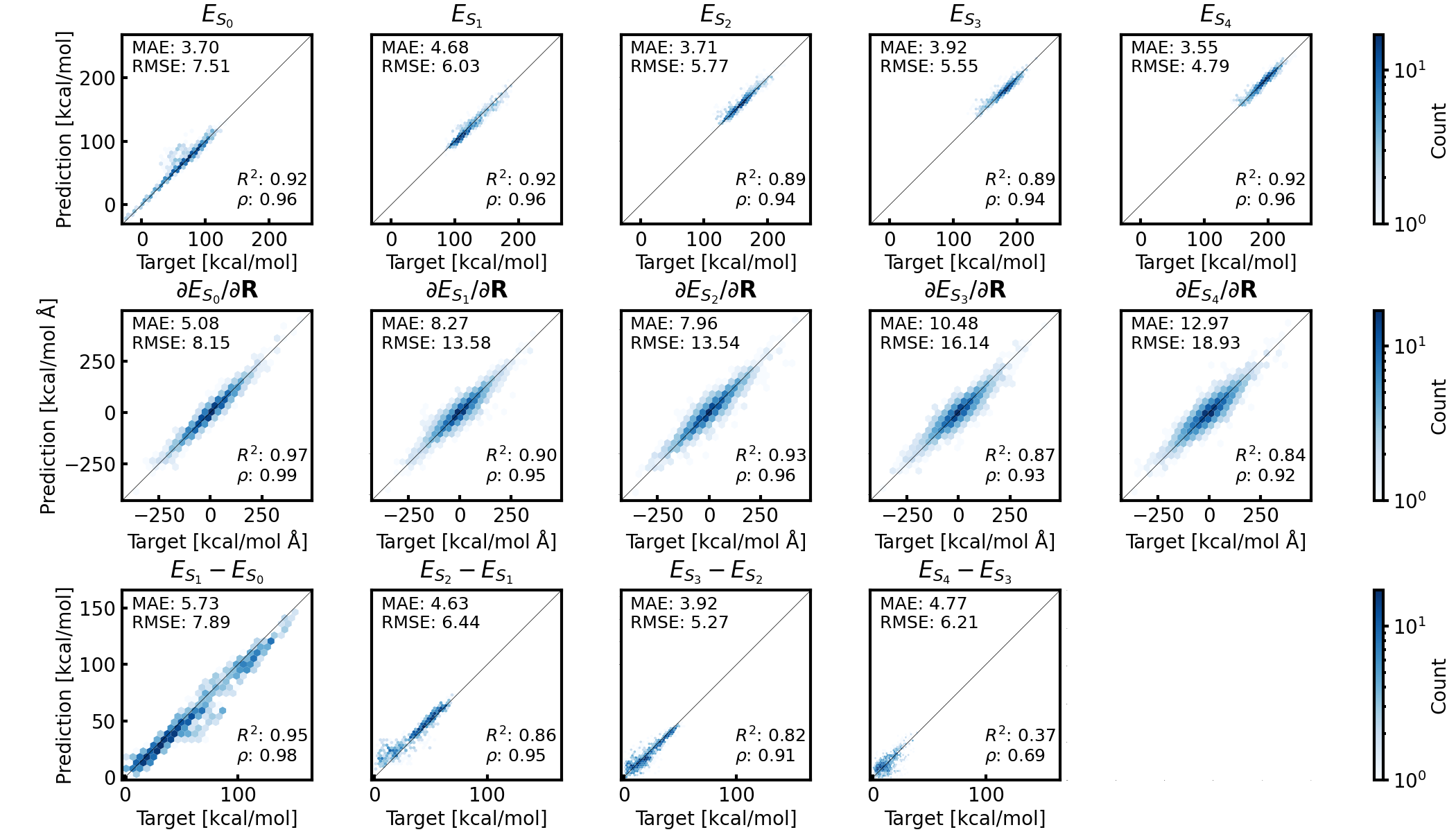} }\\
         \\
         Model 3 &
         \adjustbox{valign=t}{\includegraphics[width=0.66\linewidth]{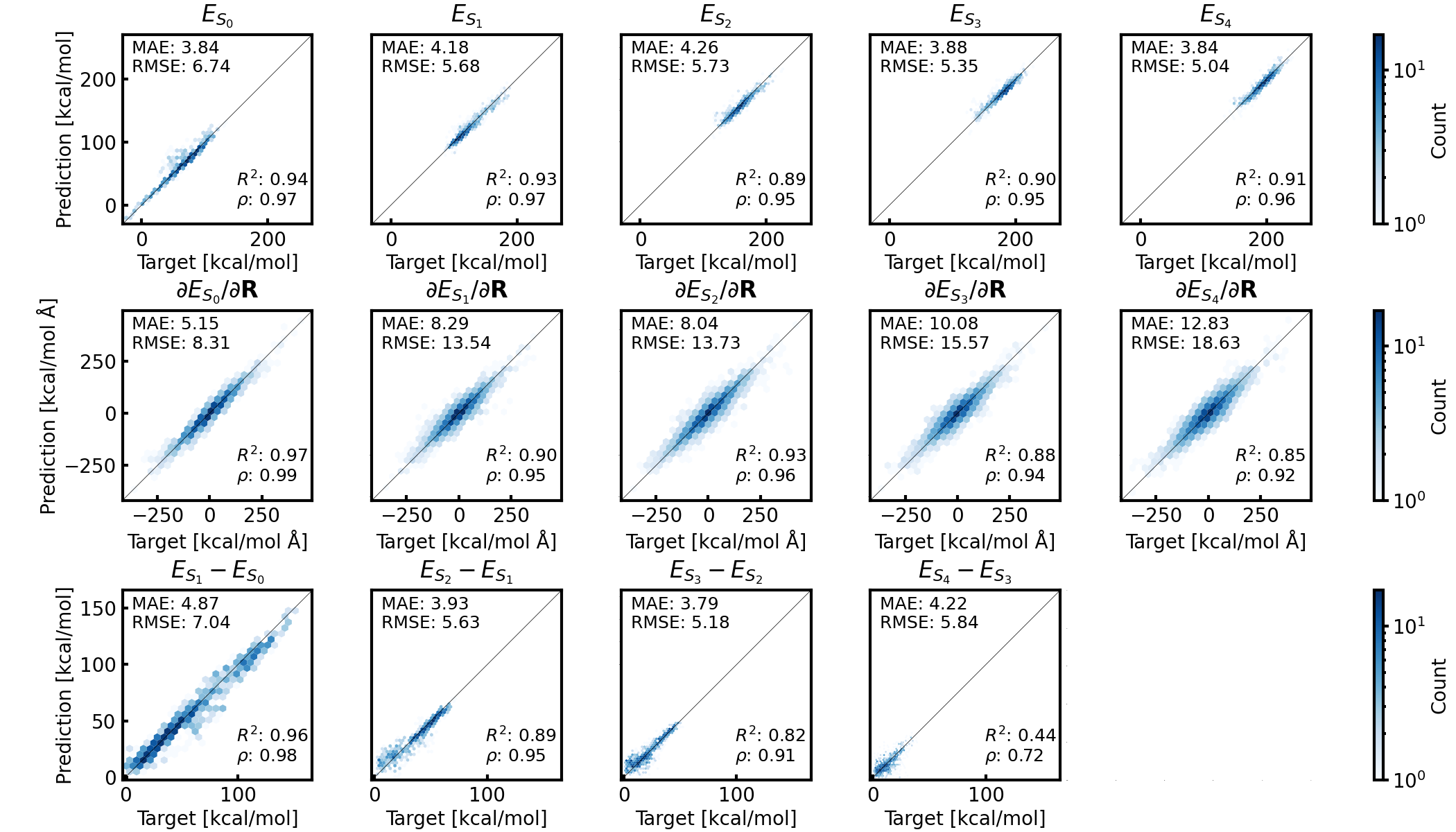} }
    \end{tabular}
    \caption{Parity plots for the set~I test set for the three "Split by Trajectory" models trained on 33~\% of the available frames (every 1.5~fs) from the 36 training trajectories.}
    \label{fig:SplitbyTraj_SetI_skip3}
\end{figure}

\clearpage
\subsubsection{Random Split; 100~\% of Available Data}

\begin{figure}[h!]
    \centering
    \begin{tabular}{ll}
         Model 1 &
         \adjustbox{valign=t}{\includegraphics[width=0.66\linewidth]{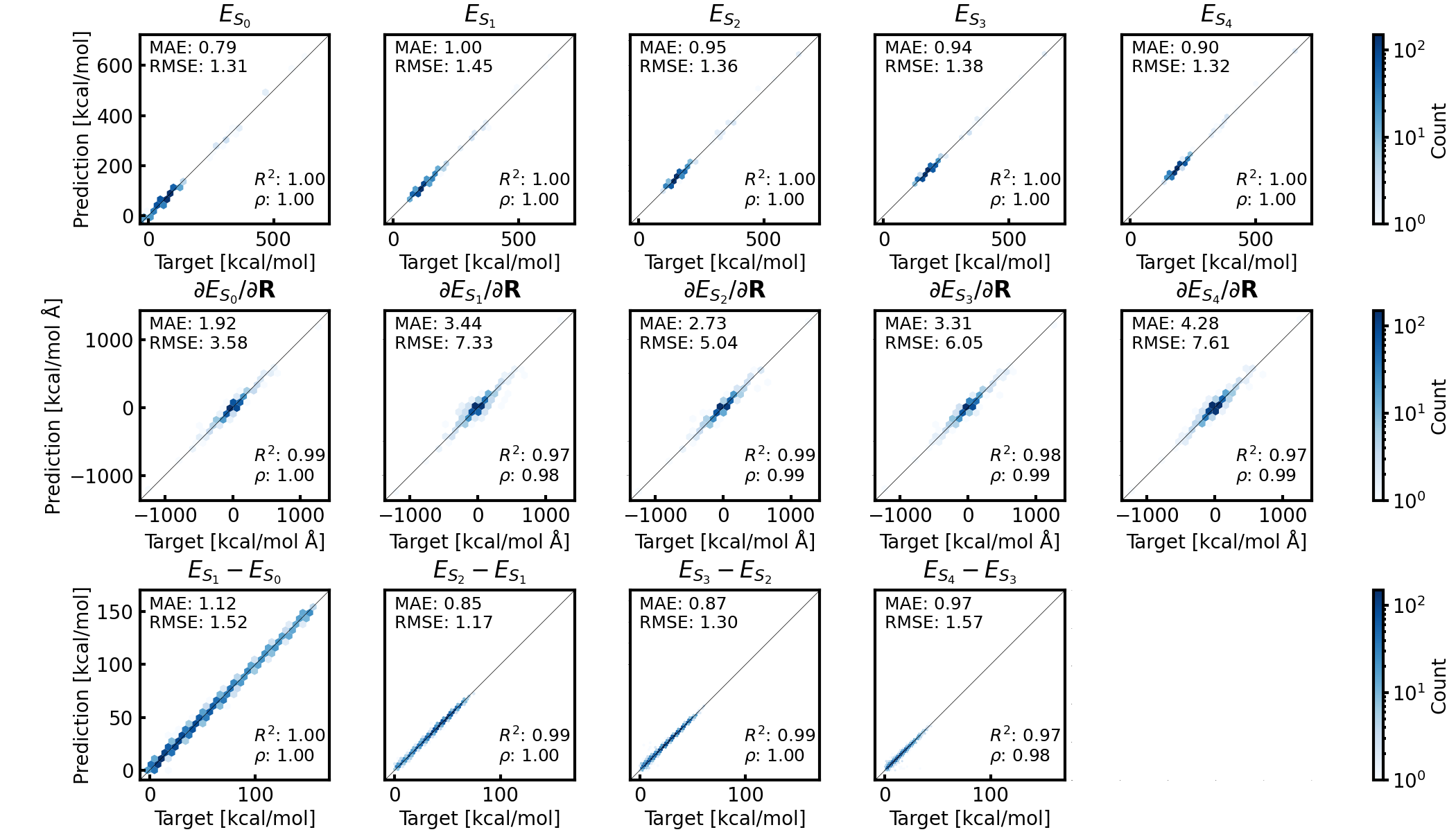} }\\
         \\
         Model 2 &
         \adjustbox{valign=t}{\includegraphics[width=0.66\linewidth]{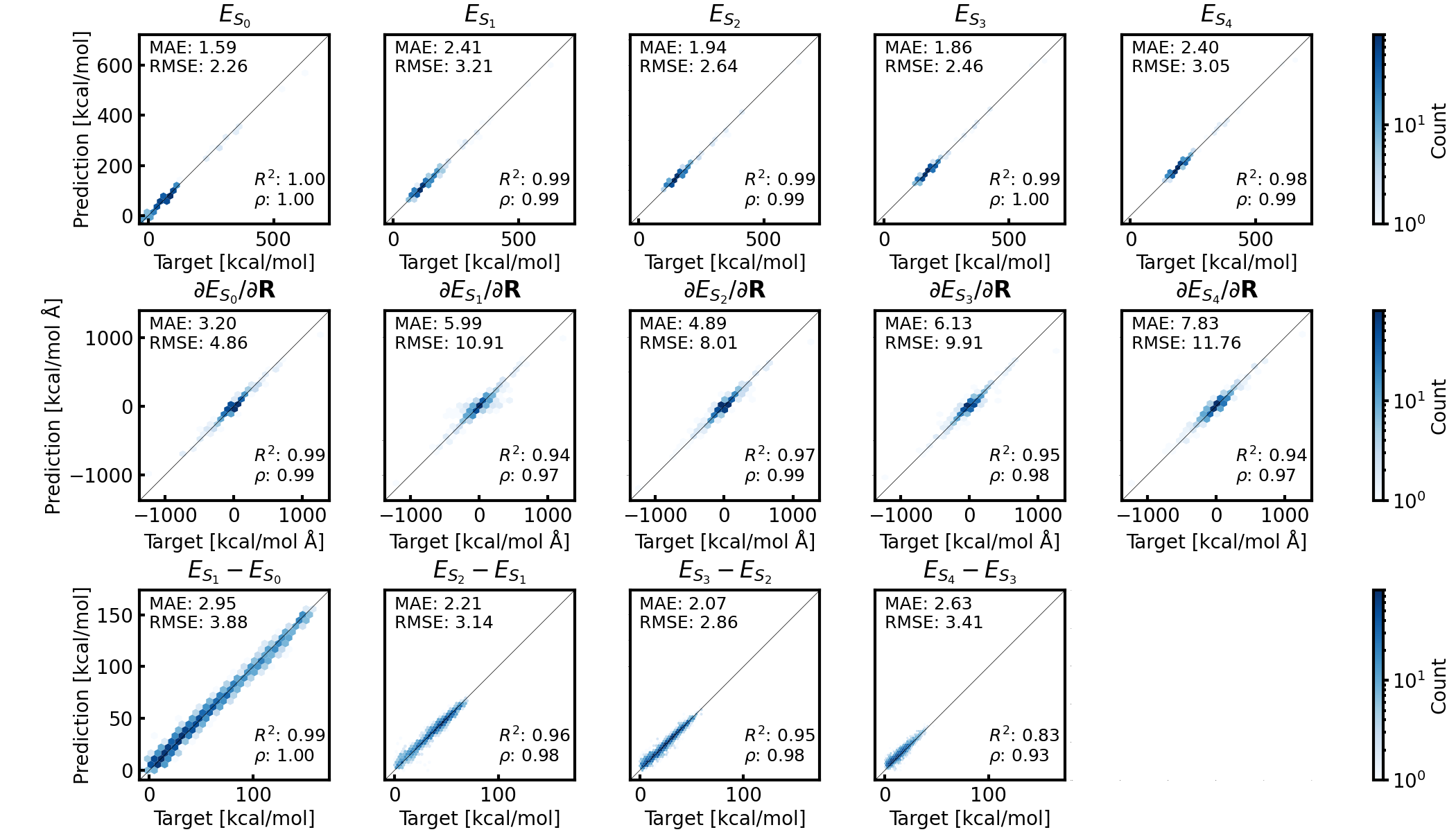} }\\
         \\
         Model 3 &
         \adjustbox{valign=t}{\includegraphics[width=0.66\linewidth]{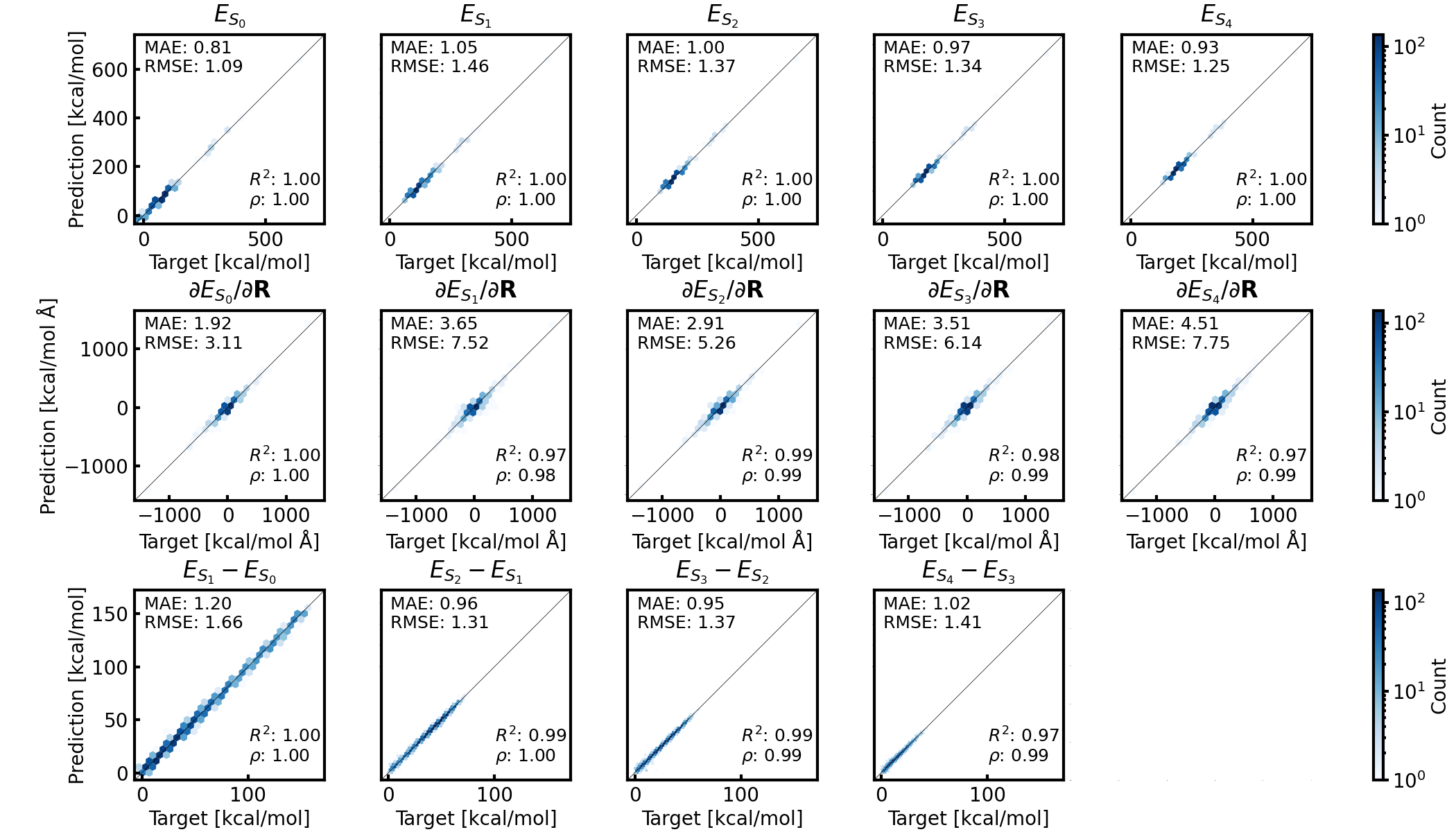} }
    \end{tabular}
    \caption{Parity plots for the set~I test set for the three "Random Split" models trained on 100~\% of the available frames (every 0.5~fs) from the 36 training trajectories.}
    \label{fig:RandomSplit_SetI_skip1}
\end{figure}

\clearpage
\subsubsection{Random Split; 33~\% of Available Data}

\begin{figure}[h!]
    \centering
    \begin{tabular}{ll}
         Model 1 &
         \adjustbox{valign=t}{\includegraphics[width=0.66\linewidth]{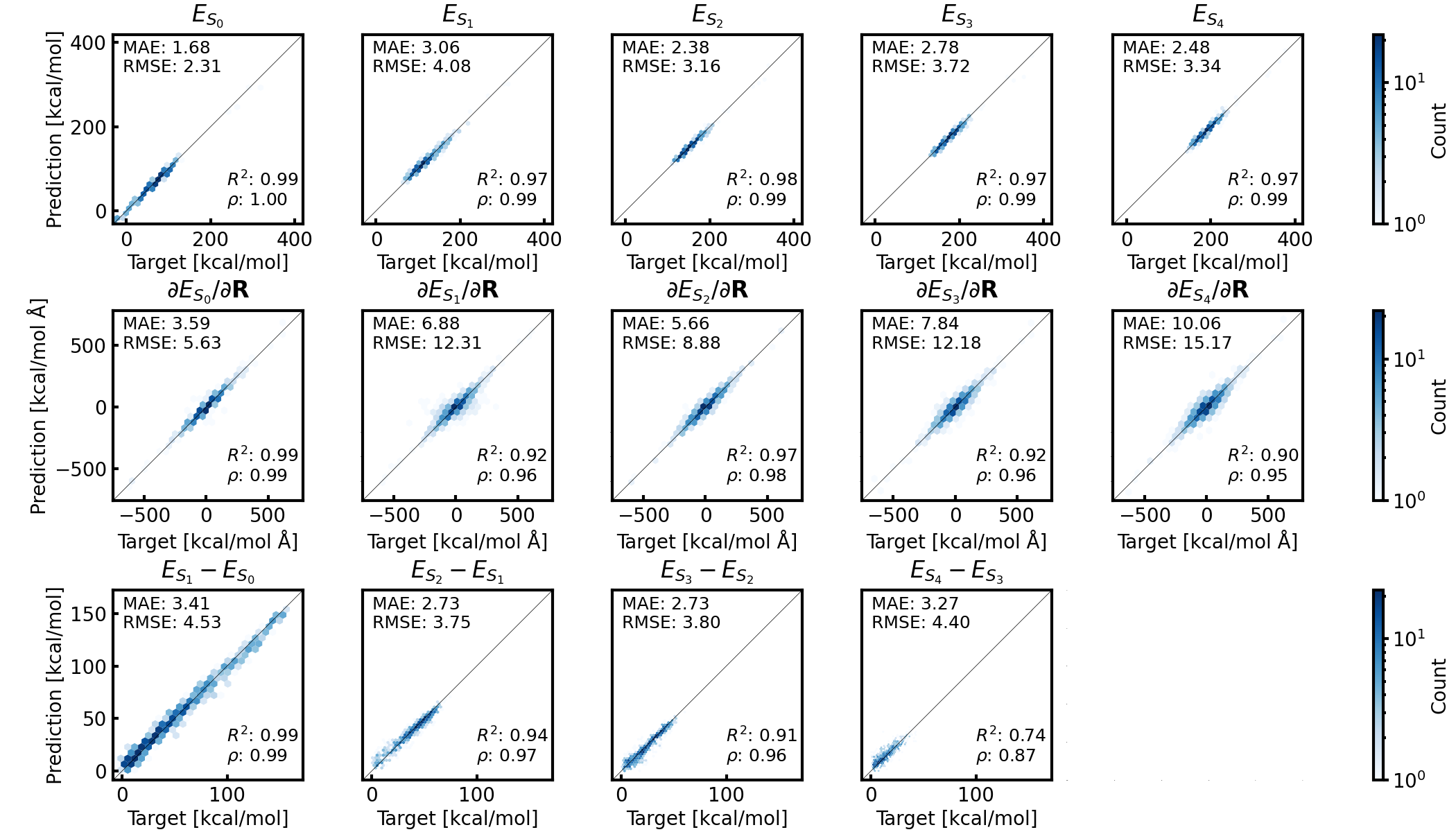} }\\
         \\
         Model 2 &
         \adjustbox{valign=t}{\includegraphics[width=0.66\linewidth]{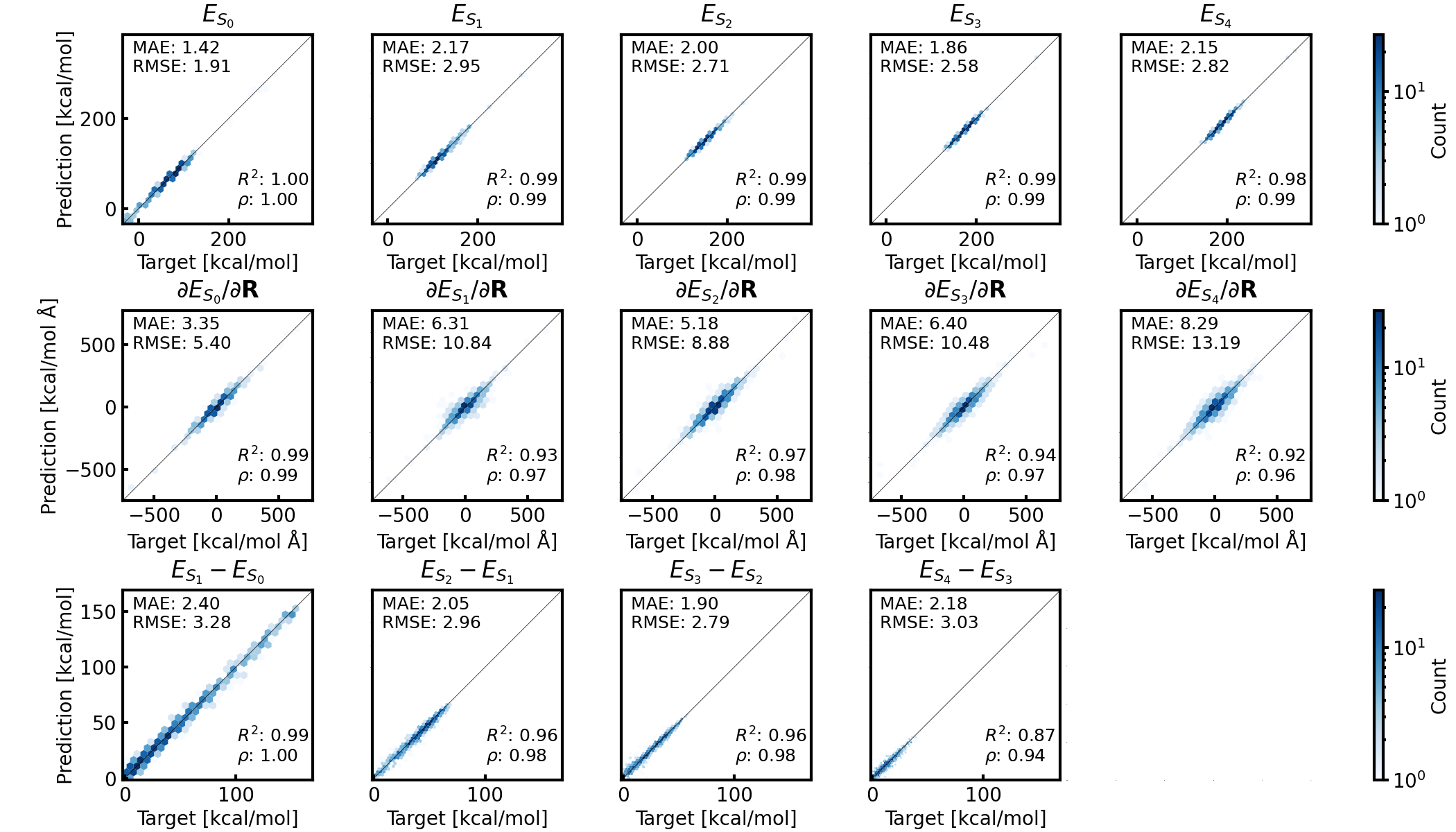} }\\
         \\
         Model 3 &
         \adjustbox{valign=t}{\includegraphics[width=0.66\linewidth]{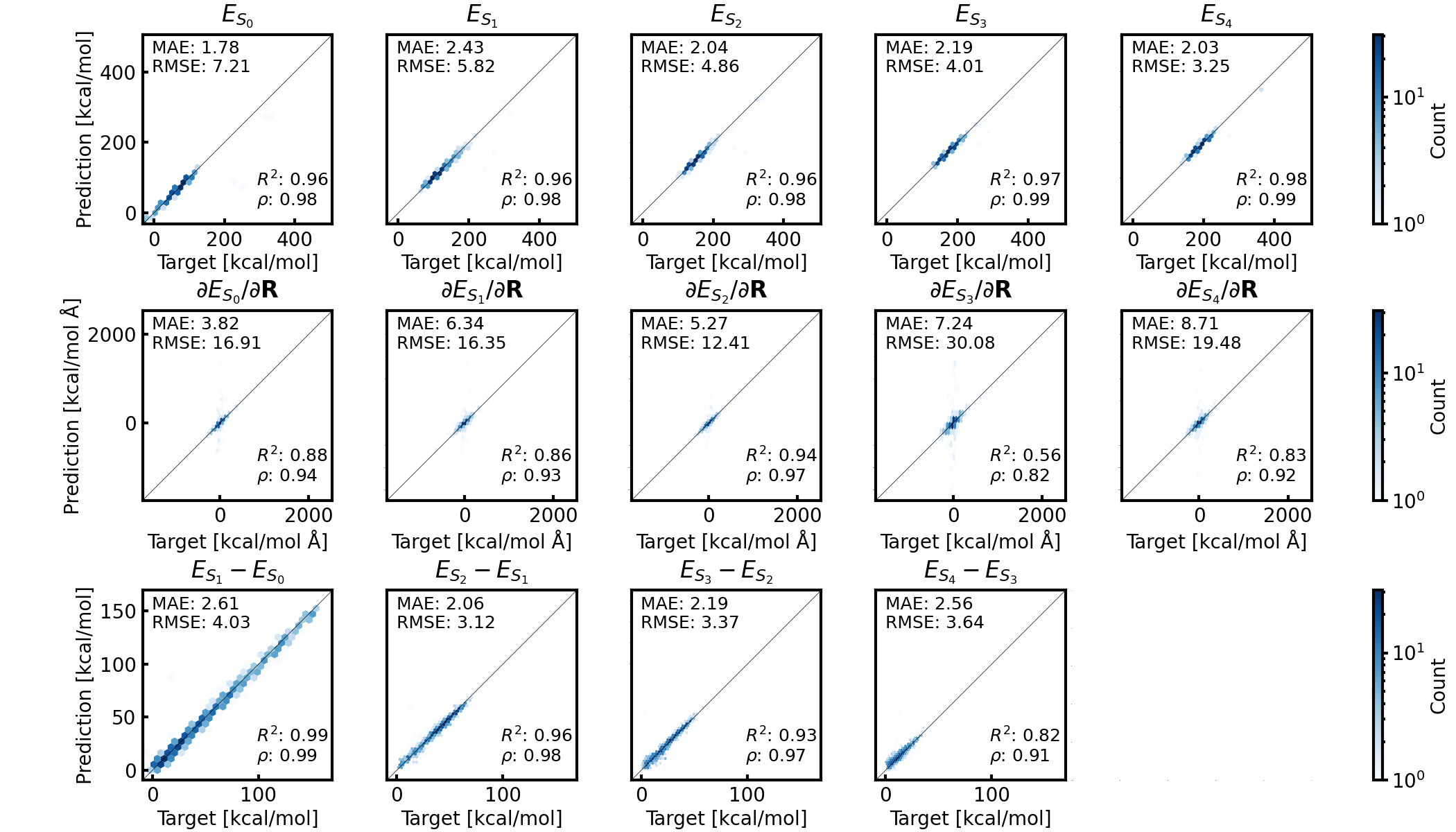} }
    \end{tabular}
    \caption{Parity plots for the set~I test set for the three "Random Split" models trained on 33~\% of the available frames (every 1.5~fs) from the 36 training trajectories.}
    \label{fig:RandomSplit_SetI_skip3}
\end{figure}

\clearpage
\subsection{Performance on Set~II}

\subsubsection{Split by Trajectory; 100~\% of Available Data}

\begin{figure}[h!]
    \centering
    \begin{tabular}{ll}
         Model 1 &
         \adjustbox{valign=t}{\includegraphics[width=0.64\linewidth]{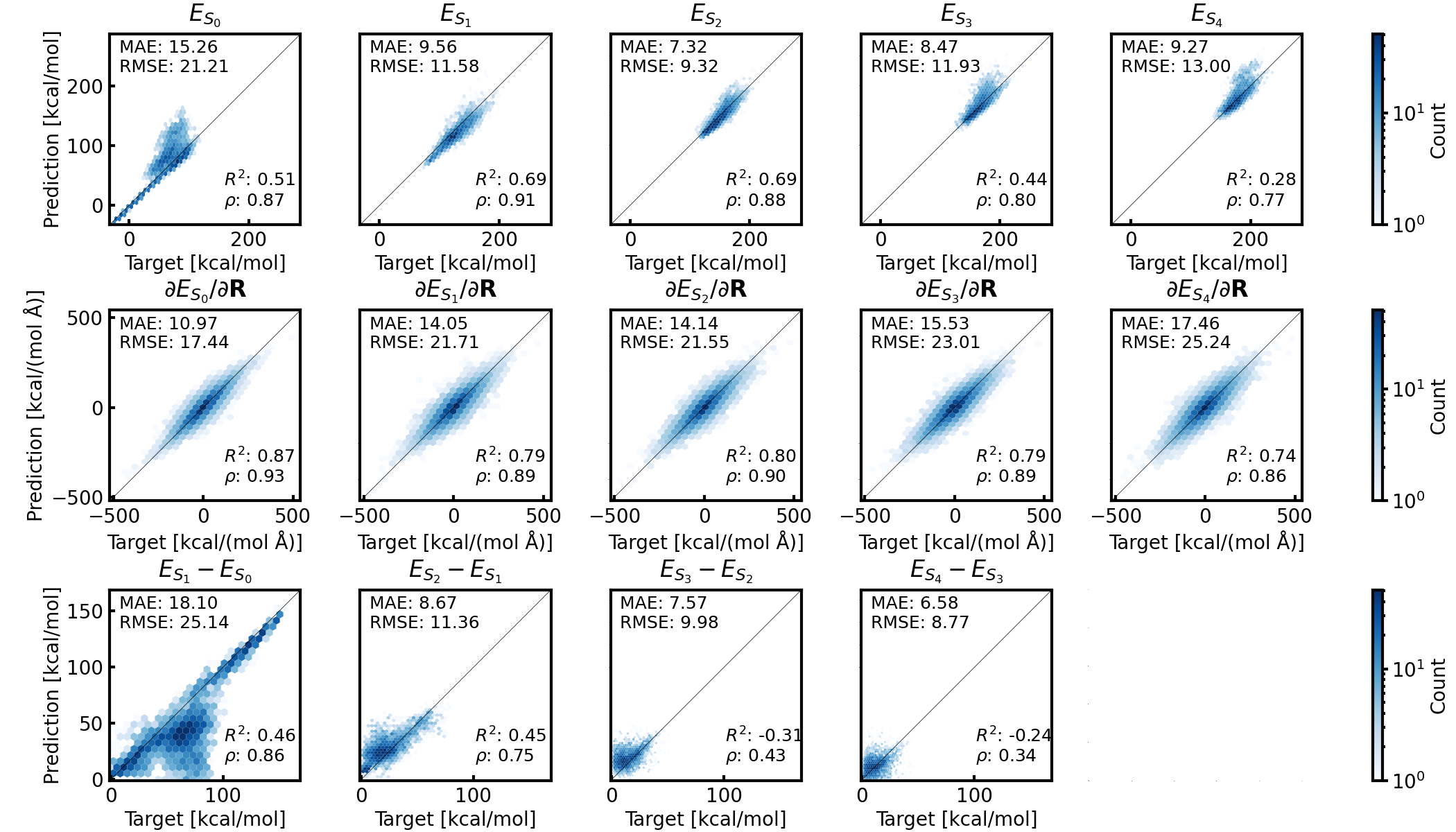} }\\
         \\
         Model 2 &
         \adjustbox{valign=t}{\includegraphics[width=0.64\linewidth]{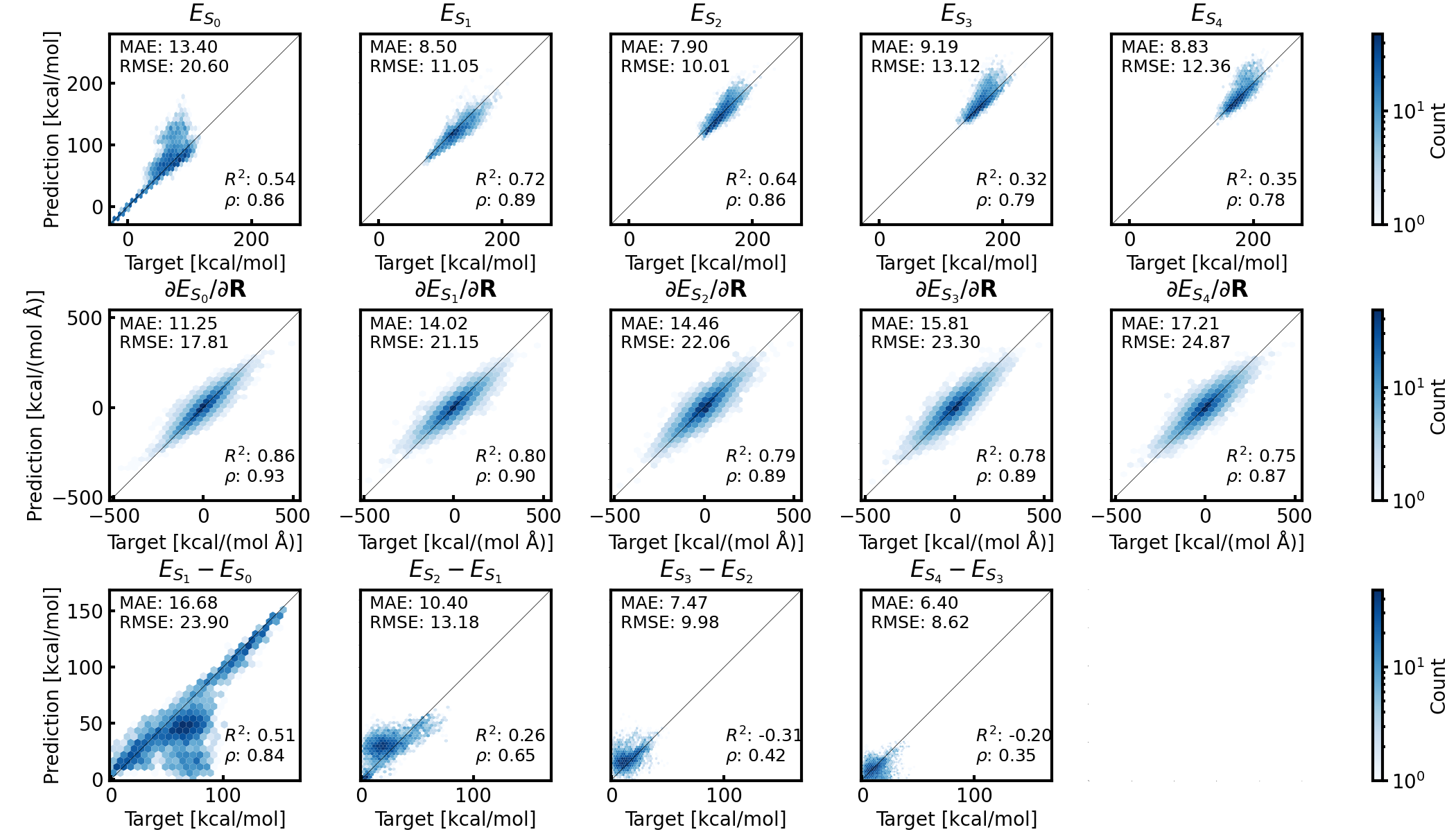} }\\
         \\
         Model 3 &
         \adjustbox{valign=t}{\includegraphics[width=0.64\linewidth]{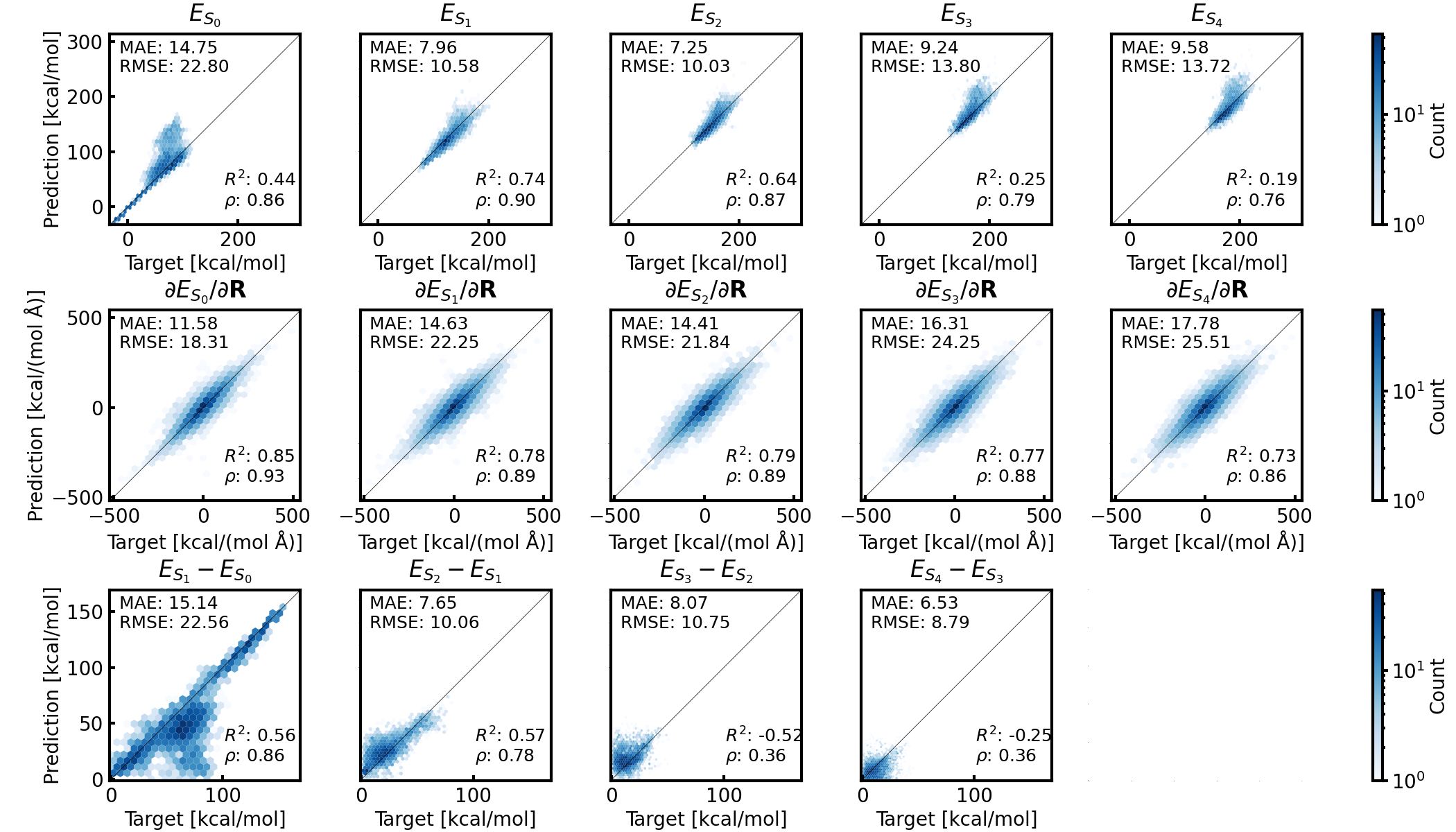} }
    \end{tabular}
    \caption{Parity plots on frames from Set~II for the three "Split by Trajectory" models trained on 100~\% of the available frames (every 0.5~fs) from the 36 training trajectories.}
    \label{fig:SplitbyTraj_SetII_skip1}
\end{figure}

\clearpage
\subsubsection{Split by Trajectory; 33~\% of Available Data}

\begin{figure}[h!]
    \centering
    \begin{tabular}{ll}
         Model 1 &
         \adjustbox{valign=t}{\includegraphics[width=0.66\linewidth]{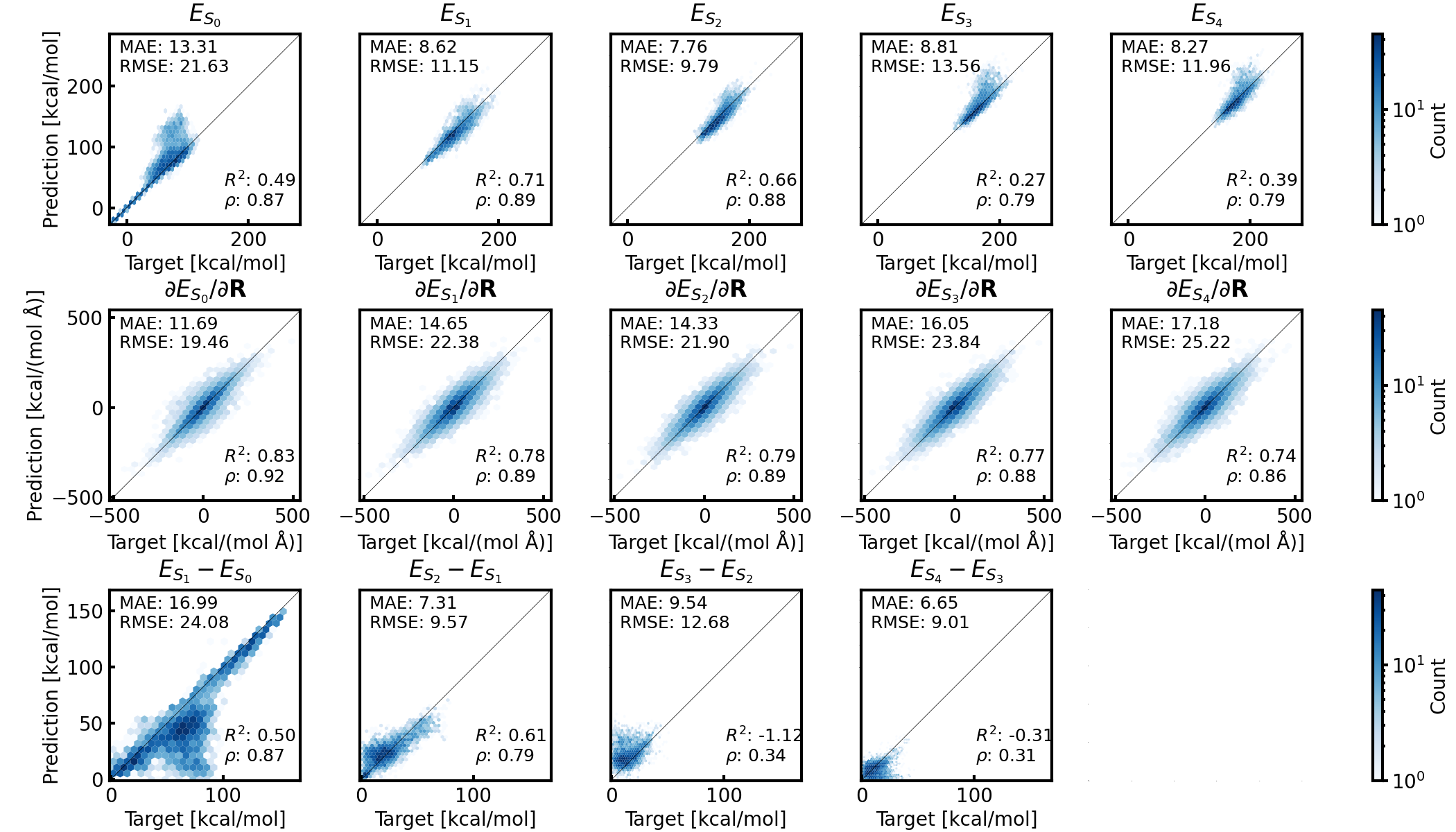} }\\
         \\
         Model 2 &
         \adjustbox{valign=t}{\includegraphics[width=0.66\linewidth]{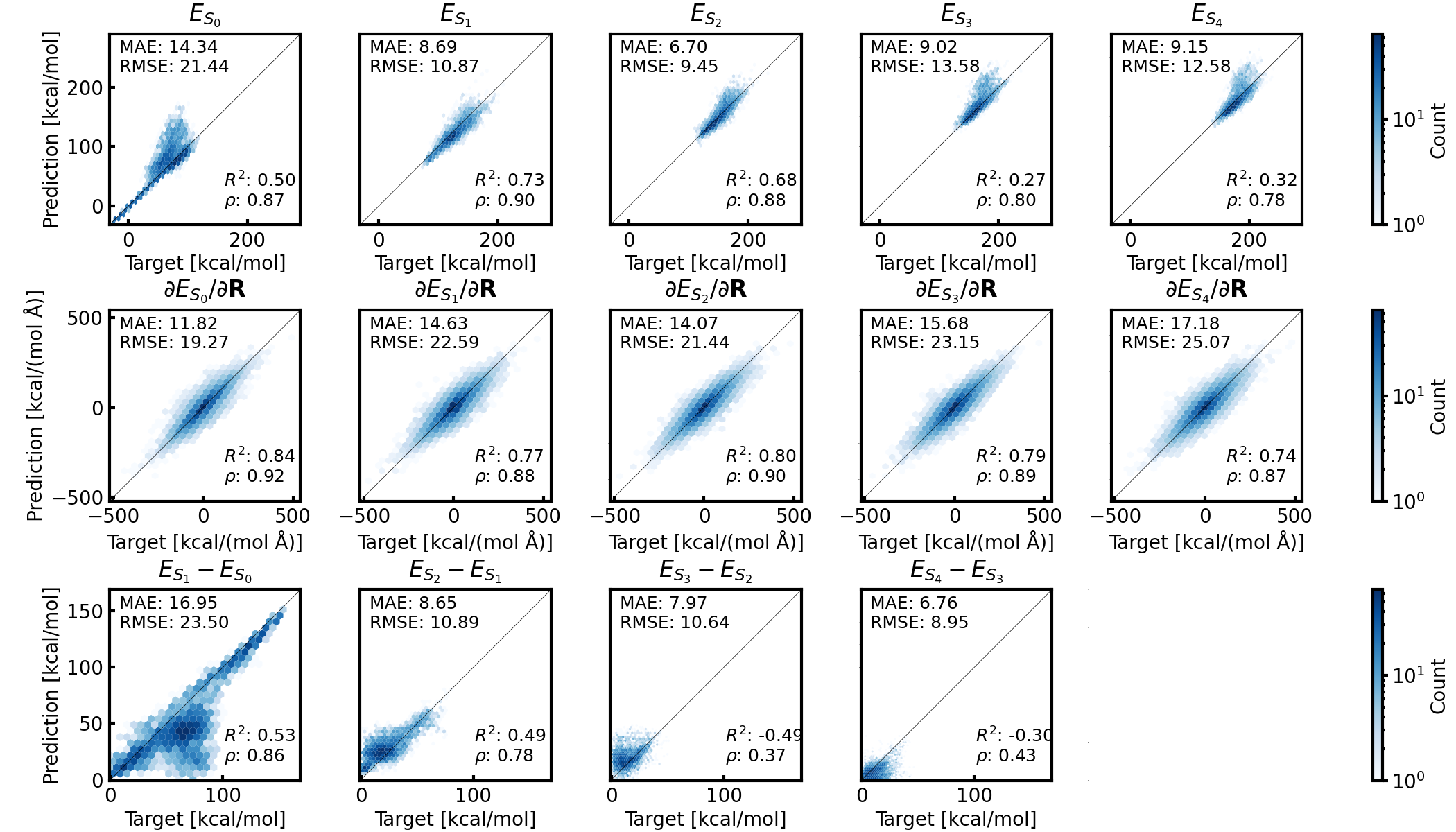} }\\
         \\
         Model 3 &
         \adjustbox{valign=t}{\includegraphics[width=0.66\linewidth]{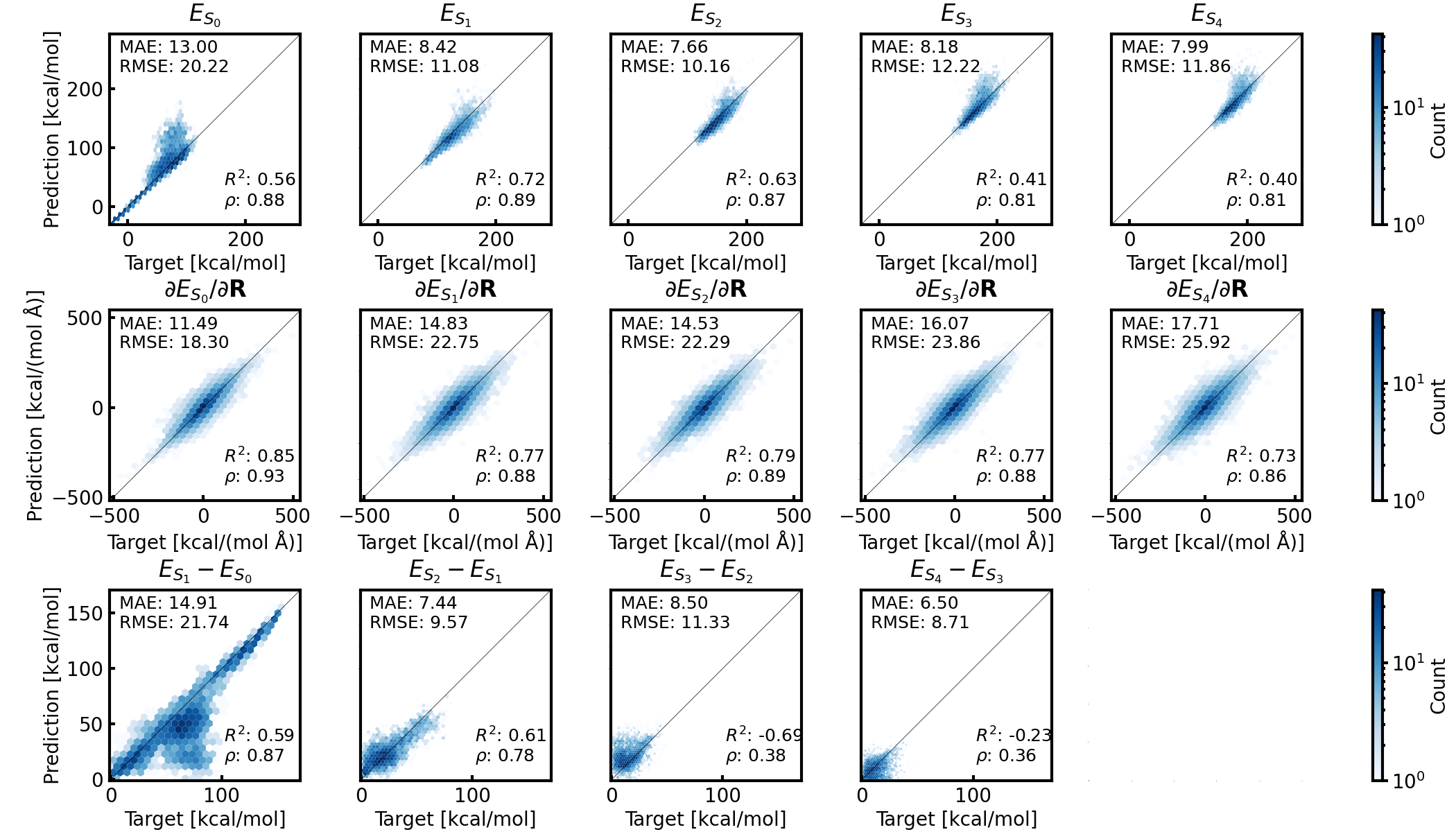} }
    \end{tabular}
    \caption{Parity plots on frames from Set~II for the three "Split by Trajectory" models trained on 33~\% of the available frames (every 1.5~fs) from the 36 training trajectories.}
    \label{fig:SplitbyTraj_SetII_skip3}
\end{figure}

\clearpage
\subsubsection{Split by Trajectory; 1~\% of Available Data}

\begin{figure}[h!]
    \centering
    \begin{tabular}{ll}
         Model 1 &
         \adjustbox{valign=t}{\includegraphics[width=0.66\linewidth]{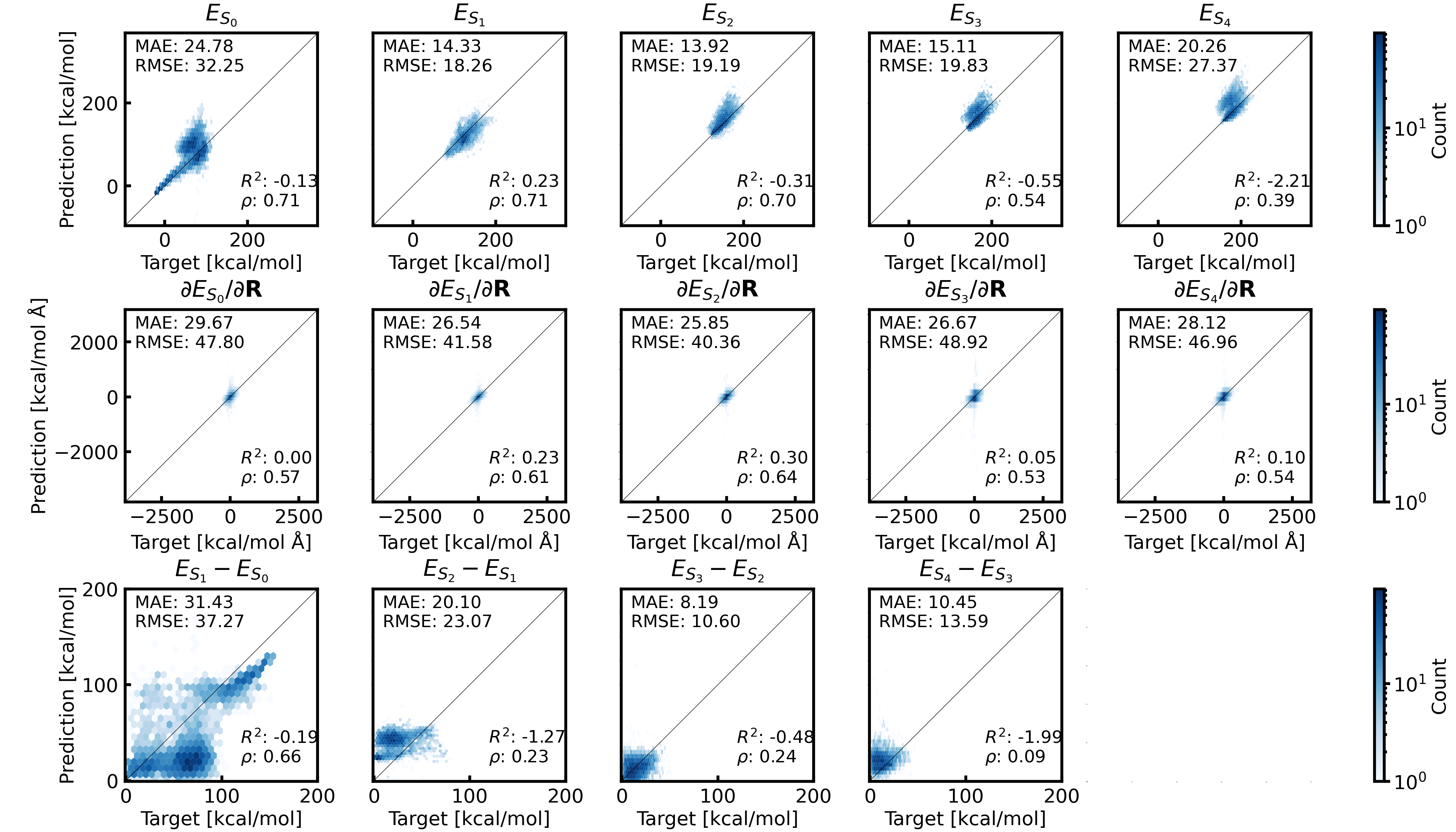} }\\
         \\
         Model 2 &
         \adjustbox{valign=t}{\includegraphics[width=0.66\linewidth]{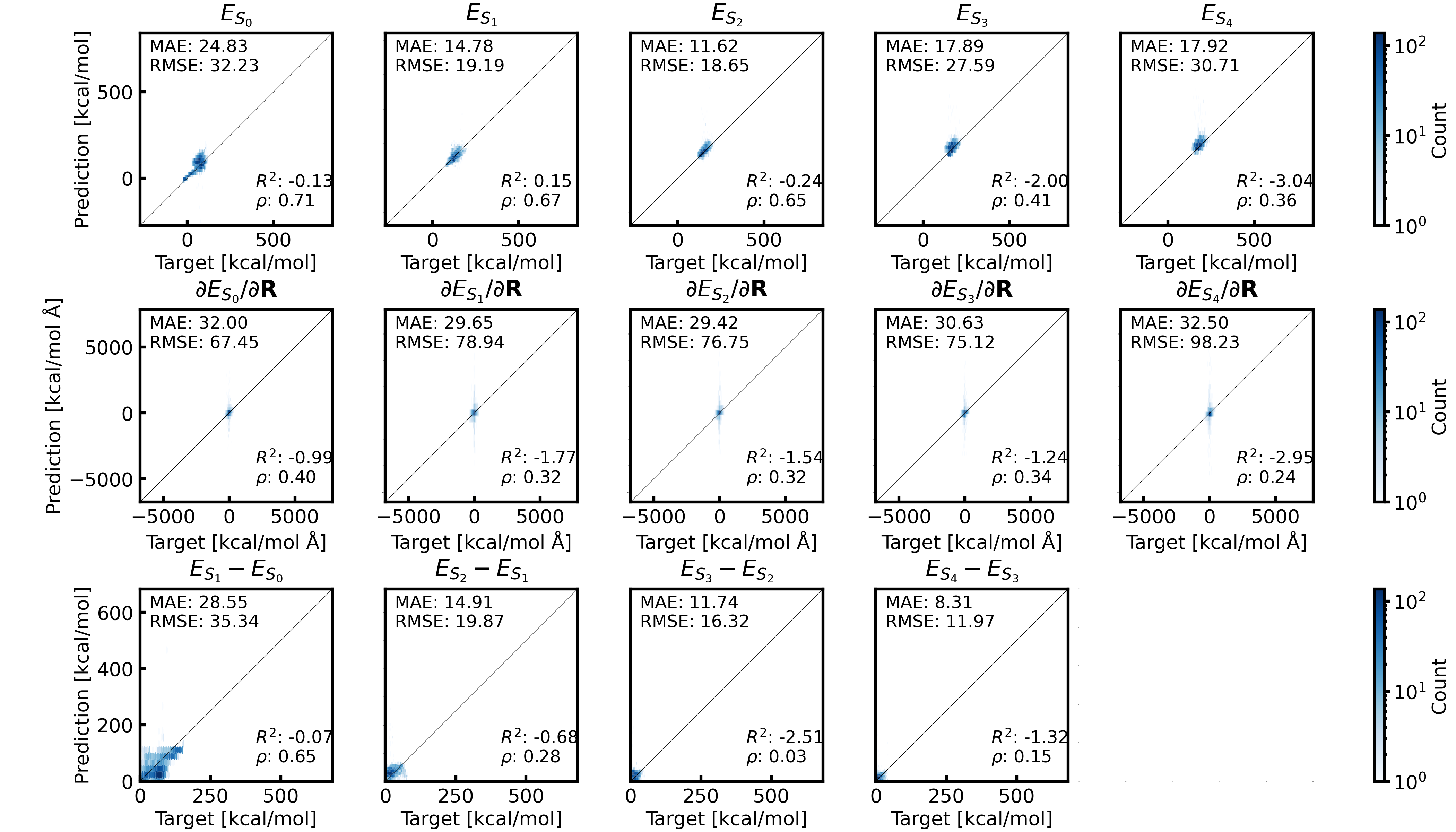} }\\
         \\
         Model 3 &
         \adjustbox{valign=t}{\includegraphics[width=0.66\linewidth]{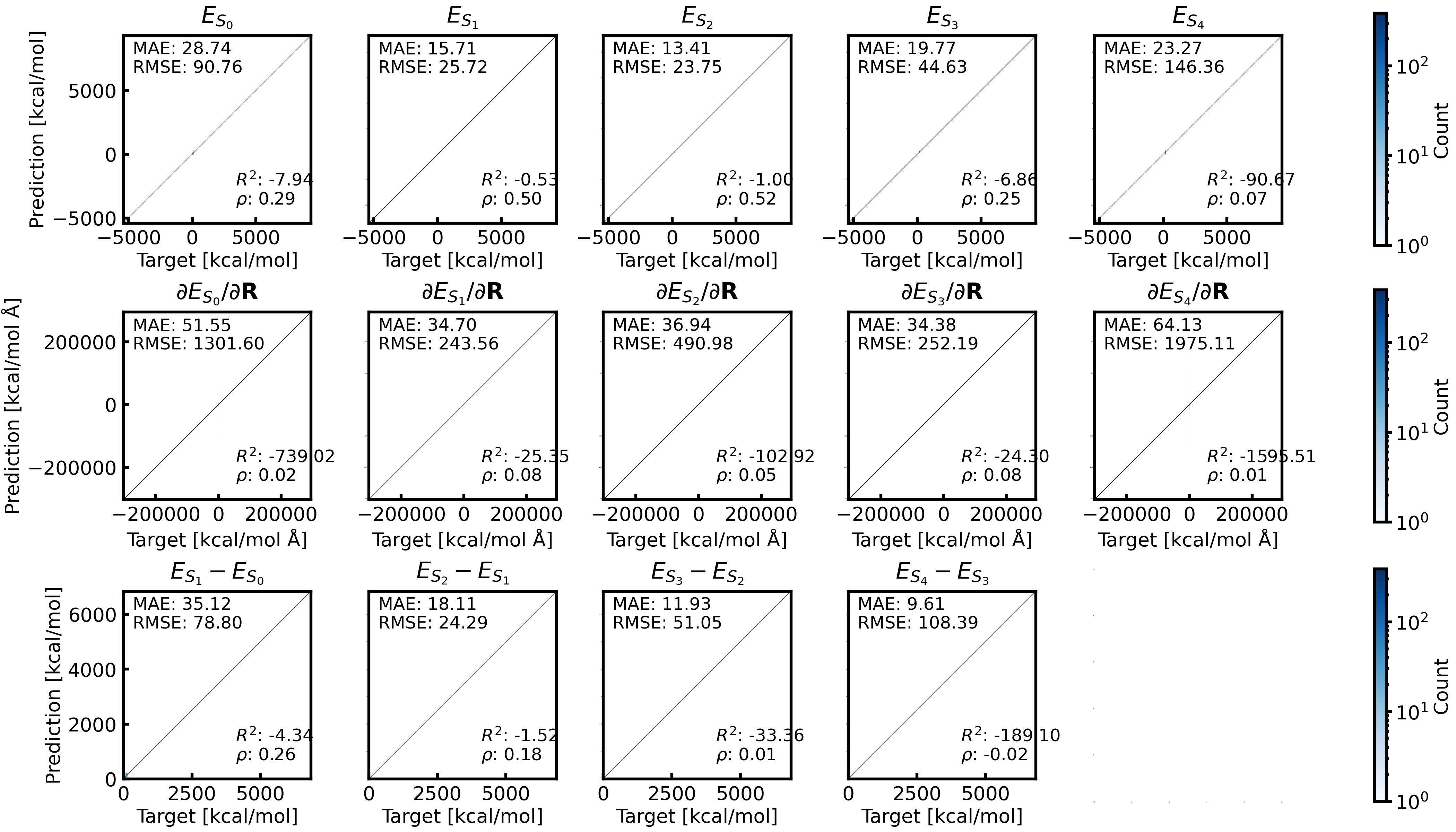} }
    \end{tabular}
    \caption{Parity plots on frames from Set~II for the three "Split by Trajectory" models trained on 1~\% of the available frames (every 50~fs) from the 36 training trajectories.}
    \label{fig:SplitbyTraj_SetII_skip100}
\end{figure}

\clearpage
\subsubsection{Random Split; 100~\% of Available Data}
\begin{figure}[h!]
    \centering
    \begin{tabular}{ll}
         Model 1 &
         \adjustbox{valign=t}{\includegraphics[width=0.66\linewidth]{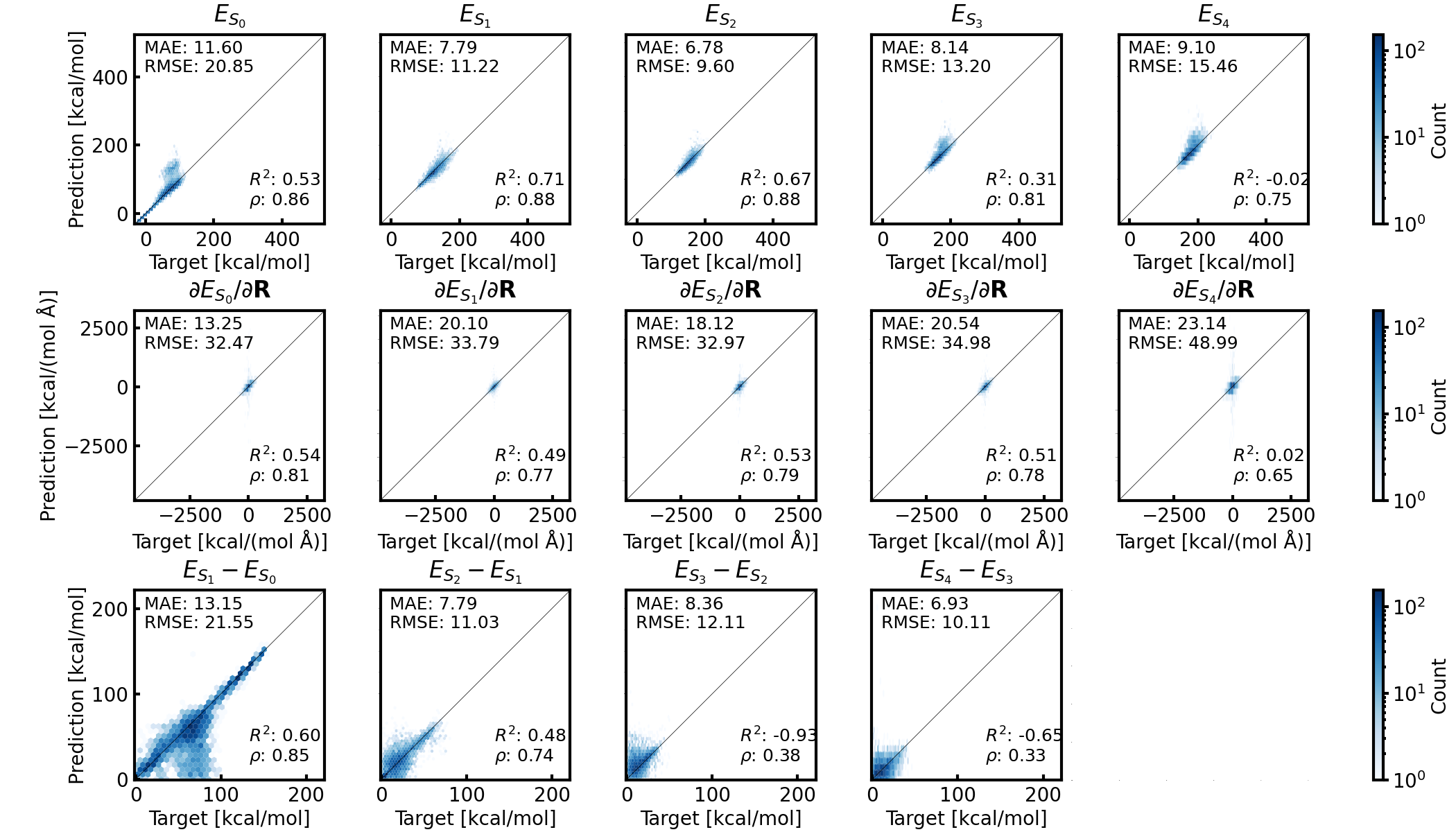} }\\
         \\
         Model 2 &
         \adjustbox{valign=t}{\includegraphics[width=0.66\linewidth]{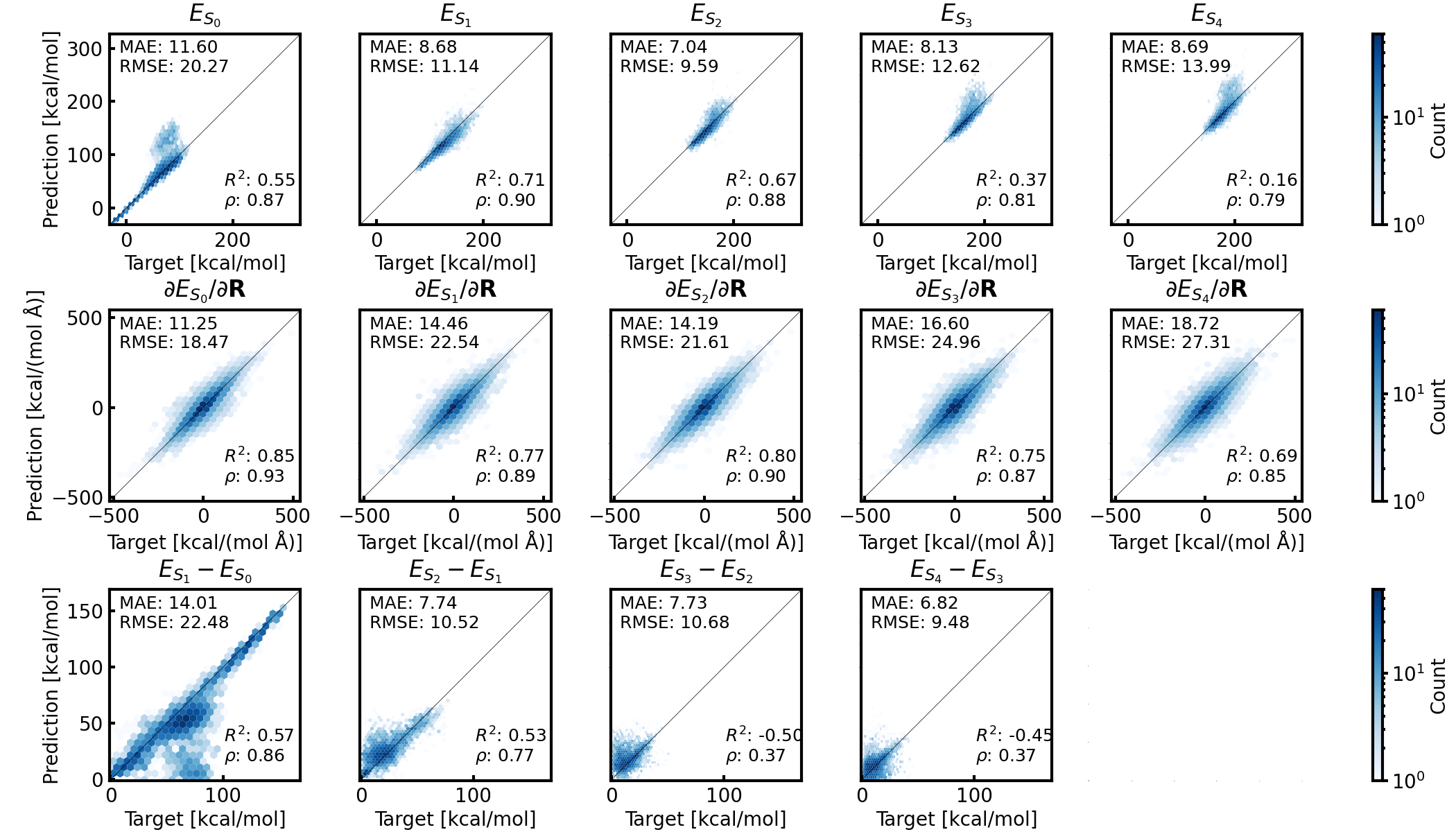} }\\
         \\
         Model 3 &
         \adjustbox{valign=t}{\includegraphics[width=0.66\linewidth]{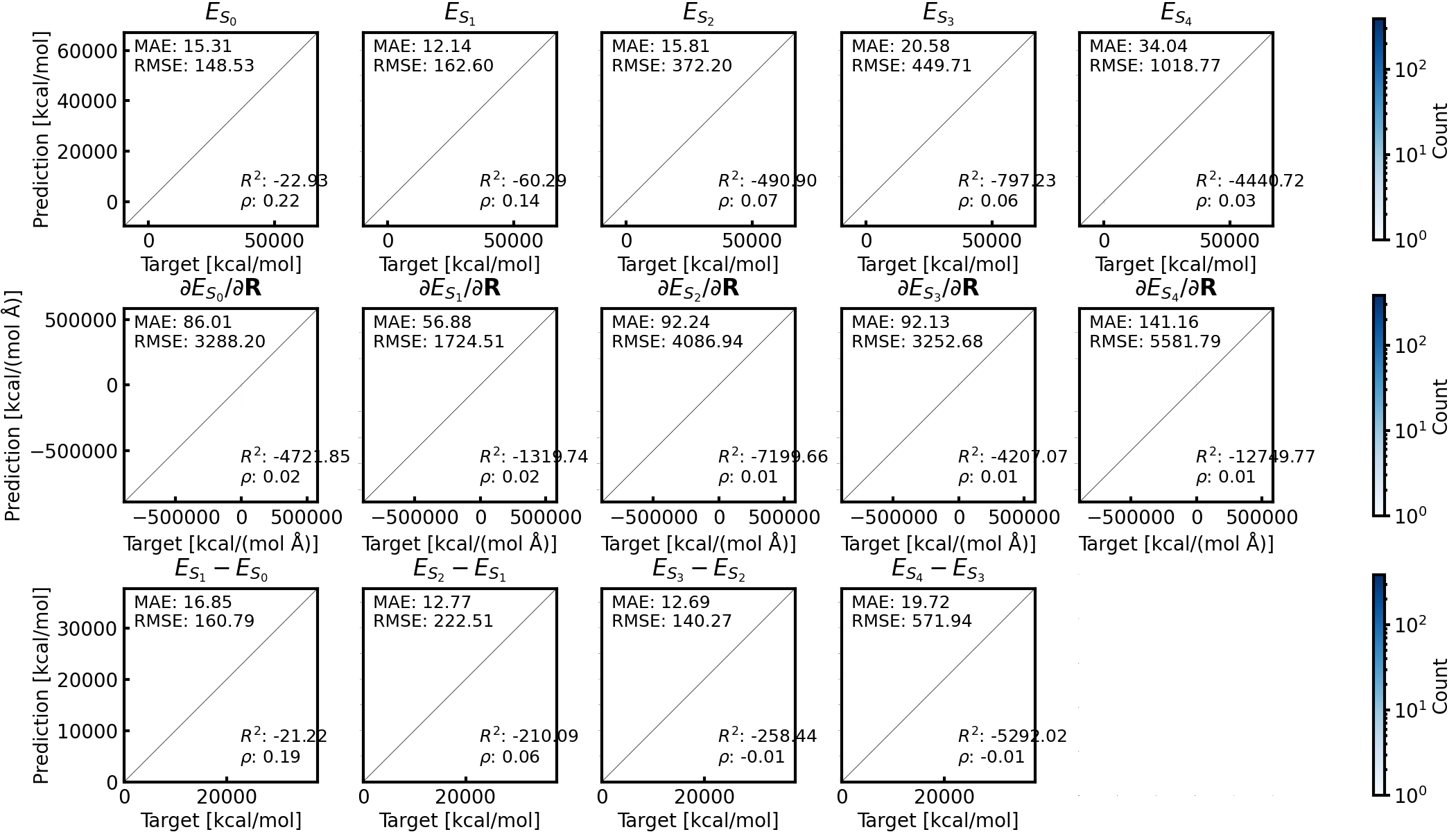} }
    \end{tabular}
    \caption{Parity plots on frames from Set~II for original test for the three "Random Split" models trained on 100~\% of the available frames (every 0.5~fs) from the 36 training trajectories.}
    \label{fig:RandomSplit_SetII_skip1}
\end{figure}

\clearpage
\subsubsection{Random Split; 33~\% of Available Data}
\begin{figure}[h!]
    \centering
    \begin{tabular}{ll}
         Model 1 &
         \adjustbox{valign=t}{\includegraphics[width=0.66\linewidth]{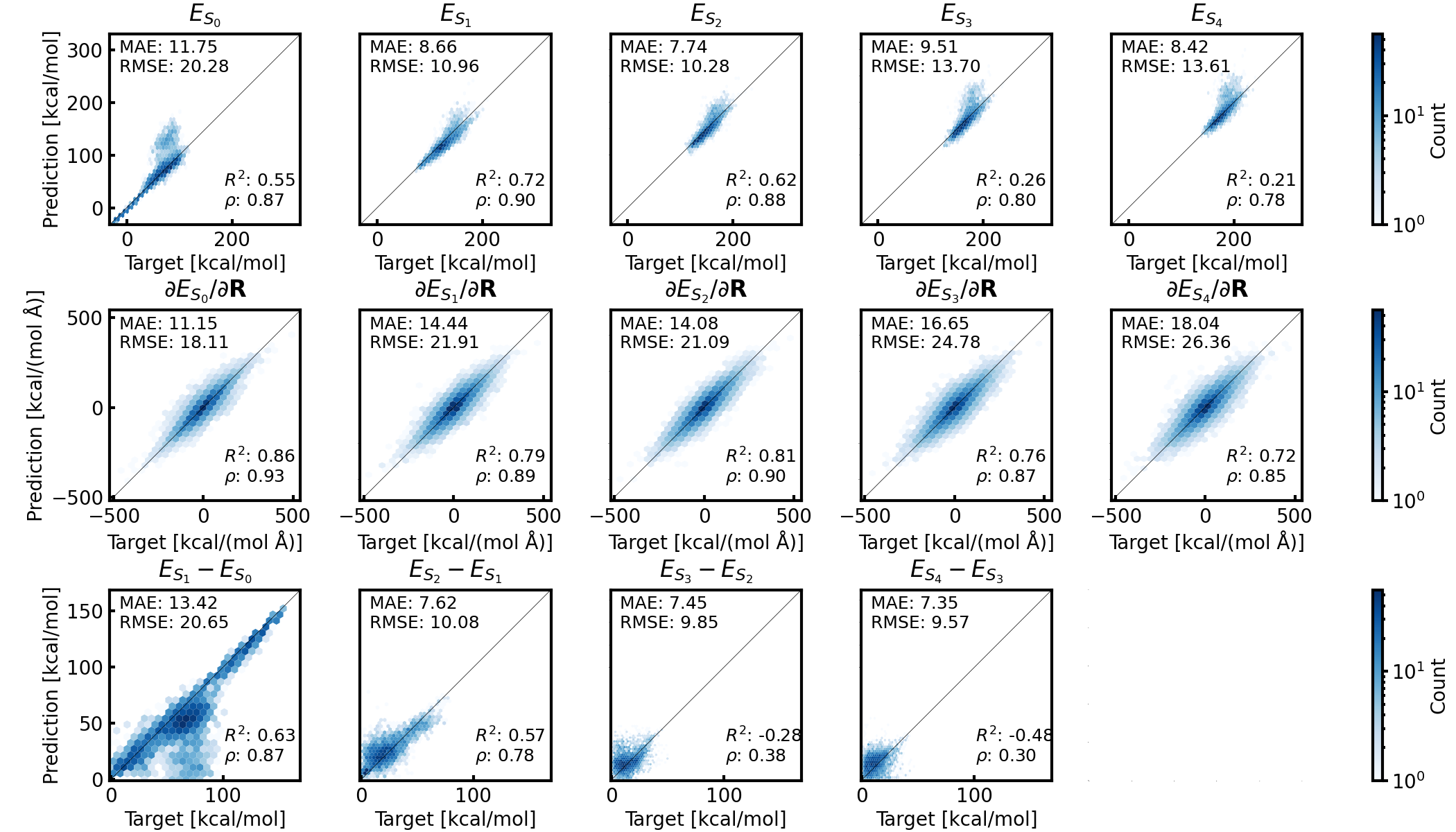} }\\
         \\
         Model 2 &
         \adjustbox{valign=t}{\includegraphics[width=0.66\linewidth]{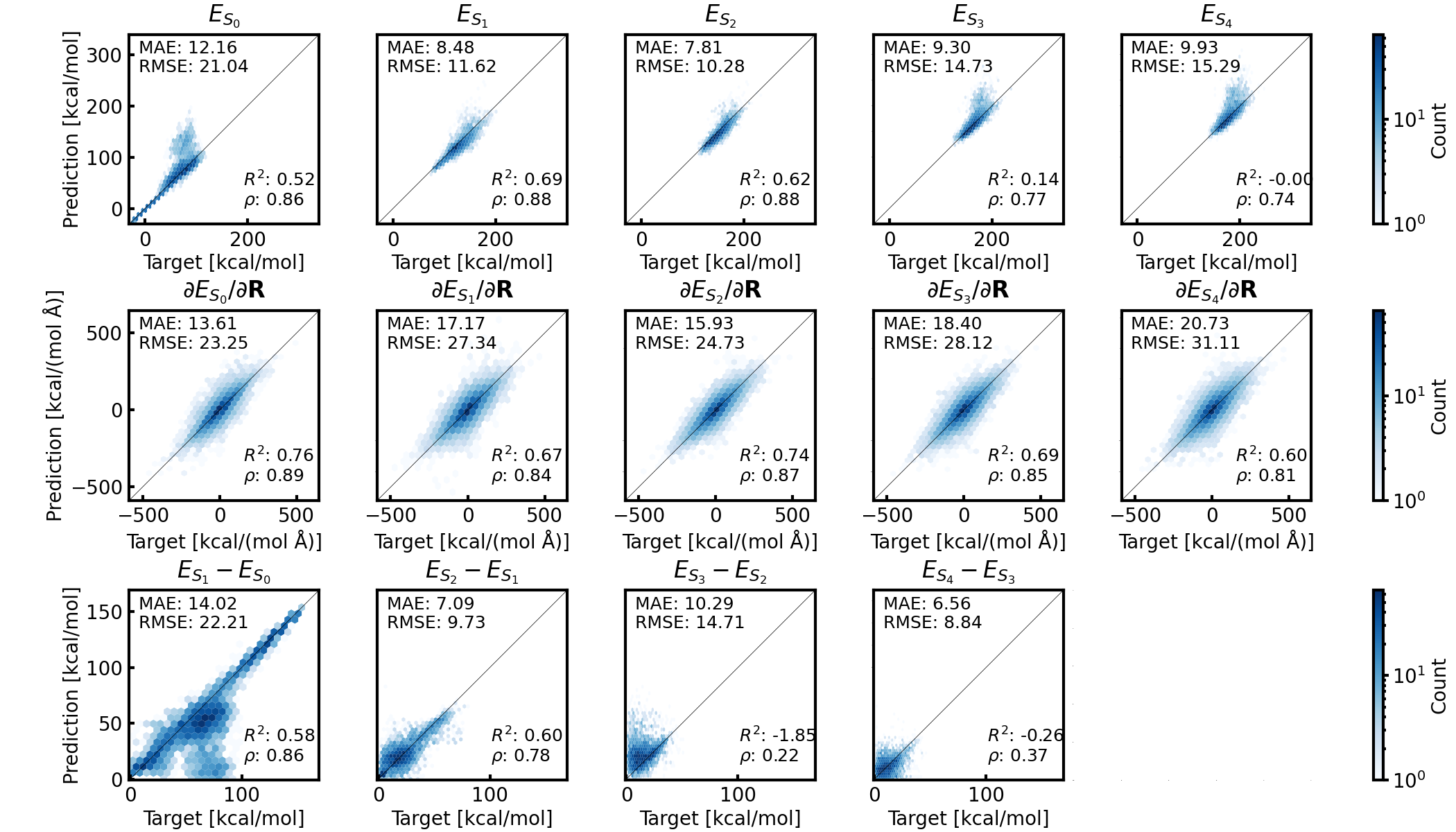} }\\
         \\
         Model 3 &
         \adjustbox{valign=t}{\includegraphics[width=0.66\linewidth]{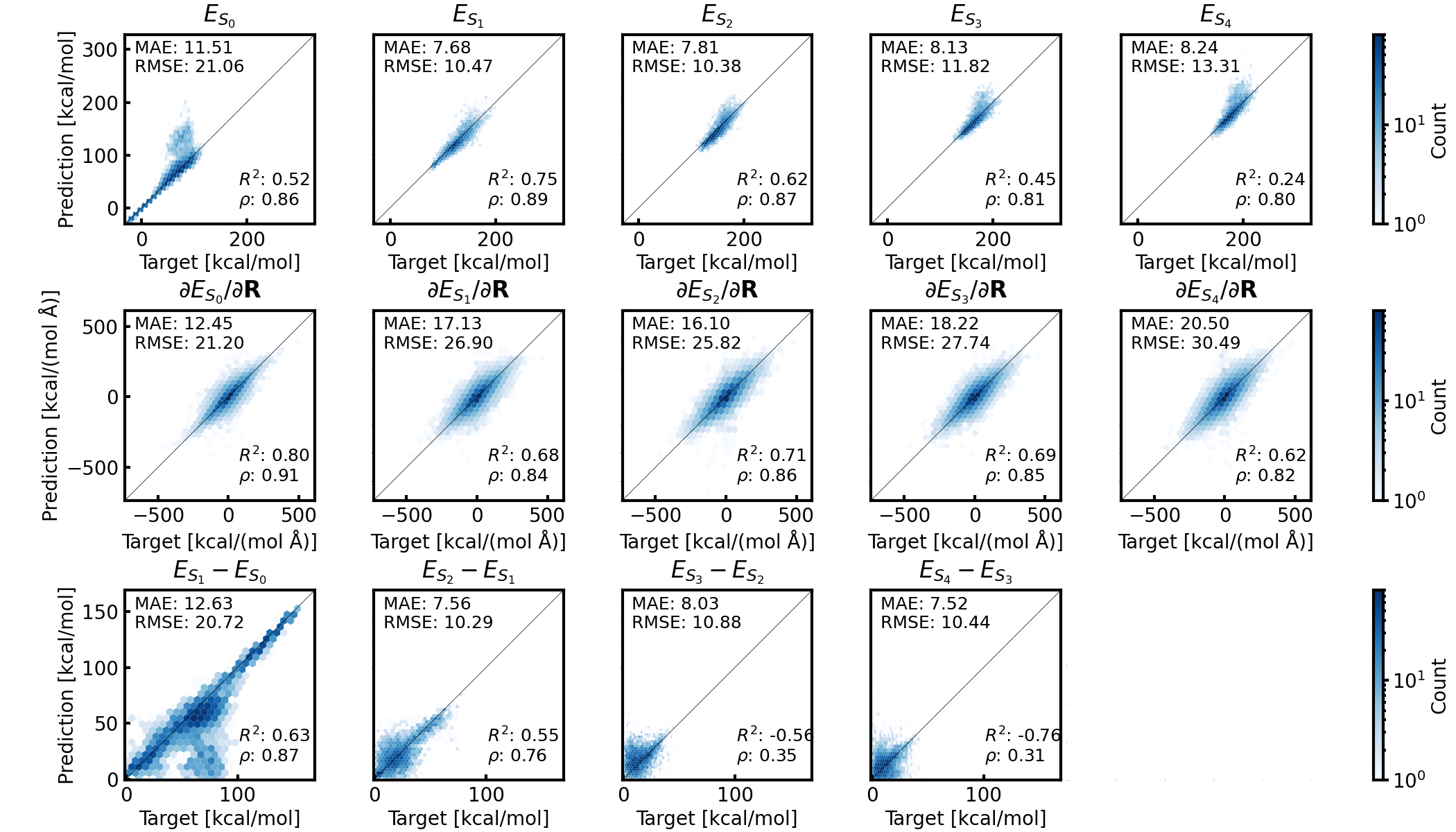} }
    \end{tabular}
    \caption{Parity plots on frames from Set~II for original test for the three "Random Split" models trained on 33~\% of the available frames (every 1.5~fs) from the 36 training trajectories.}
    \label{fig:RandomSplit_SetII_skip3}
\end{figure}

\clearpage
\subsubsection{Random Split; 1~\% of Available Data}
\begin{figure}[h!]
    \centering
    \begin{tabular}{ll}
         Model 1 &
         \adjustbox{valign=t}{\includegraphics[width=0.66\linewidth]{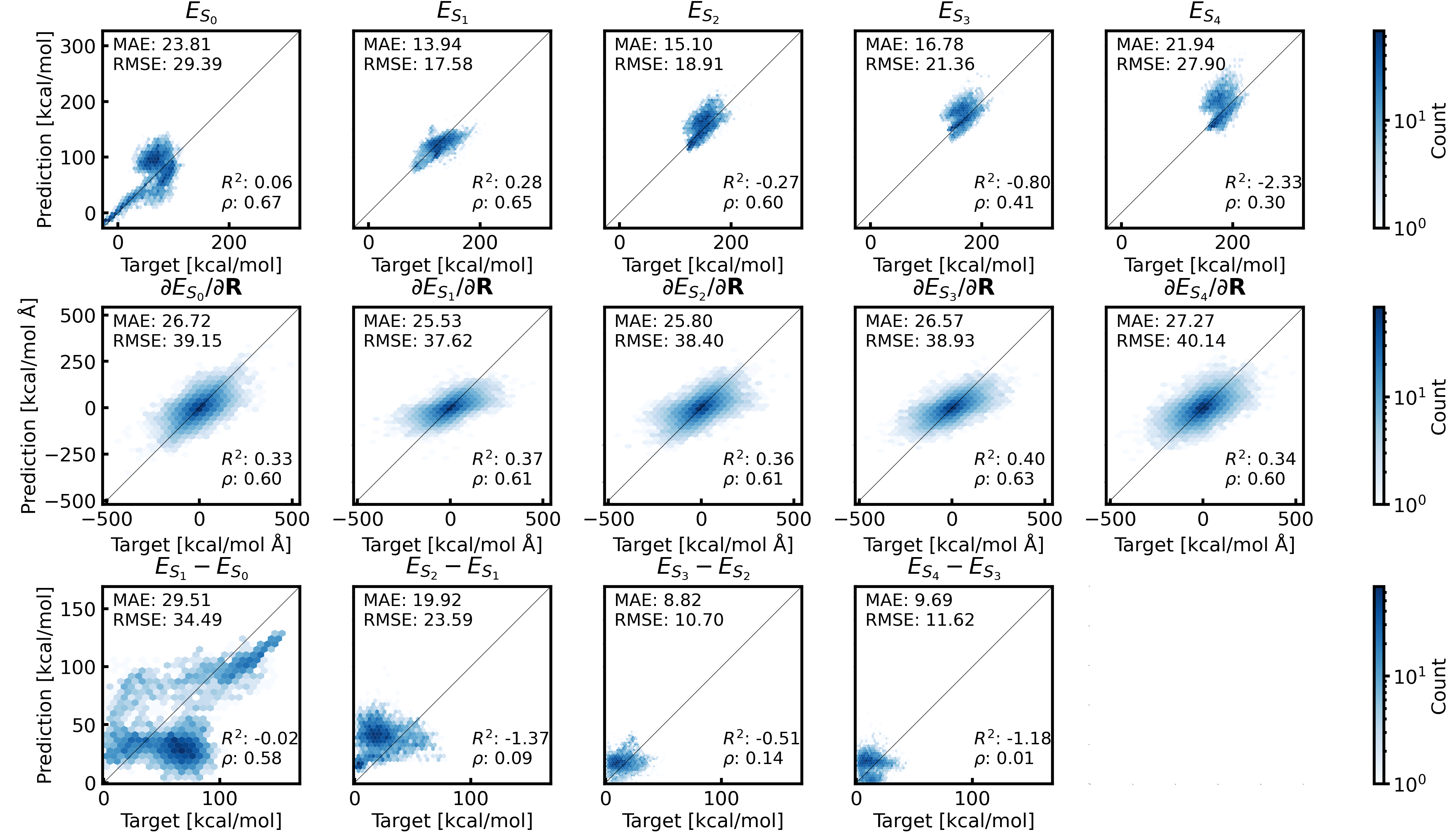} }\\
         \\
         Model 2 &
         \adjustbox{valign=t}{\includegraphics[width=0.66\linewidth]{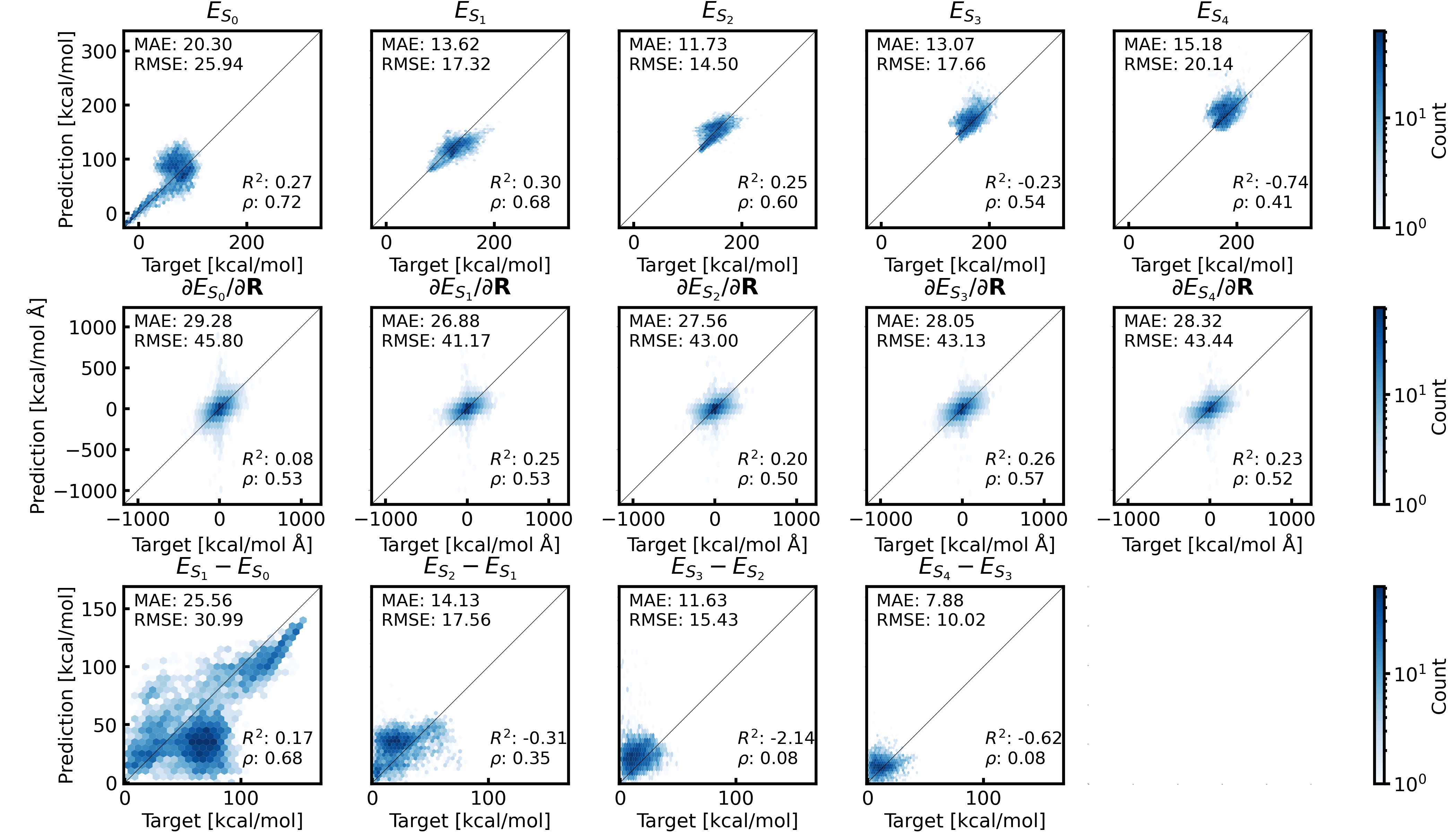} }\\
         \\
         Model 3 &
         \adjustbox{valign=t}{\includegraphics[width=0.66\linewidth]{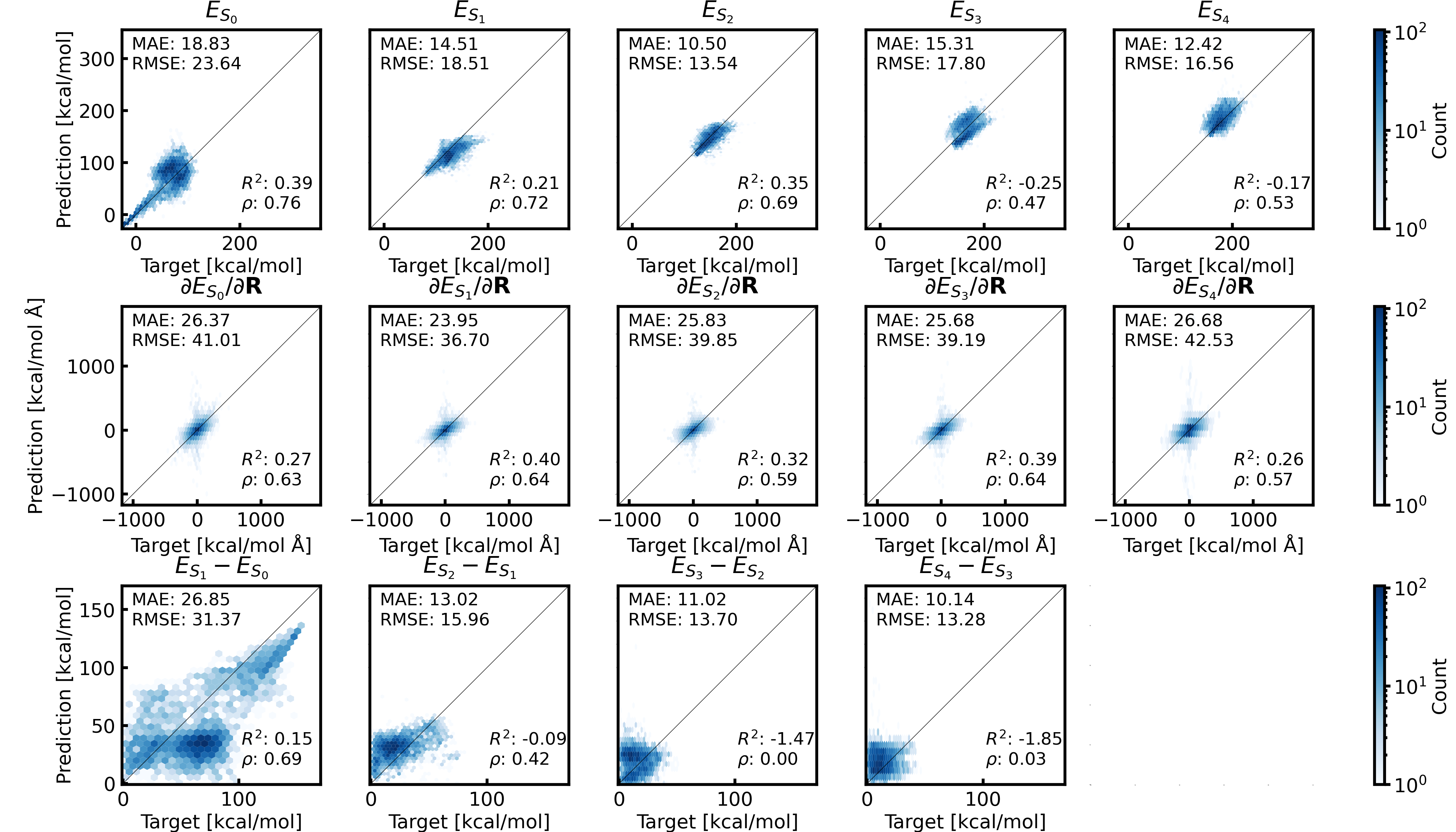} }
    \end{tabular}
    \caption{Parity plots on frames from Set~II for original test for the three "Random Split" models trained on 1~\% of the available frames (every 50~fs) from the 36 training trajectories.}
    \label{fig:RandomSplit_SetII_skip100}
\end{figure}

\clearpage
\subsection{Performance on first 75~fs of Set~II}

\subsubsection{Split by Trajectory; 100~\% of Available Data}

\begin{figure}[h!]
    \centering
    \begin{tabular}{ll}
         Model 1 &
         \adjustbox{valign=t}{\includegraphics[width=0.63\linewidth]{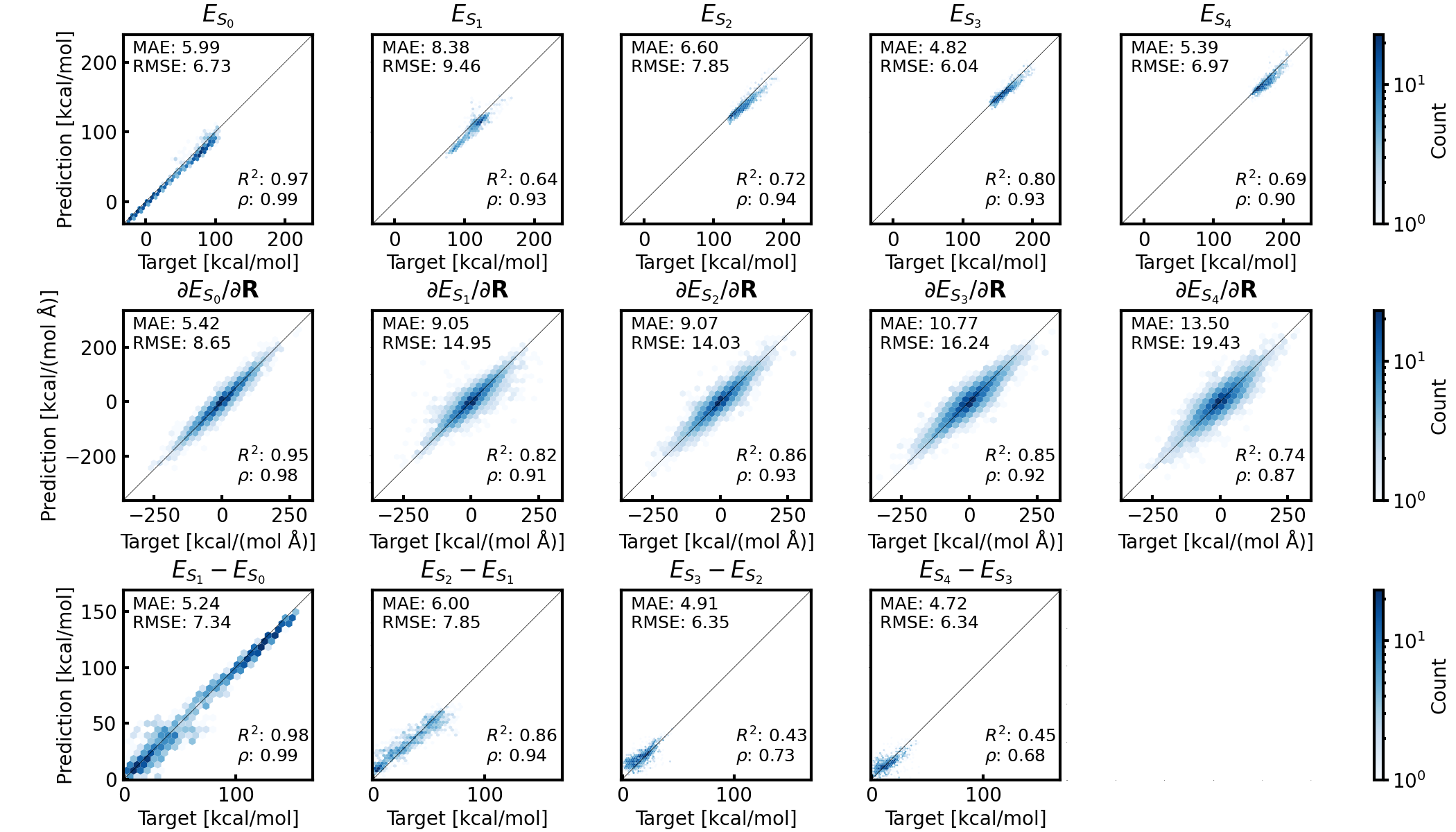} }\\
         \\
         Model 2 &
         \adjustbox{valign=t}{\includegraphics[width=0.63\linewidth]{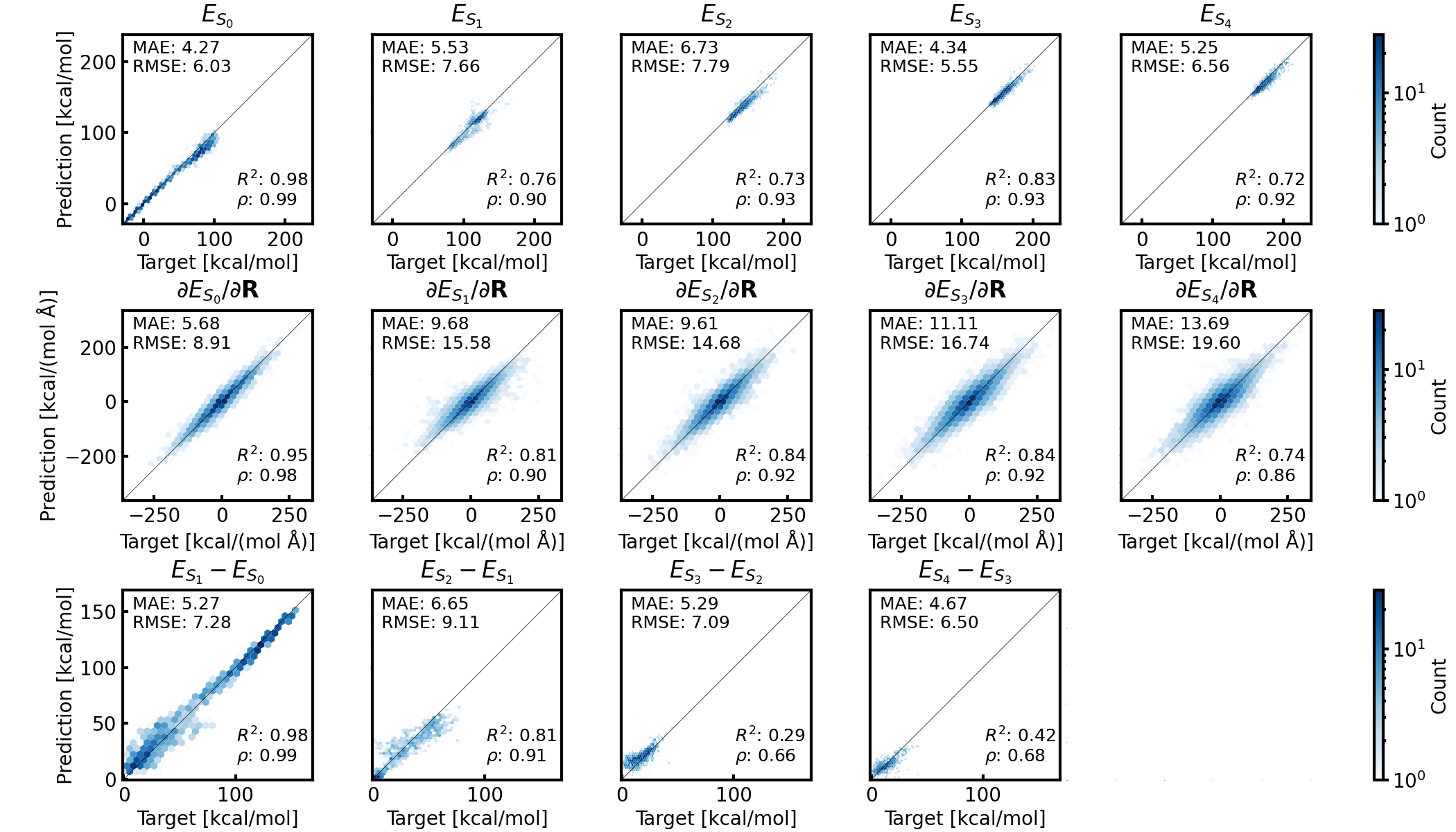} }\\
         \\
         Model 3 &
         \adjustbox{valign=t}{\includegraphics[width=0.63\linewidth]{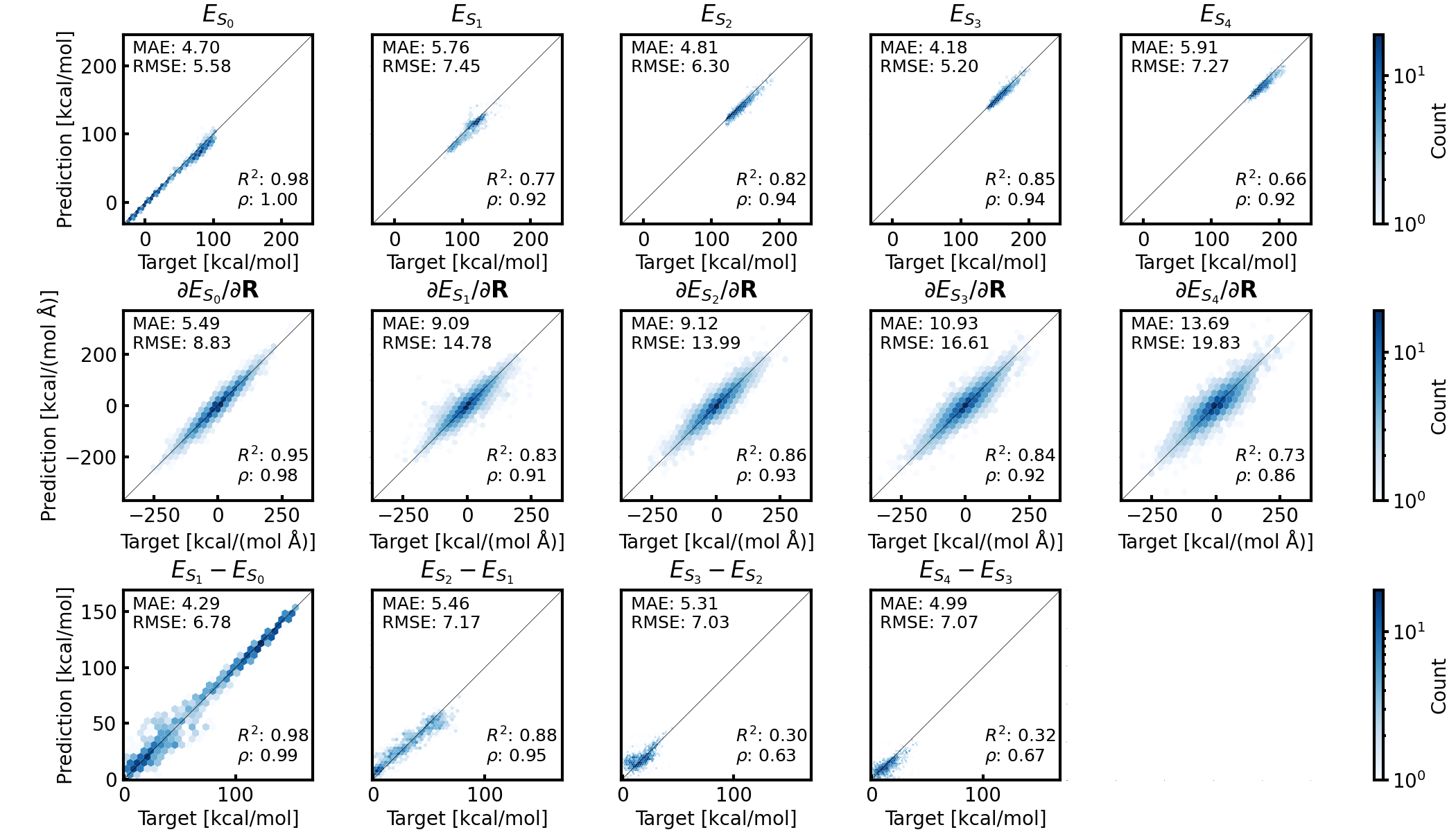} }
    \end{tabular}
    \caption{Parity plots on frames from the first 75~fs of trajectories from Set~II for the three "Split by Trajectory" models trained on 100~\% of the available frames (every 0.5~fs) from the 36 training trajectories.}
    \label{fig:SplitbyTraj_SetII_first75_skip1}
\end{figure}

\clearpage
\subsubsection{Split by Trajectory; 33~\% of Available Data}

\begin{figure}[h!]
    \centering
    \begin{tabular}{ll}
         Model 1 &
         \adjustbox{valign=t}{\includegraphics[width=0.66\linewidth]{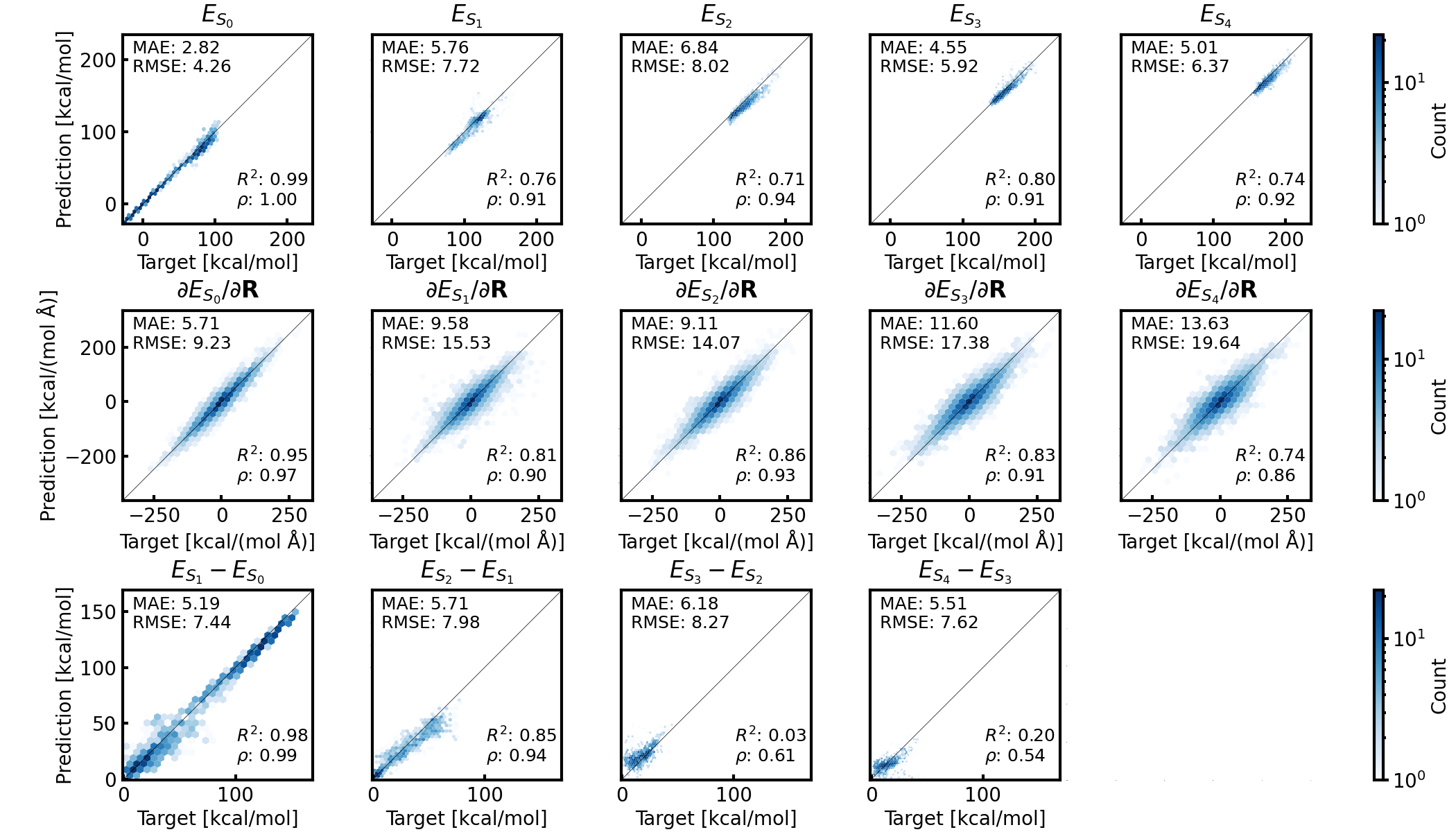} }\\
         \\
         Model 2 &
         \adjustbox{valign=t}{\includegraphics[width=0.66\linewidth]{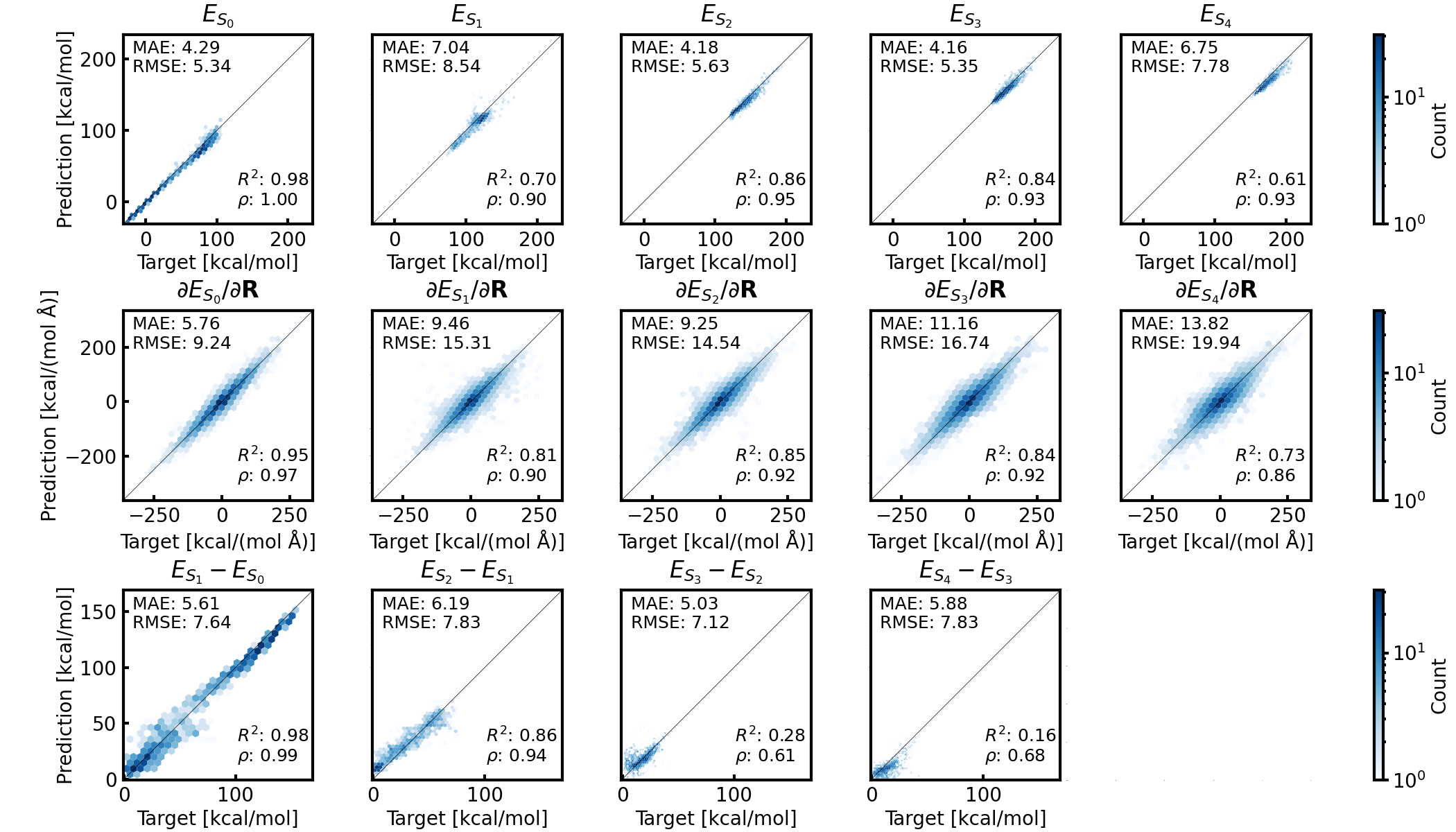} }\\
         \\
         Model 3 &
         \adjustbox{valign=t}{\includegraphics[width=0.66\linewidth]{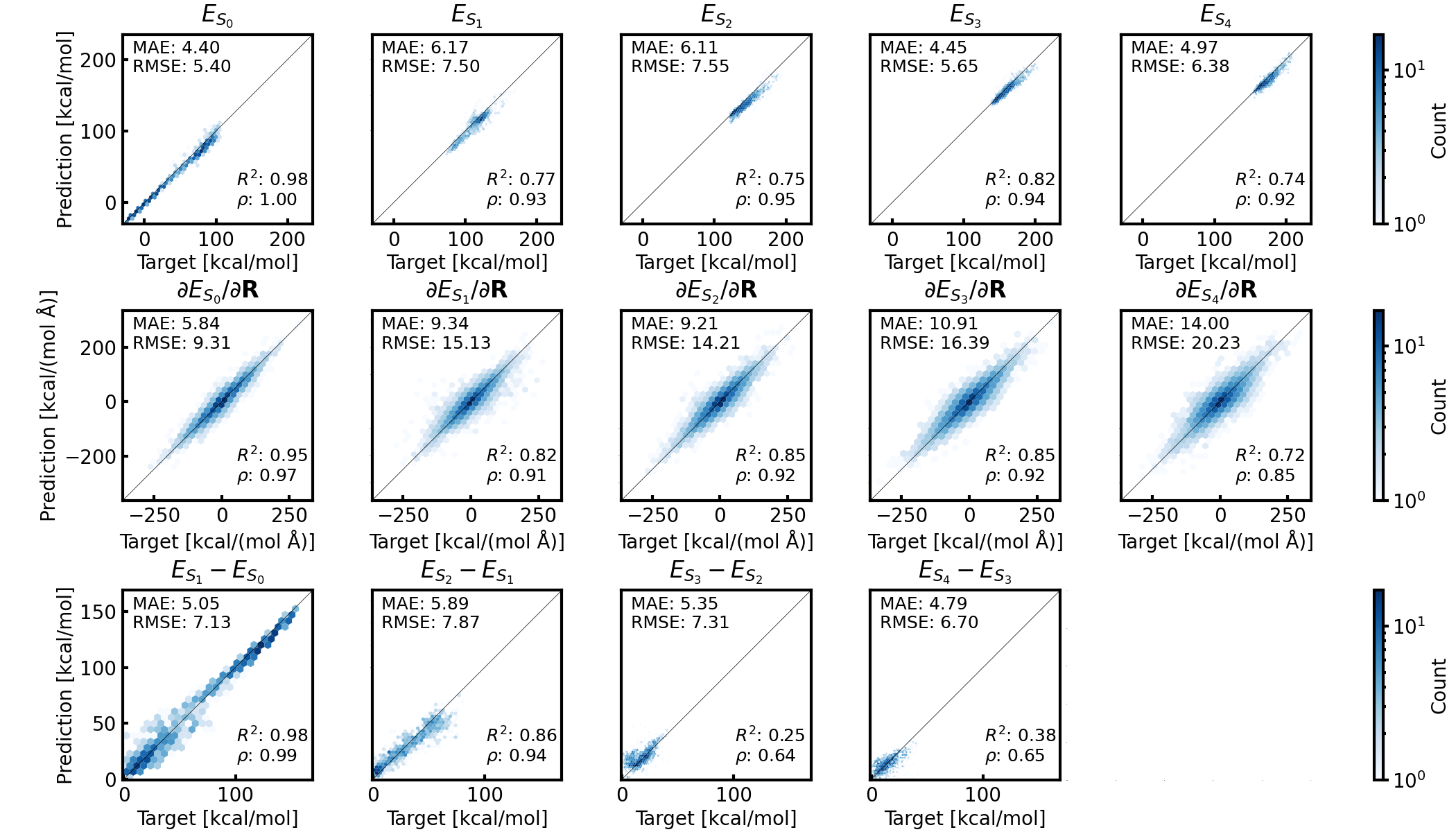} }
    \end{tabular}
    \caption{Parity plots on frames from the first 75~fs of trajectories from Set~II for the three "Split by Trajectory" models trained on 33~\% of the available frames (every 1.5~fs) from the 36 training trajectories.}
    \label{fig:SplitbyTraj_SetII_First75_skip3}
\end{figure}

\clearpage
\subsubsection{Random Split; 100~\% of Available Data}
\begin{figure}[h!]
    \centering
    \begin{tabular}{ll}
         Model 1 &
         \adjustbox{valign=t}{\includegraphics[width=0.66\linewidth]{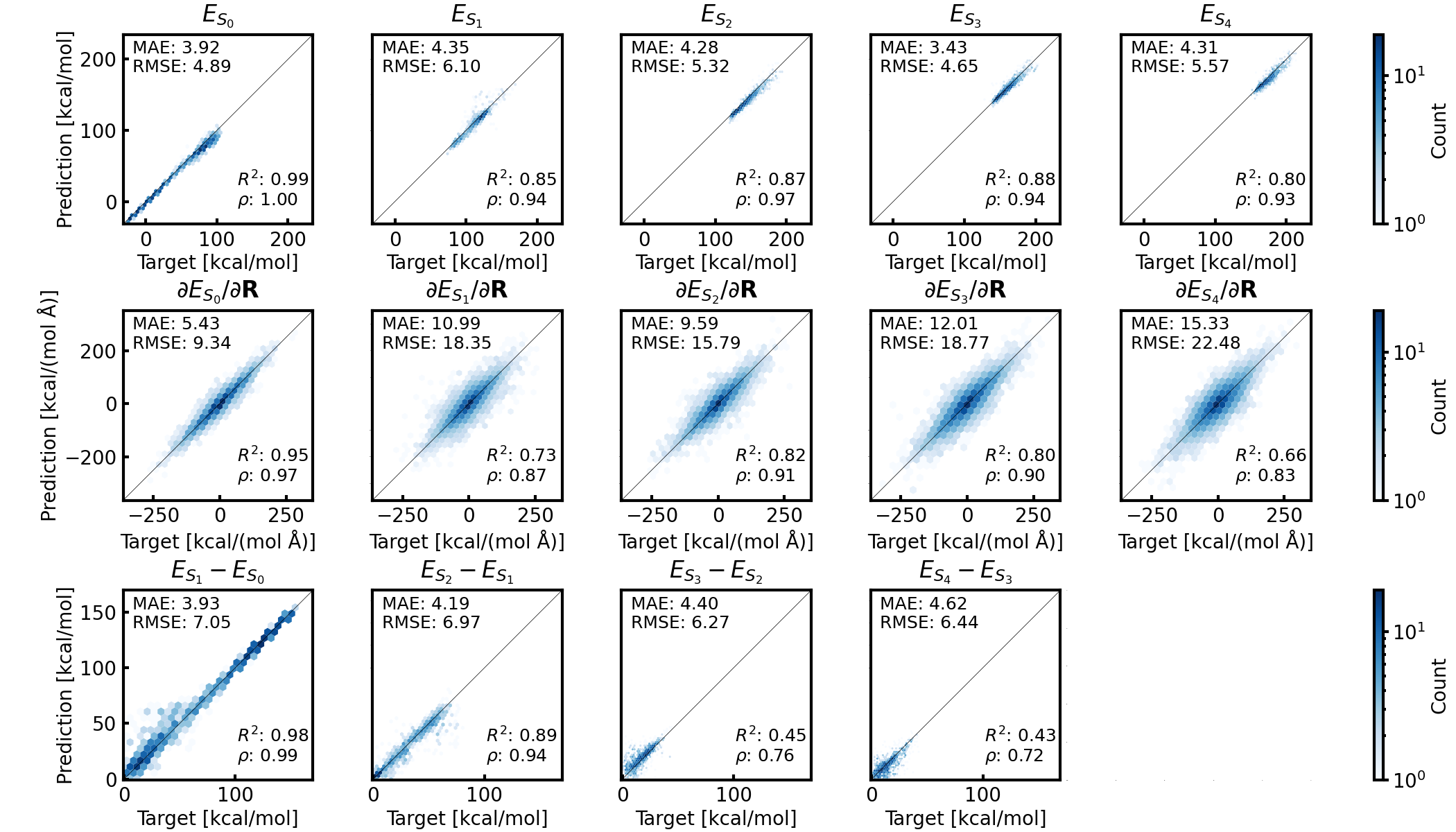} }\\
         \\
         Model 2 &
         \adjustbox{valign=t}{\includegraphics[width=0.66\linewidth]{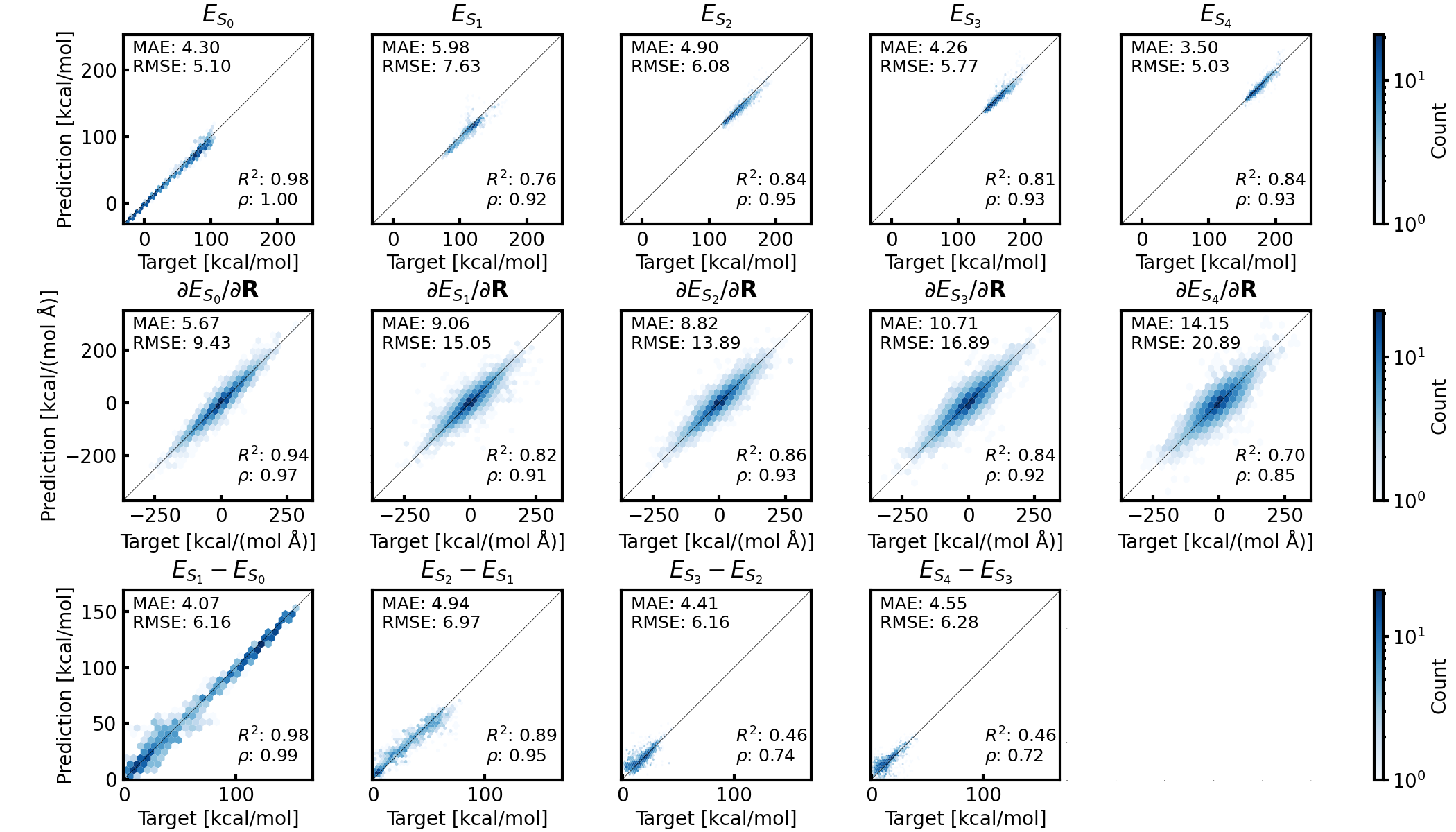} }\\
         \\
         Model 3 &
         \adjustbox{valign=t}{\includegraphics[width=0.66\linewidth]{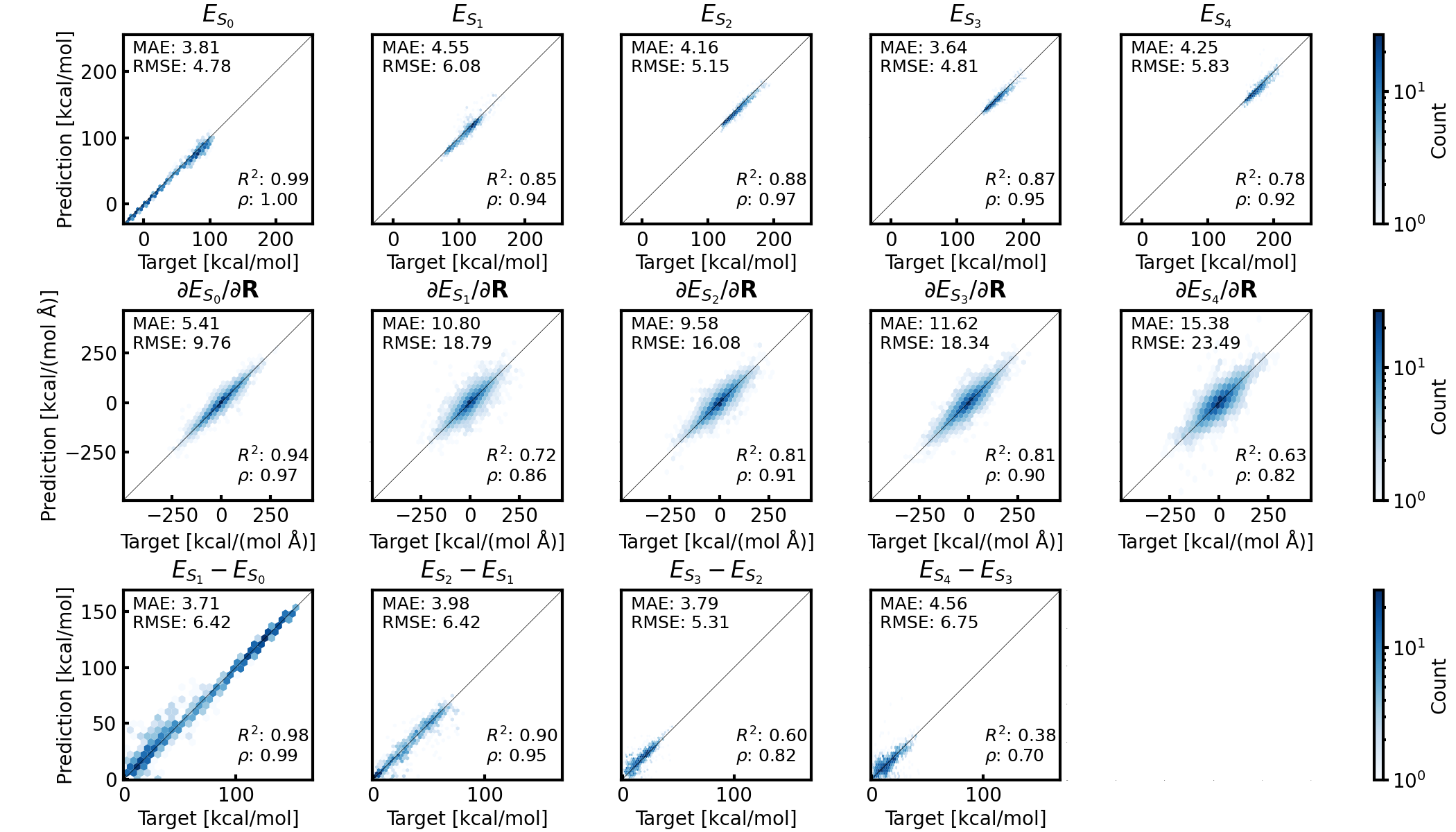} }
    \end{tabular}
    \caption{Parity plots on frames from the first 75~fs of trajectories from Set~II for original test for the three "Random Split" models trained on 100~\% of the available frames (every 0.5~fs) from the 36 training trajectories.}
    \label{fig:RandomSplit_SetII_First75_skip1}
\end{figure}

\clearpage
\subsubsection{Random Split; 33~\% of Available Data}
\begin{figure}[h!]
    \centering
    \begin{tabular}{ll}
         Model 1 &
         \adjustbox{valign=t}{\includegraphics[width=0.66\linewidth]{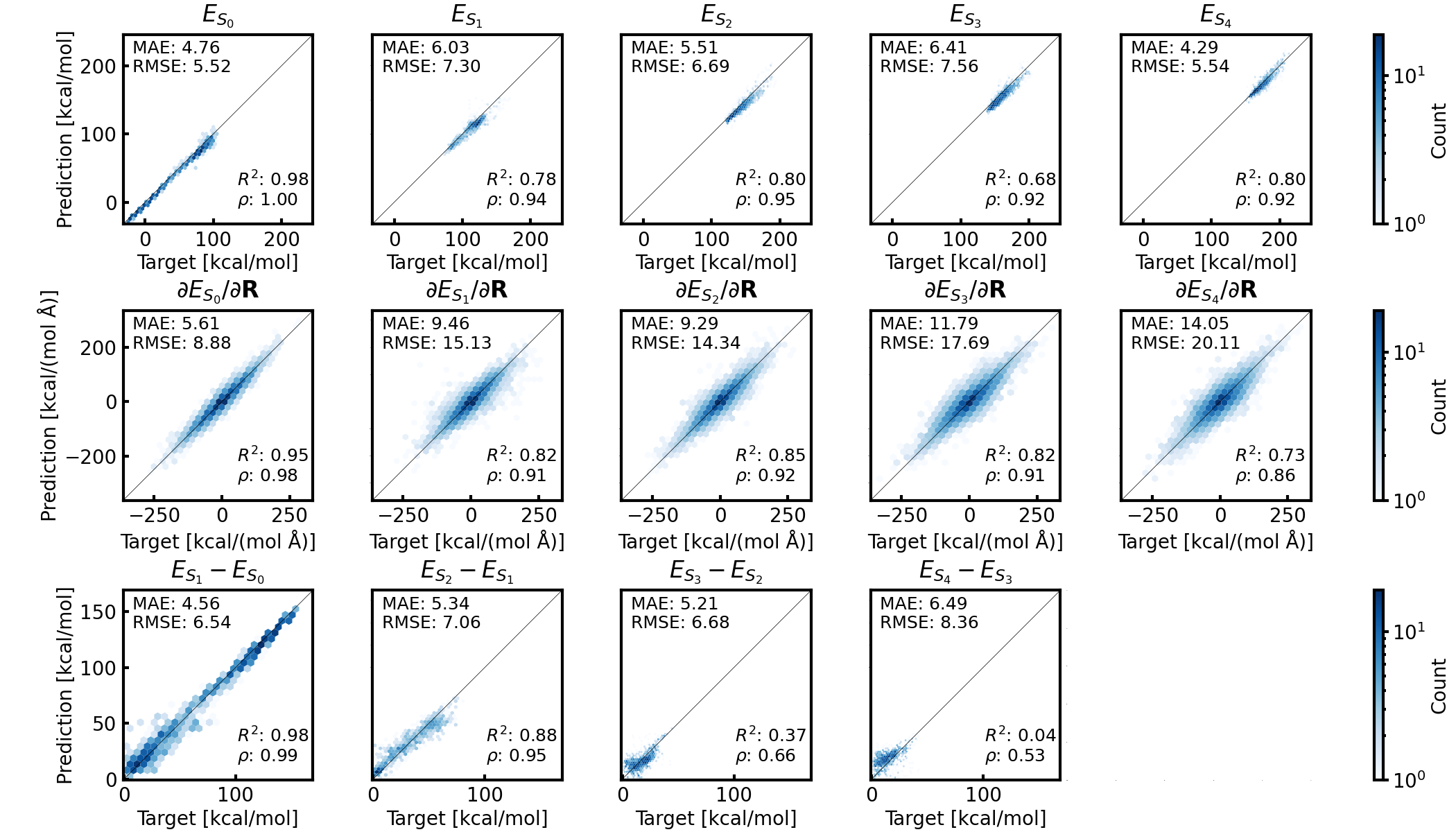} }\\
         \\
         Model 2 &
         \adjustbox{valign=t}{\includegraphics[width=0.66\linewidth]{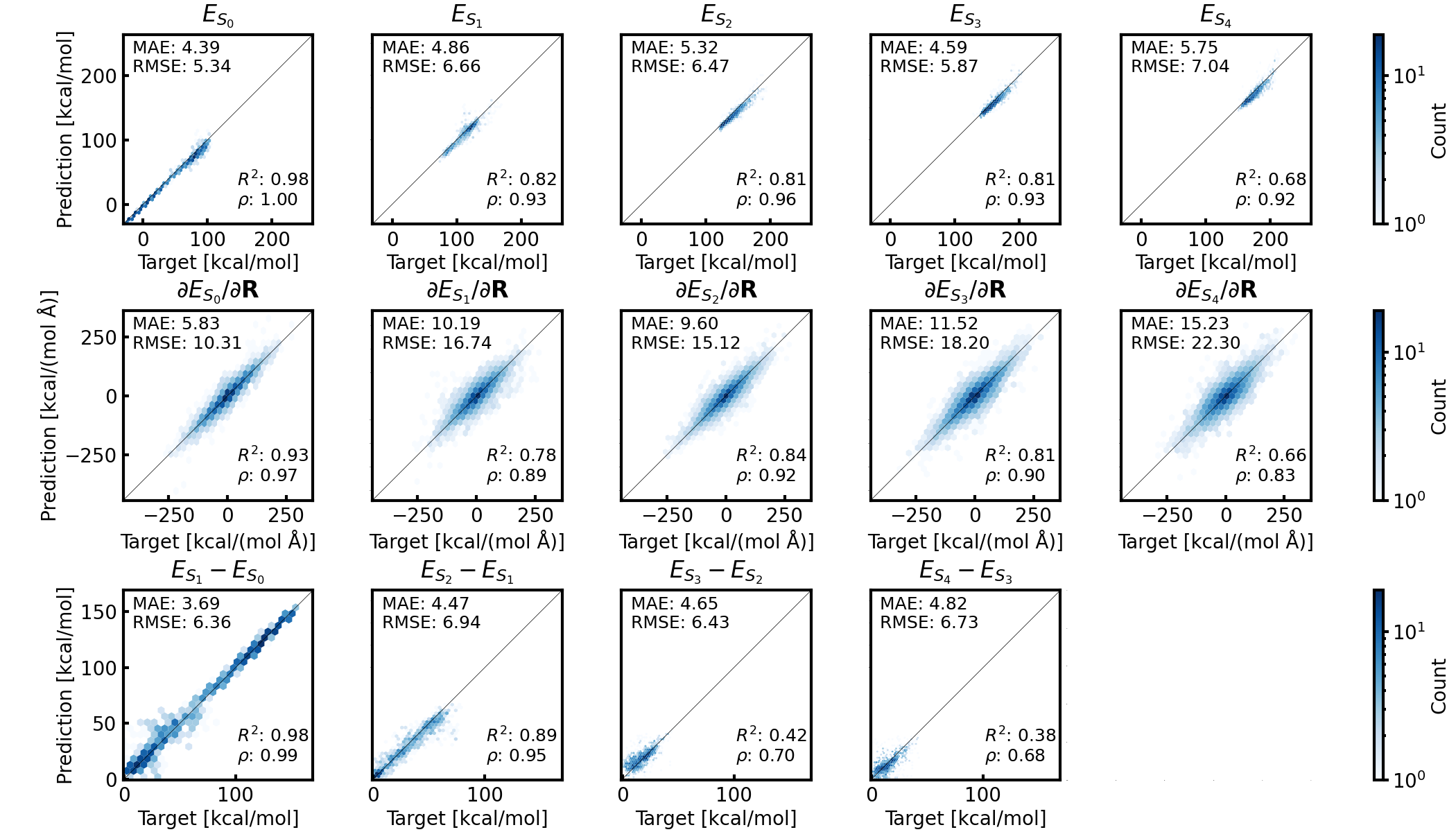} }\\
         \\
         Model 3 &
         \adjustbox{valign=t}{\includegraphics[width=0.66\linewidth]{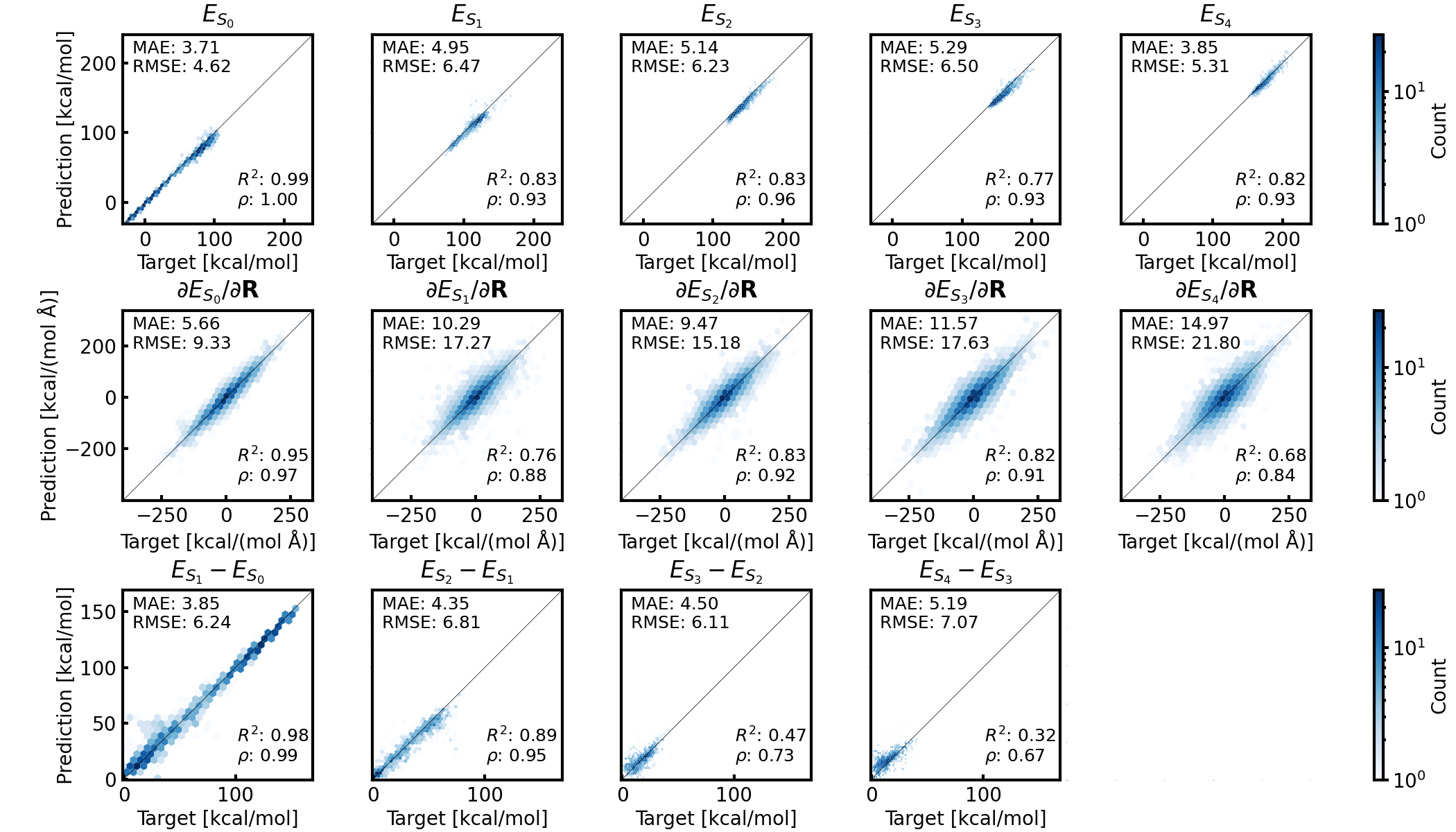} }
    \end{tabular}
    \caption{Parity plots on frames from the first 75~fs of trajectories from Set~II for original test for the three "Random Split" models trained on 33~\% of the available frames (every 1.5~fs) from the 36 training trajectories.}
    \label{fig:RandomSplit_SetII_First75_skip3}
\end{figure}

\clearpage

\section{Electronic populations}

In this section we show the average electronic population derived from the active state for all ML/MM simulations (including those not shown in the main text).
These populations are the basis used to compute the relaxation times reported in the manuscript.
All of these trajectories were performed with the initial conditions from set~II. 
The electronic populations of the reference QM/MM dynamics (set~II) are shown as dashed lines.
As a reminder, we trained 18 different models altogether: We compared two different ways to split the training data set (set~I), namely "Random Split" and "Split by Trajectory" and additionally we used three different training set sizes obtained by subsampling the trajectories (100~\%, 33~\% and 1~\% of the data).
For each combination of split and subsampling we trained three independent models, i.e. with different random weight initializations.

\begin{figure}
    \centering
    \includegraphics[width=\linewidth]{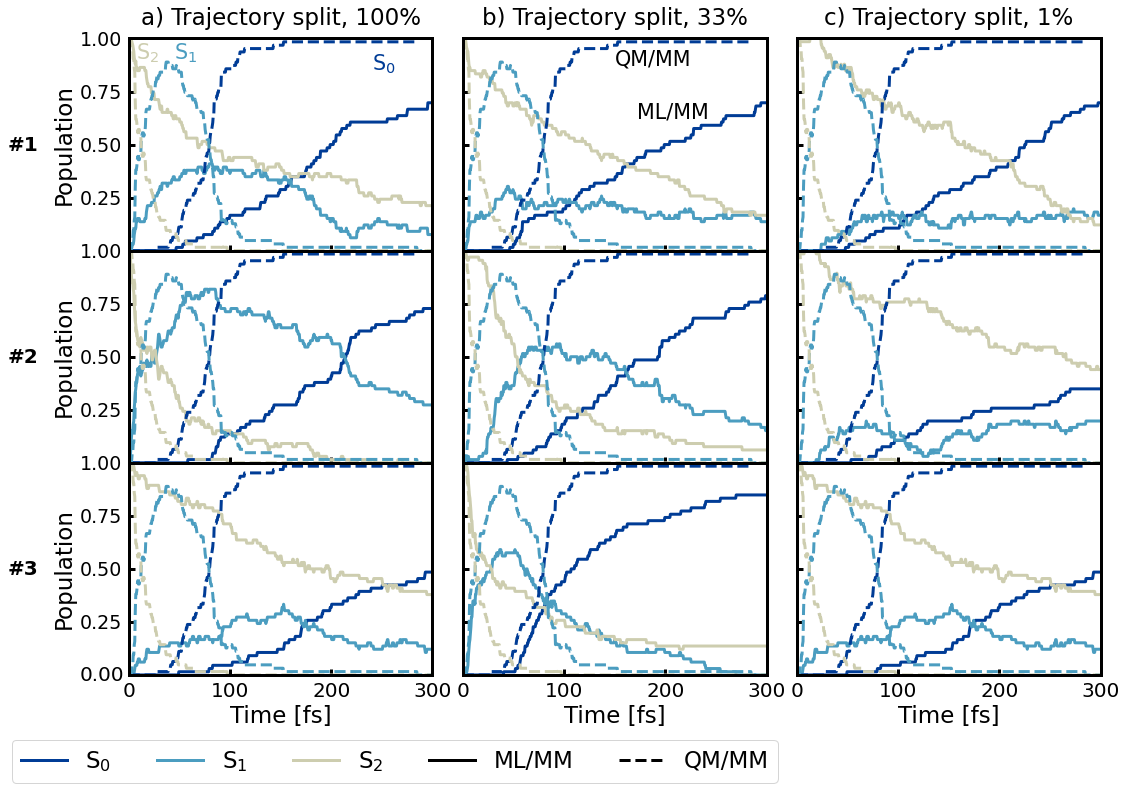}
    \caption{Electronic populations based on the active state with the initial conditions from Set II. The potentials for these populations were obtained from models trained with the "Split by Trajectory" procedure with 100~\%, 33~\% and 1~\% of the data. For each of the split three different weight initializations were used resulting in a total of nine models. }
    \label{fig:occs_traj}
\end{figure}

Figure~\ref{fig:occs_traj} shows the results for the "Split by Trajectory" models.
The 100~\% models do not agree well with the QM/MM simulations, which can also be seen in the corresponding half-lives (see Table~1 in the main text).
The S$_2$ decays too slowly (except for model no.~2) and the transitions to the ground state are also much slower than in the reference.
The 33~\% models perform better than the 100~\%, those were also shown in the main text.

For the 33~\% models, the agreement increases from model no.~1 to~3.
Model no.~1, produces dynamics with a very slow decay from the S$_2$ into S$_1$ so that the population of the S$_1$ appears constant as the two decay rates are similar (incorrectly so).
Model two produces dynamics that are qualitatively correct, only the relaxation times are too large (compare Table~1 in the main text).
Model three performs the best, the intersections of the population curves occur at almost the same time after excitation as in the QM/MM simulations.
The models trained with only 1~\% of the data show very poor performance, which was expected based on the low amount of available data.

\begin{figure}
    \centering
    \includegraphics[width=\linewidth]{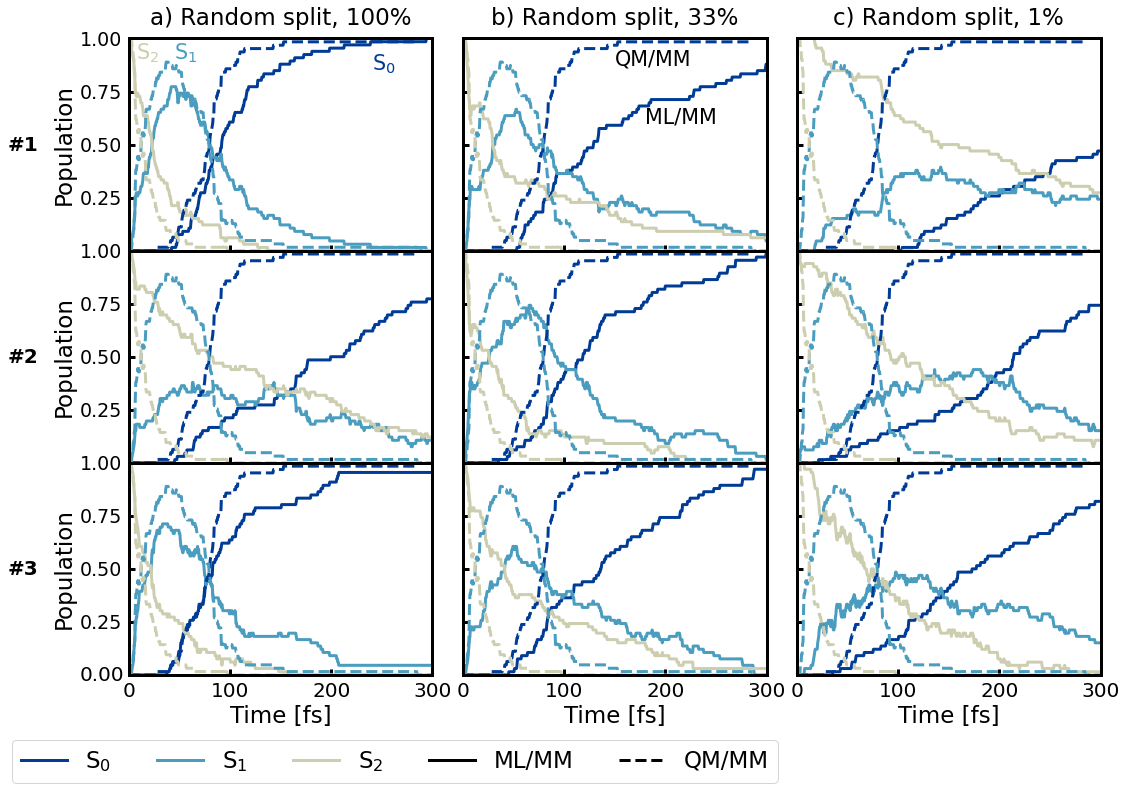}
    \caption{Electronic populations based on the active state with the initial conditions from Set II. The potentials for these populations were obtained from models trained with the "Random split" procedure with 100~\%, 33~\% and 1~\% of the data. For each of the split three different weight initializations were used resulting in a total of nine models. }
    \label{fig:occs_random}
\end{figure}

Figure~\ref{fig:occs_random} shows the results from "Random Split" procedure.
As has been mentioned in the manuscript, the occupations for models one and three with 100~\% of the data agree quite well with the QM/MM simulations.
They both only slightly overestimate the relaxation times and also show very similar qualitative agreement.
Model two on the other hand shows quite bad agreement strongly overestimating the relaxation times leading to qualitative and quantitative disagreement.
The models for 33~\% of the data perform quite similar from a qualitative point of view.
All of them have electronic population staying too long the S$_2$, lading to slower over all relaxation times to the S$_0$.
Model 2 seems to perform the best, which is also supported by the computed half-life times.
As expected, the models with only 1~\% of the data show very large deviations as was the case for the "Split by Trajectory" models.

\clearpage
\section{Structural analysis of the dynamics}

In this section, we analyze the structural changes during the dynamics simulations. 

\subsection{Ring opening}\label{sec:ring}
We start with an analysis of the distance between the oxygen atom and the neighboring carbon atoms (C$_1$ and C$_4$, see Fig.~1 in the main text) in the QM/MM simulations. 
By taking the maximum distance of these two bonds we monitor whether the furan ring opens up over the course of the simulation. 
We will refer to this distance as ``maximal C-O distance`` from now on.

\begin{figure}[h]
    \centering
    \includegraphics[width=\linewidth]{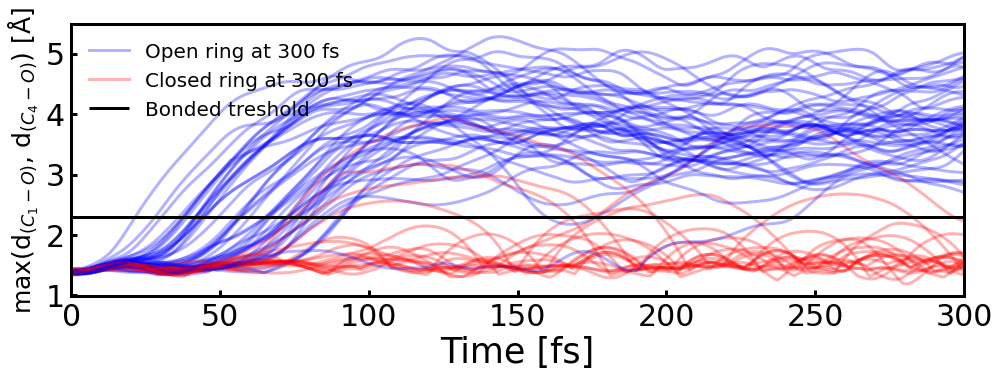}
    \caption{Maximum of C$_1$-O and C$_4$-O distances (see Fig.~1 of the main text for naming convention)  over time for all QM/MM trajectories from set~II. 
    The red colored curves have a closed ring after the 300~fs simulation time. 
    Trajectories are colored blue if the furan ring is open at the end of the simulation. 
    We consider a ring closed if the maximum distance is smaller than 2.3~\AA{}. }
    \label{fig:CO_distances}
\end{figure}

Figure~\ref{fig:CO_distances} shows the maximum of the C$_1$-O and C$_4$-O distances over time for each trajectory in set~II. 
We define a ring as open, when the maximum C-O distance is greater than 2.3~\AA{}. 
Most of the ring openings occur soon after the excitation within the first 70~fs. 
In total, 53 out of 65 trajectories (ca. 82~\%) display a ring opening (maximum of the two C-O distances above 2.3~\AA{}). 
Six of them revert back to a closed configuration within the 300~s simulation time.
One furan ring opens up at ca. 250~fs, while all other 49 trajectories do so much earlier. 
The remaining 12 trajectories do not exhibit any ring opening. 

Fuji et al.\cite{Fuji2010} reported two relaxation channels, one in which the furan remains intact and the other in which one of the C-O bonds breaks. 
Our simulations recover both of these pathways.

\subsection{Hopping geometries}

Here we investigate the hopping geometries (the configurations of furan at the time step when the system switches PESs), and compare those found in the QM/MM and ML/MM simulations.

\subsubsection{S$_1$ $\rightarrow$ S$_0$}

First, we focus on transitions between S$_1$ and S$_0$.
In total, we obtain 88 hopping geometries between these two states (including hops in both directions).
Based on our hopping scheme, hops from the ground state to higher states are not allowed, however, in regions where ground and excited state are close or fully degenerate the first excited state can have a lower energy than the ground state in TDA TD-DFT leading to additional hops, which is not a problem, as this only occurs for configurations very close to the intersection seam.

When analyzing the hopping geometries, we solely inspect the configuration of furan.
We describe the structures only with intramolecular distances by using the coulomb matrix\cite{Rupp2012} as defined in equation~\ref{eq:cmatrix}, this removes the influence of rotations and translations.
\begin{equation}
C_{ij} = 
\begin{cases}
  i=j &  0.5 Z_i^{2.4} \\
  i\neq j & \frac{Z_i Z_j}{|\mathbf{R}_i-\mathbf{R}_j|}
\end{cases}
\label{eq:cmatrix}
\end{equation}
where $Z_x$ and $\mathbf{R}_x$ are the nuclear charge and position of atom x.
To facilitate comparison between QM/MM and ML/MM hopping geometries, we perform a dimensionality reduction with principal component analysis (PCA).
Since the Coulomb matrix is symmetric, we only used the values of the upper triangle as input for the PCA.

\begin{figure}
    \centering
    \includegraphics[width=0.5\linewidth]{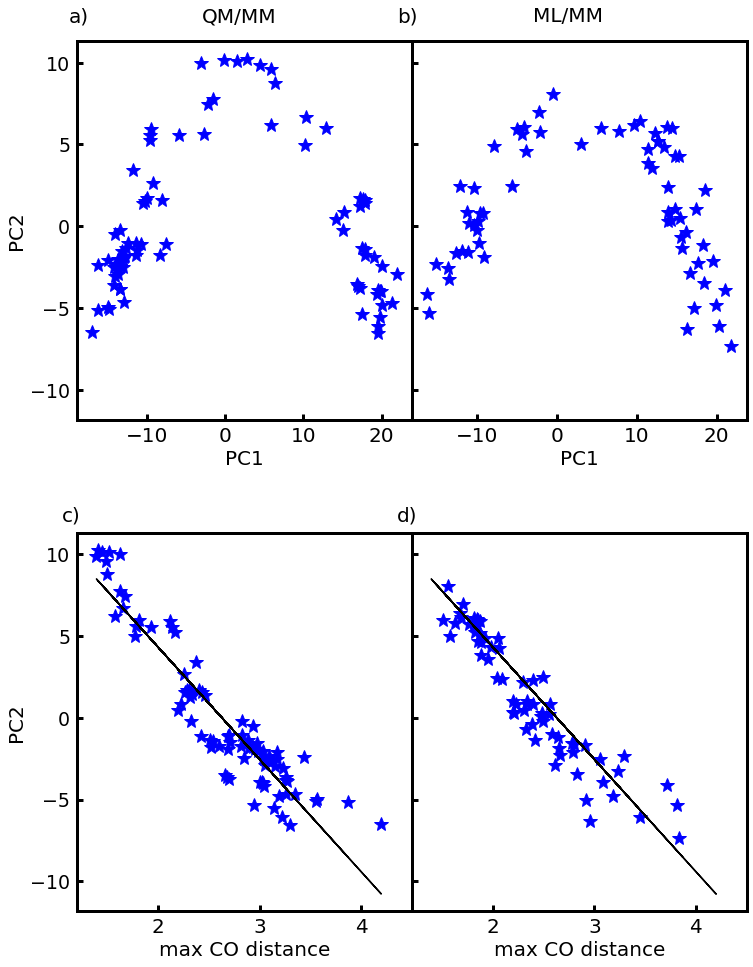}
    \caption{Projections of hopping geometries between states S$_1$ and S$_0$.
    a) Projection of the hopping geometries from the QM/MM reference simulations onto PC1 and PC2. 
    b) Projection of hopping geometries from the ML/MM simulations with the 100~\% random split model no.~1 onto the same principal components as in a). 
    c) and d) show the same projections as in a) and b) with PC1 substituted by the max CO distance.
    In c) a linear least squares fit was applied to the QM/MM reference geometries. This line is used for comparison in Fig.~S\ref{fig:AllS1S0Hoppings}.
    }
    \label{fig:PCA_10}
\end{figure}

The first and second principle components (PCs) represent 88.8~\% and 9.2~\% of the variance of data, with all other PCs comprising less than 2~\% together.
Figure~\ref{fig:PCA_10}a shows the projection of the QM/MM hopping geometries onto the first (PC1) and second (PC2) principal component.
The hopping geometries appear to be symmetric with respect to PC1.
Further analysis of PC1 shows that it is mainly a linear combination of the two C-O bonds (C$_1$-O and C$_4$-O). 
PC2 is constructed by a linear combination of several non-hydrogen interatomic distances.
When replacing PC1 with the maximum of these two bond distances, we obtain a clear diagonal line, see Fig.~\ref{fig:PCA_10}c.

When projecting the 66 hopping geometries from the ML/MM simulation with model no.~1 from the random split procedure with 100~\% of the data onto the same principal components, we obtain Figure~\ref{fig:PCA_10}b.
Analogous to Figure~\ref{fig:PCA_10}c, PC1 is substituted by the maximal C-O distance in Figure~\ref{fig:PCA_10}d. 

\begin{figure}
    \centering
    \includegraphics[width=\linewidth]{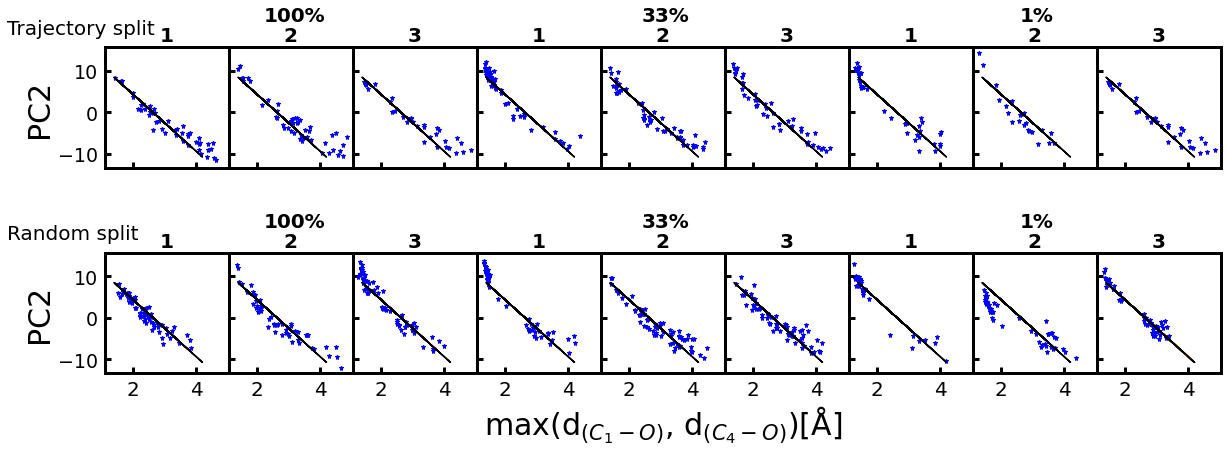}
    \caption{Projections of the hopping geometries from all ML/MM simulations onto the maximum C-O distance and the PC2 from the QM/MM data. 
    The upper row shows the geometries obtained from trajectories generated with the split by trajectory models starting with 100~\% of the data followed by 33~\% and 1~\%.
    The second row shows the same for the models trained on random splits. 
    The black line represents a linear fit projections of the QM/MM reference data.}
    \label{fig:AllS1S0Hoppings}
\end{figure}

Figure~S\ref{fig:AllS1S0Hoppings} shows the projection of all hopping geometries between states S$_1$ and S$_0$ from all ML/MM simulations onto the maximal C-O distance and the PC2 from QM/MM PCA.
The line fitted to the QM/MM reference data is included in every subplot to guide the eye. 
In the QM/MM ground truth the samples are scatter relatively equally around this line.
In Figure~S\ref{fig:AllS1S0Hoppings} we observe that some models tend to hop at significantly larger C-O distances or produce more clustered sets of hopping geometries, indicating that the fitted PES did not reproduce the same seam.
In general, the more similar the projection of the hopping geometries for a model is the better the dynamics seems to reproduce the QM/MM ground truth as well.

\clearpage
\subsubsection{S$_2$ $\rightarrow$ S$_1$}

\begin{figure}[h]
    \centering
    \includegraphics[width=0.5\linewidth]{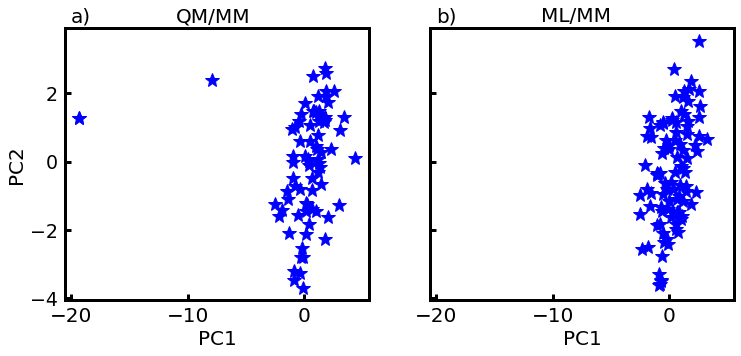}
    \caption{Projections of hopping geometries between states S$_2$ and S$_1$.
    a) Projection of the hopping geometries from the QM/MM reference simulations onto PC1 and PC2 (different components than in Fig.~S\ref{fig:PCA_10}). 
    b) Projection of hopping geometries from the ML/MM simulations with the 100~\% random split model no.~1 onto the same principal components as in a).
    }
    \label{fig:PCA_21}
\end{figure}

We applied the same workflow to the hopping geometries collected for hops between the states S$_2$ and $S_1$.
Performing PCA on the Coulomb matrices of these geometries yields a PC1 that explains 74~\%  and a PC2 that explains 15~\% of the total variance.
All other PCs contribute less than 10~\% each.
Figure~\ref{fig:PCA_21} shows the QM/MM and ML/MM hopping geometries projected onto the first two PCs obtained from the QM/MM geometries, in panel a and b respectively. 
The geometries form one cluster with two clear outliers for the QM/MM simulations.
Both of the outliers stem from the same trajectory and represent a hop from the S$_1$ to the S$_2$ with a very low likelihood and a hop back one time step later, where one of the C-O bonds is breaking.
PC1 and PC2 are again mostly determined by the two C-O bonds.
As all other geometries for this transition are extremely similar (see Fig.~8a in the main text), the PCs are extremely sensitive to changes in these bonds.

\begin{figure}
    \centering
    \includegraphics[width=\linewidth]{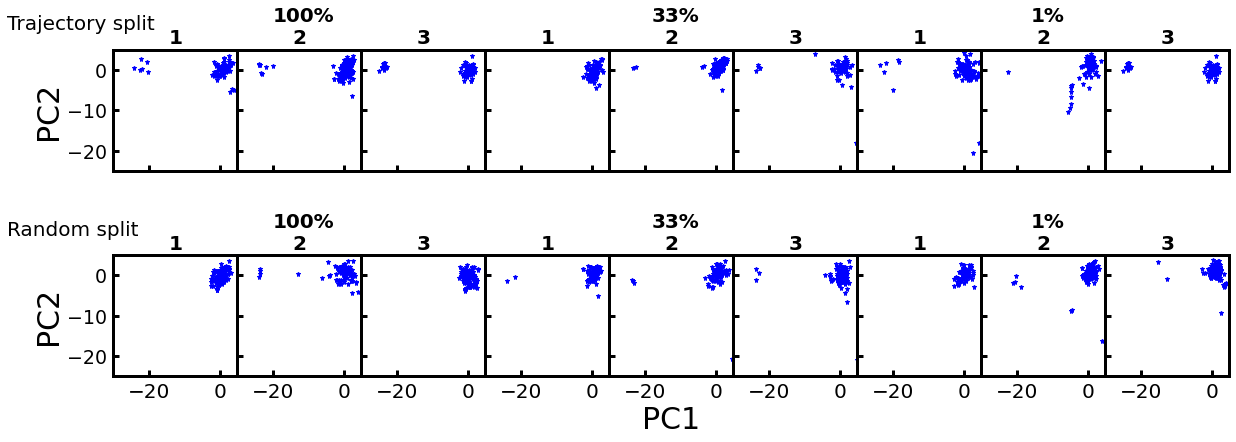}
    \caption{Projections of the hopping geometries from all ML/MM simulations onto PC1 and PC2 obtained from the QM/MM data. 
    The upper row shows the geometries obtained from trajectories generated with the split by trajectory models starting with 100~\% of the data followed by 33~\% and 1~\%.
    The second row shows the same for the models trained on random splits.}
    \label{fig:AllS21Hoppings}
\end{figure}

Figure~\ref{fig:AllS1S0Hoppings} shows the hopping geometries between S$_2$ and S$_1$ for all ML/MM simulations projected into the first and second principal components obtained from the PCA of the reference QM/MM hopping geometries between the same states.
The majority of geometries form a cluster centered at (0/0).
In general, we observe that the larger the disagreement between the model and the QM/MM dynamics with respect to population curves and lifetimes, the more points deviate from this cluster and spread to lower PC1 values.
Since the transition from the S$_2$ to the S$_1$ happens shortly after the excitation in the QM/MM simulations, furan cannot be very distorted in the hopping frames.
Therefore, the slower the relaxation from the S$_2$ to the S$_1$ in the ML/MM dynamics, the more time there is for structural changes and the greater the deviation from the cluster at (0/0) we observe.
Although this is a general trend, the absence of "outliers" does not mean that all hops occurred at the beginning of the simulation.

\end{document}